\documentclass[12pt,preprint]{aastex}
\usepackage{amsmath}
\usepackage{txfonts}
\usepackage{color}
\usepackage{fancybox}

\newcommand{\mr}[1]{\ensuremath{\mathrm{#1}}}
\newcommand{\ux}[2]{\mr{\ensuremath{^{#1} #2}}}
\newcommand{\pen}{\mr{p(e^{-},\nu_{e})n}}
\newcommand{\nep}{\mr{n(e^{+},\overline{\nu_{e}})p}}

\received{2010 June}
\begin{document}

\title{Trends in $^{44}$Ti and $^{56}$Ni from Core-Collapse Supernovae}

\author{
Georgios Magkotsios\altaffilmark{1,2,5},
Francis X. Timmes\altaffilmark{2,5},
Aimee L. Hungerford\altaffilmark{3,4},
Christopher L. Fryer\altaffilmark{3,4},
Patrick A. Young\altaffilmark{2},
Michael Wiescher\altaffilmark{1,5}
       }
\altaffiltext{1}{University of Notre Dame,
                 Department of Physics,
                 Notre Dame, IN 46556}
\altaffiltext{2}{School of Earth and Space Exploration,
                 Arizona State University,
                 Tempe, AZ  85287}
\altaffiltext{3}{Los Alamos National Laboratory,
                 Los Alamos, NM 87545}
\altaffiltext{4}{University of Arizona,
                 Physics Department,
                 Tucson, AZ 85721}
\altaffiltext{5}{The Joint Institute for Nuclear Astrophysics}

\begin{abstract}

We compare the yields of $^{44}$Ti and $^{56}$Ni produced from
post-processing the thermodynamic trajectories from three
dif\-ferent core-collapse models -- a Cassiopeia A progenitor, a
double shock hypernova progenitor, and a rotating 2D explosion --
with the yields from exponential and power-law trajectories. The
peak temperatures and densities achieved in these core-collapse
models span several of the distinct nucleosynthesis regions we
identify, resulting in dif\-ferent trends in the $^{44}$Ti and
$^{56}$Ni yields for dif\-ferent mass elements. The $^{44}$Ti and
$^{56}$Ni mass fraction profiles from the exponential and power-law
profiles generally explain the tendencies of the post-processed
yields, depending on which regions are traversed by the model. We
find integrated yields of $^{44}$Ti and $^{56}$Ni from the
exponential and power-law trajectories are generally within a factor
2 or less of the post-process yields. We also analyze the influence
of specific nuclear reactions on the $^{44}$Ti and $^{56}$Ni
abundance evolution. Reactions that af\-fect all yields globally are
the $3\alpha$, \pen\ and \nep. The rest of the reactions are ranked
according to their degree of impact on the synthesis of \ux{44}{Ti}.
The primary ones include \mr{\ux{44}{Ti}(\alpha,p)\ux{47}{V}},
\mr{\ux{40}{Ca}(\alpha,\gamma)\ux{44}{Ti}},
\mr{\ux{45}{V}(p,\gamma)\ux{46}{Cr}},
\mr{\ux{40}{Ca}(\alpha,p)\ux{43}{Sc}},
\mr{\ux{17}{F}(\alpha,p)\ux{20}{Ne}},
\mr{\ux{21}{Na}(\alpha,p)\ux{24}{Mg}},
\mr{\ux{41}{Sc}(p,\gamma)\ux{42}{Ti}},
\mr{\ux{43}{Sc}(p,\gamma)\ux{44}{Ti}},
\mr{\ux{44}{Ti}(p,\gamma)\ux{45}{V}}, and
\mr{\ux{57}{Ni}(p,\gamma)\ux{58}{Cu}}, along with numerous weak
reactions. Our analysis suggests that not all $^{44}$Ti need be
produced in an $\alpha$-rich freeze-out in core-collapse events, and
that reaction rate equilibria in combination with timescale
ef\-fects for the expansion profile may account for the paucity of
\ux{44}{Ti} observed in supernovae remnants.

\end{abstract}


\section{Introduction}
\label{sec:intro}

Core-collapse supernovae inject energy and material enriched freshly
synthesized isotopes into the interstellar medium. Some of this
material is short to medium lived radioactivities with half-lives
ranging from several days to several million years.  Detecting
$\gamma$-rays from the decay chains of such isotopes, either in
individual supernova remnants or through the accumulation of material
in interstellar medium, provides a direct calibration of the
nucleosynthesis in core-collapse events.  For example, the radioactive
decay $^{44}$Ti and $^{56}$Ni has significant observational
consequences for the light curves of core-collapse supernovae
\citep{arnett_1989_aa,timmes_1996_ab,vink_2001_aa,hungerford_2005_ab,renaud_2006_aa,
  young_2006_aa}, isotopic patterns measured in primitive meteorites
\citep{wadhwa_2007_aa} and presolar grains \citep{zinner_1998_aa},
anomalies in a deep-sea crust \citep{knie_2004_aa}, and the solar
abundances of $^{44}$Ca and $^{56}$Fe \citep{lodders_2003_aa}.

The past decade has brought substantial progress to the theory of
core-collapse supernovae.  There now seems to be general agreement
that hydrodynamic instabilities above the proto-neutron star play a
crucial role in, not only achieving an explosion, but also in
determining critical properties such as the timing, strength and
asymmetry of this explosion \citep{buras_2006_aa,bruenn_2006_aa,
kifonidis_2006_aa,fryer_2007_aa,messer_2008_aa,ott_2008_aa,lunardini_2008_aa}.
For explosion scenarios where the growth of these instabilities is
suf\-ficiently long, neutrino transport through this region seems
capable of resetting the electron fraction $Y_e$ of at least some
material from being neutron-rich to being proton-rich
\citep{pruet_2005_aa,pruet_2006_ab,buras_2006_aa,frohlich_2006_aa}.
While the details are sensitive to the numerical techniques and
physical approximations employed in dif\-ferent simulations, the
range of explosion strengths and timings obtained imply significant
variations in the evolutions of the temperature, density and $Y_e$
in the tumultuous inner regions.

Observations of $^{44}$Ti and $^{56}$Ni in individual core-collapse
supernova may provide the best probes for constraining aspects of the
explosion mechanism precisely because the production of these two
isotopes are sensitive to the temperature, density and $Y_e$
evolution.  Perhaps most compelling are abundance determinations of
the Cassiopeia A remnant from Compton Gamma Ray Observatory, BeppoSAX,
INTEGRAL, and Chandra measurements.  The inferred ratio of $^{44}$Ti
to $^{56}$Ni in Cas A is higher than that predicted by standard,
spherical supernova explosion models
\citep{young_2006_aa,young_2007_aa}. The solar abundance ratio of
$^{44}$Ca to $^{56}$Fe is similar to Cas A's ratio of $^{44}$Ti to
$^{56}$Ni, suggesting that spherical models are simply falling short
in their synthesis of $^{44}$Ti.  Of course, it can be argued that Cas
A was simply a peculiar event \citep{the_2006_aa}.

Multi-dimensional ef\-fects may play some role in resolving this
discrepancy \citep[e.g.,][]{arnett_2008_aa}.  Explosions with
artificially imparted asymmetries in 2D were modeled by
\citet{nagataki_1997_aa} to show that bipolar explosion scenarios
could account for enhanced $^{44}$Ti synthesis along the poles of
model supernova explosions.  Simulations of core-collapse and
hypernovae, where high energies and large asymmetries are imparted
to launch the explosion, can reproduce the trends in the abundances
of metal poor stars and imply larger masses of $^{44}$Ti are ejected
\citep{tominaga_2007_aa,umeda_2008_aa}.  If ef\-fects from
asymmetries are important for setting the nucleosynthesis of
$^{44}$Ti and $^{56}$Ni, then quantifying the physical and numerical
uncertainties which determine those asymmetries becomes important.

However, multi-dimensional explosion simulations are resource
intensive, and thus run primarily to address hydrodynamic and
transport aspects and uncertainties of supernovae. Such models have
not yet been run to assess the sensitivity of isotopic yields
\citep{young_2007_aa} to the nuclear physics input.  Parameterized
expansion profiles bypass these dif\-ficulties by simplifying the
hydrodynamics in favor of focusing on nucleosynthesis.  A motivation
for this paper is to begin the process of examining the interplay
between these two modes of analysis. Thus, in this paper we focus on
the production of $^{44}$Ti and $^{56}$Ni from classic adiabatic
freeze-out thermodynamic trajectories, power-law thermodynamic
trajectories suggested by 2D explosion models, and core-collapse
supernova models. We explore in detail the sensitivity of the
$^{44}$Ti and $^{56}$Ni produced to variations in the reaction
rates, electron fraction, and nuclear network size with the simple
thermodynamic trajectories.  We assess how yields determined from
the simple thermodynamic trajectories compare to the post-process
yields from complex simulations of core-collapse supernovae. This
assessment of\-fers a calibration of where simple trajectories
provide a reasonable approximation to the final yields, and allows
discovery of which regions in the explosion models deviate from the
simple trajectories and why they dif\-fer. Previous ef\-forts along
these lines explored the sensitivity of $^{44}$Ti synthesis to the
assumed reaction rates or the electron fraction
\citep{woosley_1973_aa,woosley_1992_aa,the_1998_aa,hoffman_2010_aa}.
In this paper we study the sensitivity of $^{44}$Ti and $^{56}$Ni
synthesis for both dependencies over an extended parameter space.

In \S\ref{s.si_burn} we briefly discuss equilibrium states and in
\S\ref{s.thermo_profiles} we present the exponential and power-law
thermodynamic trajectories to be interrogated. Section
\ref{s.thermo_trends} considers general trends of $^{44}$Ti and
$^{56}$Ni from these trajectories in the peak temperature-density
plane. We also show where in this plane multi-dimensional models of
asymmetric supernovae and hypernovae tend to reside.  In
\S\ref{s.ti44_nucleo} we discuss the nucleosynthesis of $^{44}$Ti
and $^{56}$Ni in material with dif\-ferent $Y_e$, while the
sensitivities to reaction rate values and network size are discussed
in \S\ref{s.rate_sensitivity} and \S\ref{s.network_size}
respectively. Section \ref{s.postprocessed_yields} describes the
yields of $^{44}$Ti and $^{56}$Ni from post-processing core-collapse
trajectories, compared with the yields from the parameterized
profiles. We conclude with a summary of our main results in
\S\ref{s.summary}.

We establish our nomenclature and conventions.  Let isotope $i$ have
$Z_i$ protons, $A_i$ nucleons (protons + neutrons), and an atomic
weight $W_i$.  We shall assume $W_i$ = $A_i$.  Let the aggregate
total of isotope $i$ have a baryon number density $n_i$ (in
cm$^{-3}$) in material with a temperature $T$ (in K) and a baryon
mass density $\rho$ (in g cm$^{-3}$). Define the dimensionless mass
fraction of isotope $i$ as $X_i = \rho_i/\rho = n_i A_i/(\rho N_A$),
where $N_A$ is the Avogadro's number, and the molar fraction of
isotope $i$ as $Y_i = X_i / A_i$.  The electron fraction, or more
properly, the total proton to nucleon ratio is $Y_{e} =
\sum_{i}Z_{i}Y_{i} = \sum_{i}{Z_{i}}/{A_{i}}\ X_{i}$.  We define
``nuclear flow'' to mean the instantaneous rate of change of isotope
$i$'s molar abundance with time, $dY_{i}/dt$, due to a given nuclear
reaction \citep{iliadis_2007_aa}.  For any single reaction linking
isotope $i$ with isotope $j$ there is a forward flow, a reverse
flow, and a relative net flow $\phi_{i}$=(forward $-$
reverse)/max(forward,reverse) that measures the equilibrium state of
the reaction.

\section{Silicon burning and equilibrium states}
\label{s.si_burn}

Silicon burning is the last exothermic burning stage and produces
the Fe-peak nuclei. Due to Coulomb repulsion, it is rather
improbable that two \ux{28}{Si} nuclei will fuse to \ux{56}{Ni}.
Instead, a photodisintegration driven rearrangement of the
abundances takes place, originating from equilibria established
among individual reactions with their reverses
\citep{bodansky_1968_aa}.  When such equilibria happen among many
reactions, the plasma reaches an equilibrium state where nuclei
merge into clusters. Units of interaction are no longer nuclei, but
the clusters themselves, which adapt their properties according to
the local thermodynamic conditions. In general, not all reactions
are in equilibrium. Consequently, this state is named quasi-static
equilibrium (henceforth QSE). The special case where all strong and
electromagnetic reactions are balanced by their reverses is called
nuclear statistical equilibrium (henceforth NSE), because all mass
fractions may be described in terms of statistical properties of
excited nuclear states (partition functions) and nuclear structure
variables (masses and $Q$ values). Weak interactions are always
excluded from these definitions, since for conditions relevant to
hadronic physics they never attain equilibrium.  Hence, equilibrium
notions are related only with strong and electromagnetic
interactions. In practice,there is either one cluster in NSE or QSE,
or two QSE clusters, one for the Si-group and one for the Fe-group
nuclei.

An NSE state may be completely described by a triplet of
macroscopic parameters such as temperature, density and electron
fraction $Y_{e}$. A QSE state requires additional parameters, one
for each equilibrium cluster, which may be chosen to be the number
of nuclei in each cluster \citep{meyer_1998_aa,
wallerstein_1997_aa}.  The mass fractions of nuclei in such
equilibria states are completely described as functions of these
parameters. A benefit from this property, is that the choice of
initial composition has no impact on the equilibrium state, as
long as it remains consistent with the equilibria parameters.
Thus, if an equilibrium state is established, the details how the
plasma attained that equilibrium are not necessary to model
aspects of the continuing evolution. This feature is the basis for
reliable results from parameterized expansion profiles, whose
starting point is the moment the explosion shock strikes the inner
stellar layers.

Reaction cross sections have, in general, an asymptotic trend
towards a saturation value at high energies. Reactions tend to get
balanced by their reverses in this regime. A reaction rate is
connected to its reverse according to the detailed balance theorem
\citep{iliadis_2007_aa}. The dominant term in this relationship is
$\exp(-Q/kT)$, implying that the reaction $Q$ value and the
temperature are the foremost magnitudes related with the ability of
a reaction to reach and maintain equilibrium. There is a linear
density dependence to this relationship only for reactions involving
photons. Large $Q$ values result in sensitive reaction equilibria,
which are the first to break for decreasing temperature. However,
these reactions become the most ef\-ficient flow carriers once they
break equilibrium, since they release the largest amounts of energy
per reaction. Figure \ref{fig:Q_value_charts} shows the $Q$ value
distribution for alpha particle captures within our base network
containing 489 isotopes, which includes all reactions that may
directly af\-fect \ux{44}{Ti}. Table \ref{tab:nuclear_networks}
gives a complete specification of all networks used in this study.

\begin{figure}[htp]
\includegraphics[width=0.9\textwidth]{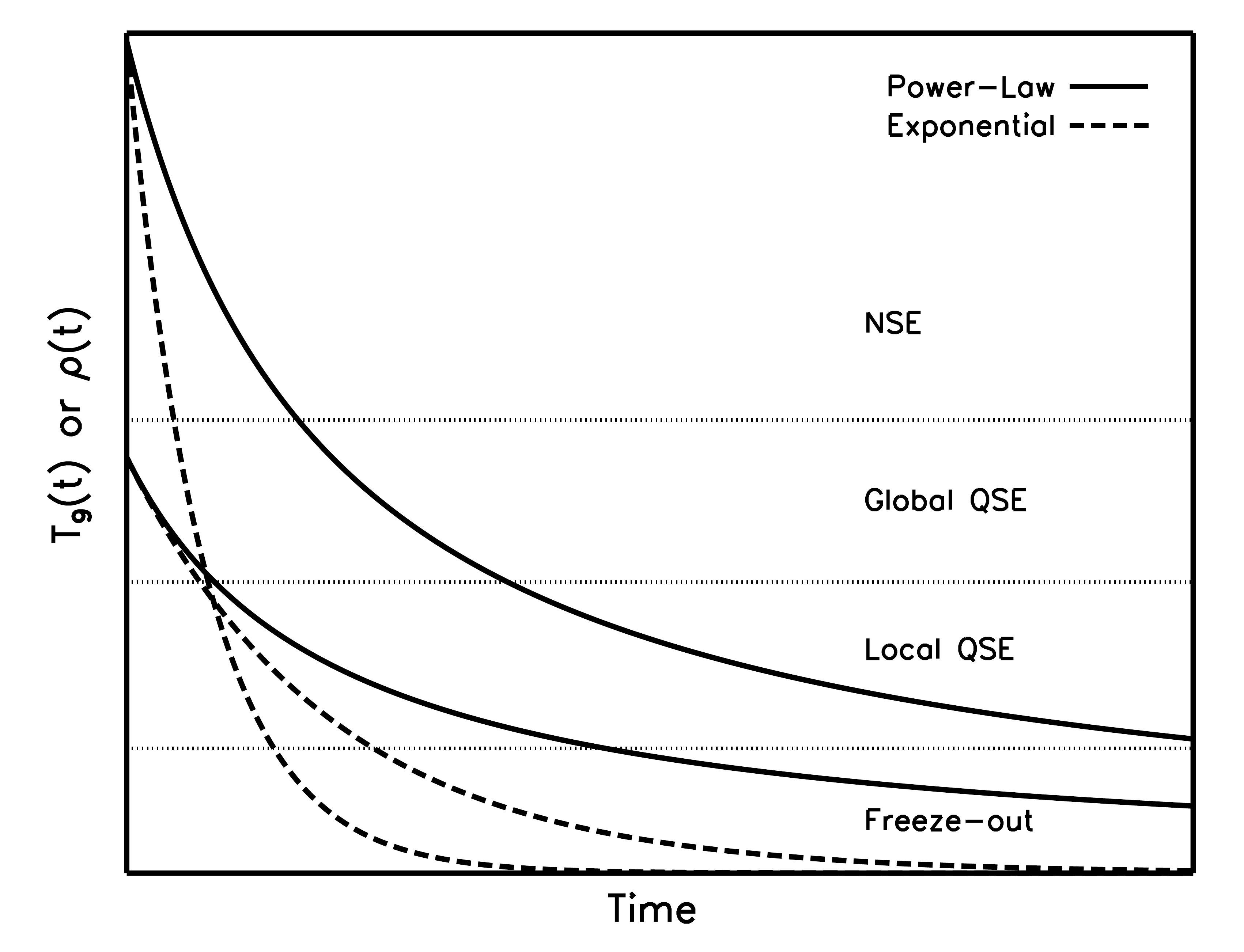}
\caption{Schematic temperature or density evolution for the
exponential (dashed) and power-law (solid) profiles. For either of
the two initial values illustrated, the exponential profile declines
faster than the power-law profile. Passages through dif\-ferent
burning regimes are indicated.} \label{fig:regimes_cartoon}
\end{figure}

The temperature expresses the internal energies of the nuclei, while
the density is related to their availability for reactions. As a
result, the larger the temperatures and densities are, the more
equilibria exist. This ef\-fect allows NSE to be established for
large temperatures and densities. For smaller temperatures and
densities, certain equilibria start breaking, but a large scale QSE
structure in the plasma may still exist. For yet lower thermodynamic
conditions, the large scale clusters dissolve into smaller clusters.
Very low temperatures and densities are not adequate to establish a
significant amount of equilibria. A few isolated equilibria may
exist, but without any specific connection between them. Depending
on the initial peak temperature and density, the plasma may
experience one or more of these states during the expansion of the
ejecta (see Figure \ref{fig:regimes_cartoon}). Unfortunately, the
threshold conditions to border each regime cannot be known
accurately, since they are sensitive to the number of species
involved (network size), the $Q$ values of the associated reactions
and the reaction rates used. Such borders exist in nature though,
and the time spent by the plasma in each regime may af\-fect the
final yields.

External flow supply to reactions may also result in equilibria
breaks, even for constant temperature and density. When there are no
external abundance flows to reaction $i(j,k)m$, the condition
$\phi_{i}=-\phi_{m}$ means the forward flow for isotope $i$ is the
reverse flow for isotope $m$ and vice versa. Due to this condition,
this reaction may achieve equilibrium, a case where both $\phi_{i}$
and $\phi_{m}$ are equal to zero. Assume now an external abundance
flow, say from another reaction that is not in equilibrium, that
supplies flow only to isotope $i$. As long as the external flow is
significant in magnitude, $|\phi_{i}|\ne|\phi_{m}|$ because the
external flow is additive only to $\phi_{i}$. That is, the external
flow term is applied to the equation for $dY_{i}/dt$ but not to the
equation for $dY_{m}/dt$, and the reaction $i(j,k)m$ breaks
equilibrium. If the external abundance flow is applied long enough,
a new equilibrium state may be established. Turning of\-f the
external flow causes the reaction to be driven back to equilibrium
with a new abundance ratio, depending on the flow transfer by the
external agent. In general, starting from an equilibrium state, a
transition to any other state means some $|\phi|$'s must be greater
than other $|\phi|$'s during the transition. External flows may also
af\-fect the equilibrium state of small clusters of nuclei, or even
large scale QSE clusters (see \S\ref{s.ti44_nucleo}).

\section{Parameterized Thermodynamic Trajectories}
\label{s.thermo_profiles}

We use two parameterized expansion profiles to identify robust
trends and uncertainties in the yield of $^{44}$Ti and $^{56}$Ni.
Both profiles assume that a passing shock wave heats material to a
peak temperature $T_{0}$ and compresses the material to a peak
density $\rho_{0}$.  This material then expands and cools down
(freezes out) under the assumption of a constant $T^3/\rho$
evolution (radiation entropy in suitable limits) until the
temperature and density are reduced to the extent that nuclear
reactions cease. Our adiabatic freeze-out trajectories
\citep{hoyle_1964_aa,fowler_1964_aa}
\begin{equation}
\frac{dT}{dt}    = -\frac{T}{3\tau} \quad
\frac{d\rho}{dt} = - \frac{\rho}{\tau}
\label{eq:ad_ode}
\end{equation}
\begin{equation}
T(t)=T_{0} \exp(-t/3 \tau) \quad
\rho(t) = \rho_{0} \exp(-t/\tau)
\label{eq:Tad_rhoad}
\end{equation}
are used with a static free-fall timescale for the expanding ejecta
\begin{equation}
\tau=(24\pi G\rho_{0})^{-1/2} \approx 446/\rho_{0}^{1/2} \ {\rm s}
\label{eq:AD_timescale}
\end{equation}
In this formulation the temperature and density evolutions are
decoupled.  If one uses $\rho(t)$ in the expansion timescale instead
of the peak density $\rho_0$, then the temperature ordinary
dif\-ferential equation becomes coupled to the density evolution.

The second thermodynamic profile we use is based on homologous
expansion.  For a fixed expansion velocity $\upsilon$, the distances
increase as $R(t)=R_{0}+\upsilon t$, the density scales as
$\rho(t)\sim 1/R(t)^{3}=1/(R_{0}+\upsilon t)^{3}$ and the temperature
scales through $\rho \varpropto T^{3}$.  Specifically we use
\begin{equation}
\frac{dT}{dt} = \frac{-T_{0}}{1/2(2t+1)^{2}} \quad
\frac{d\rho}{dt} = \frac{-3\rho_{0}}{1/2(2t+1)^{4}}
\label{eq:Tpl_ode}
\end{equation}
\begin{equation}
T(t)=\frac{T_{0}}{2t+1} \quad
\rho(t)=\frac{\rho_{0}}{(2t+1)^{3}}
\enskip ,
\label{eq:Tpl_rhopl}
\end{equation}
where the coef\-ficients in the denominator are chosen to mimic
trajectories taken from core-collapse simulations. Substituting the
power-law solution into the ordinary dif\-ferential equations they
originate from and eliminating the direct time dependence
\begin{equation}
\frac{dT}{dt}    = -\frac{T_0}{0.5 (\rho/\rho_0)^{2/3}} \quad
\frac{d\rho}{dt} = -\frac{3 \rho_0}{0.5 (T/T_0)^4}
\label{eq:Tpl_coupling}
\end{equation}
shows the temperature and density evolutions are coupled for the
power-law trajectories.

Figure \ref{fig:regimes_cartoon} compares the general properties of
these two parameterized profiles.  For a given initial condition,
the power-law evolution is always slower than the exponential one.
Moreover, the power-law evolution becomes slower for increasing
initial values.  The dif\-ferences in these two profiles af\-fect
the final yields of $^{44}$Ti and $^{56}$Ni as material traverses
dif\-ferent burning regimes on dif\-ferent timescales. The figure
also depicts the NSE, global QSE, local QSE, and final freeze-out
burning regimes. The exponential and power-law trajectories are
chosen so that they bound in general the temperature and density
trajectories of particles from the $^{44}$Ti and $^{56}$Ni producing
regions of spherically symmetric and 2D explosion models.

For any given peak temperature, peak density, and initial electron
fraction $Y_e$ we want to know the mass fraction of $^{44}$Ti and
$^{56}$Ni produced by nuclear burning from the exponential and
power-law profiles.  We chose peak temperatures, peak densities, and
$Y_e$ values spanning the range of $4\times 10^9 \leqslant T_{0}
\leqslant 10 \times 10^9$ K, $10^{4} \leqslant \rho_0 \leqslant
10^{10}$ g cm$^{-3}$, and $0.48 \leqslant Y_e \leqslant 0.52$.  This
parameter space covers the conditions encountered in most
core-collapse supernova models which produce some $^{44}$Ti or
$^{56}$Ni.  When sampling this parameter space between these limits
we use 121 points, equally spaced in base 10 logarithm, for the peak
temperature or density and increments of 0.002 in $Y_e$. That is,
for any value of $Y_e$ we compute the final nucleosynthesis at
121x121 points in the peak temperature-density plane using mature
reaction networks \citep{timmes_1999_ab,fryxell_2000_aa}. Using a
larger number of sample points does not alter our main results and
conclusions.  Our initial composition for any starting ($T_{0}$,
$\rho_0$, $Y_e$) triplet is pure \ux{28}{Si} for symmetric matter
($Y_{e}$=0.5).  We then added protons or neutrons to make initial
composition either proton or neutron-rich respectively.
Specifically, we used X($^{28}$Si) = 1 - $|$2 $Y_e$ - 1$|$ and
either X(p) = $|$2 $Y_e$ - 1$|$ for proton-rich compositions ($Y_{e}
> 0.5$) or X(n) = $|$2 $Y_e$ - 1$|$ for neutron-rich compositions
($Y_{e} < 0.5$) to set the initial $^{28}$Si, proton or neutron mass
fractions. As we show below in \S\ref{s.ti44_nucleo}, the choice of
$^{28}$Si is not important for vast regions of the chosen parameter
space.

\section{Trends in the Peak Temperature-Density Plane}
\label{s.thermo_trends}

Figure \ref{fig:contour_ti44_ni56_AD1_PL2_ye0500_sph} shows the mass
fractions of $^{44}$Ti, $^{56}$Ni and $^{4}$He produced in the peak
temperature-density plane for the exponential and power-law profiles
and an initial $Y_e$=0.5.  Each point in the plane represents one
set of initial conditions, which are evolved forward in time
according to Equations \ref{eq:ad_ode} or \ref{eq:Tpl_ode}, with the
final freeze-out abundance of $^{44}$Ti and $^{56}$Ni recorded.  The
color map is logarithmic, spanning mass fractions from 10$^{-2}$ to
10$^{-10}$ for $^{44}$Ti and from 1 to 10$^{-10}$ for $^{56}$Ni. The
overlaid colored triangles correspond to the temperature and density
of particles from a suite of supernovae and collapsar simulations in
the region where $^{44}$Ti and $^{56}$Ni are produced.  Not all
particles have an initial $Y_e$=0.5, but are relatively close to it.
Each supernova model generally spans the full range of peak
temperature, but the peak density is confined to a strip of one or
two orders of magnitude.

Several striking patterns emerge from these contour plots.  The
first is \ux{44}{Ti} seems to be produced overall with an average
mass fraction \mr{X(\ux{44}{Ti})\sim 10^{-5}}, except in certain
regions where it gets depleted.  The depletion region extends along
a thin line for low temperatures (oriented approximately
$70^{\circ}$ with respect to the temperature axis), smoothly bending
over into a wider, more horizontal band for relatively high
temperatures and densities. We name this depletion region the
``chasm''. The chasm separates the peak temperature-density plane
into distinct regions controlled by dif\-ferent burning processes.

The second pattern is that the $^{44}$Ti contour plots for the two
thermodynamic profiles have the same general structure, except that
the chasm for the power-law profile is located at lower densities
and slightly wider compared to the exponential profile.  Hence, the
power-law chasm begins to encompass the majority of the overlaid
particles. It is possible that the existence of the chasm region is
why so few supernova remnants have been observed in the glow of
radioactive $^{44}$Ti. The total mass of \ux{44}{Ti} ejected by an
individual core-collapse supernova depends critically on (i) the
location of its thermodynamic points in the peak temperature-density
plane and (ii) the exact expansion profile that the ejecta follow
past the explosion. The impact of the latter is expressed as the
chasm's ability to ``shift'' and ``widen'' itself from exact profile
to exact profile. The third pattern is $^{56}$Ni has large mass
fractions and is relatively featureless in the peak
temperature-density plane. Large variations in observed $^{44}$Ti to
$^{56}$Ni ratios are primarily due to variations in $^{44}$Ti.

The chasm's formation and trends with thermodynamic history are the
primary motivation for using two parameterized profiles. Our
analysis to uncover the nuclear physics controlling the chasm is
two-fold. First, we ascertain the basic synthesis mechanisms of
\ux{44}{Ti} in distinctive thermodynamic regions through a series of
nuclear reaction network calculations (section \ref{s.ti44_nucleo}).
Second, we identify reactions crucial to \ux{44}{Ti} in each
thermodynamic region via a three-stage process based roughly on the
methodology established by \citet{the_1998_aa}, but modified because
we are interested in more than one normalization point in each peak
temperature-density plane (section \ref{s.rate_sensitivity}).

\section{Nucleosynthesis of $^{44}$Ti and $^{56}$Ni}
\label{s.ti44_nucleo}

\begin{figure}[htp]
\includegraphics[width=0.9\textwidth]{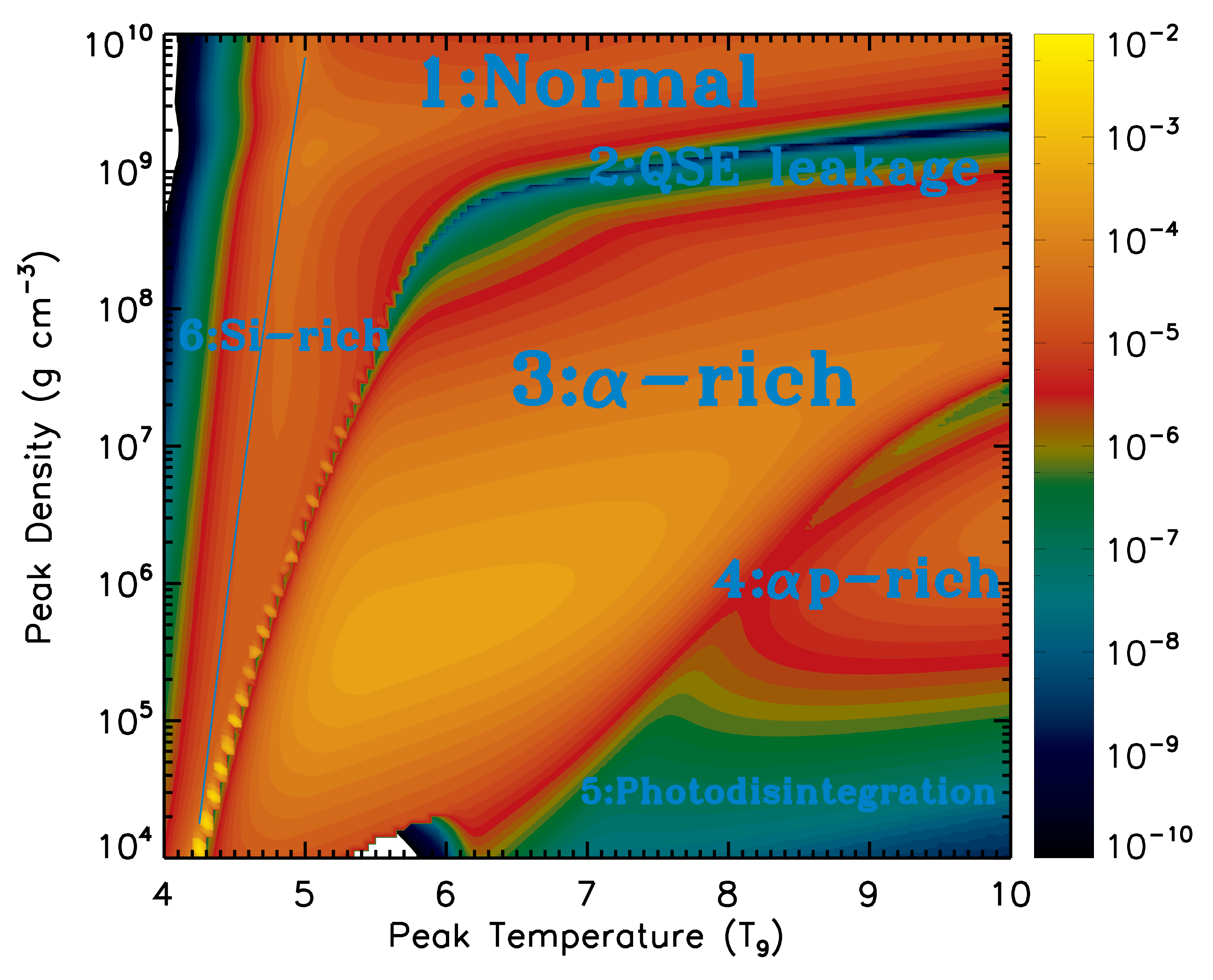}
\caption{ Final mass fraction of $^{44}$Ti in the peak
temperature-density plane for the exponential thermodynamic profile
at $Y_e$=0.5. Six distinct regions of $^{44}$Ti synthesis are
labeled. Region 1: normal freeze-out from NSE, abundance largely
determined from $Q$ values. Region 2: Chasm region, passage from 1
QSE cluster to 2 QSE clusters. Region 3: $\alpha$-rich freeze-out.
Region 4: $\alpha$$p$-rich freeze-out. Region 5: Photodisintegration
regime, neutrons, protons, and $\alpha$ dominate. Region 6:
Incomplete silicon burning, \ux{28}{Si} rich. The thin cyan line is
the locus of points where $\tau_{{\rm QSE}}$ = 0.012 $\tau_{{\rm
freeze}}$. } \label{fig:contour_ti44_AD1_ye0500_regimes}
\end{figure}

Figure \ref{fig:contour_ti44_AD1_ye0500_regimes} shows the mass
fraction of $^{44}$Ti in the peak temperature-density plane for the
exponential profile and $Y_e$=0.5. Each point in this plane
represents one set of initial conditions which evolve forward in
time with the final freeze-out abundance of $^{44}$Ti cataloged. Six
dif\-ferent regions are labeled which are characterized by specific
nuclear burning patterns controlling the production of \ux{44}{Ti}.
Despite the dif\-ferences in timescale, the corresponding
temperature-density contour plot following the power-law profile
contains the same number of regions, with similar physics associated
between regions of the same label. Thus, the duration of the
hydrodynamic expansion does not explain the underlying structure of
the contour plots for the final \ux{44}{Ti} and \ux{56}{Ni} yields.
Instead, the entropy during the expansion drives the
nucleosynthesis, by af\-fecting the strengths of key nuclear
reactions, and causing phase transitions in the burning process for
certain critical temperatures and densities. The phase transitions
are followed by a change in the burning state. On the other hand,
the expansion timescale affects the locus of the borders among
different regions on the contour plot (Figure
\ref{fig:contour_ti44_ni56_AD1_PL2_ye0500_sph}). For increasing
expansion timescale the plasma spends more time within each burning
state. Depending on the region in the peak temperature-density
plane, the evolution may include some or all states between NSE and
non-equilibrium nuclear burning (see Figure
\ref{fig:regimes_cartoon}). Timescale differences between profiles
result in different density values, when the temperature acquires a
threshold value indicative of a phase transition. Since both
profiles attain constant radiation entropy, different densities at
threshold temperatures translate to different peak densities and
hence, border shifts between regions on the temperature-density
plane (e.g. see \S\ref{s.alpha_p} for the explanation of the chasm
shift).

For most of the regions in Figure
\ref{fig:contour_ti44_AD1_ye0500_regimes}, the peak conditions are
suf\-ficiently large that the plasma reaches a large scale
equilibrium state (NSE or QSE) on timescales much shorter than the
freeze-out timescale. During the first time steps of a reaction
network calculation the initial composition rearranges to an NSE or
a QSE distribution well before the temperature and density begin
evolving. This rapid rearrangement appears as vertical line in many
of our plots. As the plasma subsequently cools and rarefies the
first transition occurs when the NSE state can no longer be
maintained. The threshold temperature for NSE is usually taken to be
$T_9 = T/(10^9 \ {\rm K}) \sim 5$.  The density at this threshold
temperature determines the subsequent burning process by prescribing
both the available amount of nuclear fuel and the dominant flows
that consume the fuel.

\begin{figure}[htp]
\includegraphics[height=0.8\textheight]{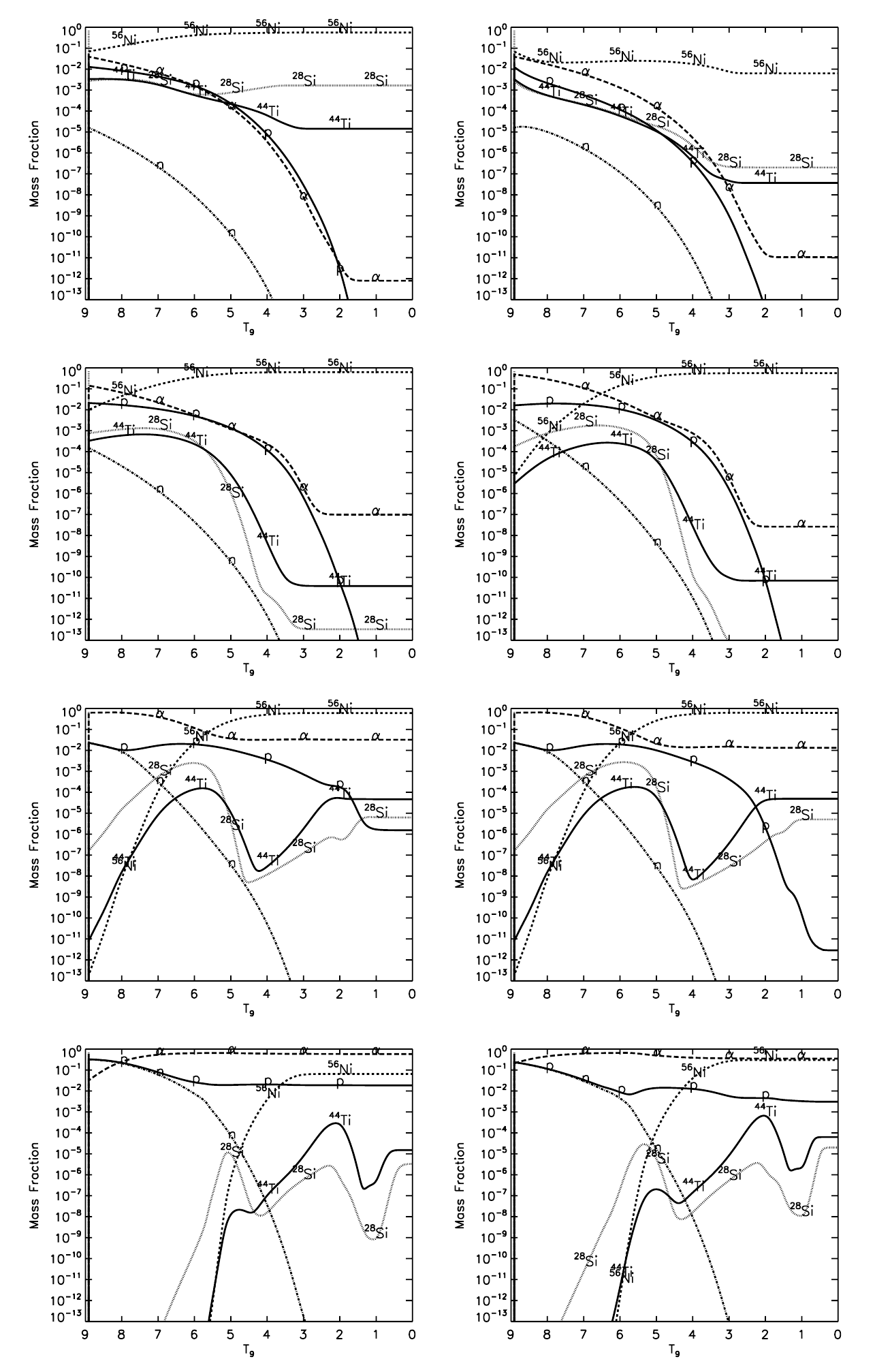}
\caption{ Mass fraction evolution of $^{28}$Si, $^{44}$Ti,
$^{56}$Ni,  neutrons, protons, and $\alpha$-particles for the
exponential (left column) and power-law (right column) expansions at
$Y_e$=0.5. In each panel the peak temperature is 9$\times$10$^9$ K.
The top row corresponds to region 1 in Figure
\ref{fig:contour_ti44_AD1_ye0500_regimes} ($\rho=5\times10^9$ g
cm$^{-3}$), second row to region 2 ($\rho=1.75\times10^9$ g
cm$^{-3}$), third row to region 3 ($\rho=1\times10^8$ g cm$^{-3}$),
and fourth row to region 4 ($\rho=1\times10^6$ g cm$^{-3}$). }
\label{fig:mass_fractions_AD1_PL2_temp9}
\end{figure}

Region 1 is essentially a freeze-out from NSE, henceforth termed a
``normal freeze-out''
\citep{woosley_1973_aa,meyer_1994_aa,meyer_1998_aa,hix_1999_aa}.
When the temperature falls to the $T_9=5$ threshold temperature, the
density is $\sim$1.0$\times$10$^{9}$ g cm$^{-3}$ for the high peak
density region above the horizontal band of the chasm. At this
density an NSE distribution is dominated by \ux{56}{Ni}, contains a
significant amount of Si-group and Fe-group nuclei, but a relatively
small amount of free alpha particles (\mr{X_{\alpha}\sim10^{-3}}).
This density is large enough to favor particle captures, but the
temperature is such that photodisintegration reactions are not
negligible either. The large scale equilibrium structure is
maintained until complete freeze-out, since \mr{X_{\alpha}\ll1} for
the majority of equilibrium states and the 3$\alpha$ reaction is
always dominated by its inverse photodisintegration. Since normal
freeze-out is a dynamic process though, some individual equilibria
are broken as the plasma cools and rarefies and QSE estimates become
progressively more accurate compared to NSE estimates
\citep{woosley_1973_aa}. Yields for the isotopes plotted in Figure
\ref{fig:mass_fractions_AD1_PL2_temp9} for region 1 (first row of
plots) are not far from NSE or QSE yields. Thus, network
calculations in this region may be avoided and accurate estimates
for yields may be determined only by nuclear properties such as
masses and $Q$ values.

Equilibrium estimates for Si-group and Fe-group nuclei during the
initial stages of the expansion remain relevant for region 2. When
the temperature falls to the $T_9=5$ threshold temperature the
density is $\sim$1.0$\times$10$^{8}$ g cm$^{-3}$. The low
availability of alpha particles at this density does not allow the
3$\alpha$ reaction to dominate its inverse, preventing significant
flow from the light nuclei to the equilibrium cluster. Compared to
region 1 though, not all of the capture reactions have the same
ef\-ficiency. The large $\alpha$ thresholds in nuclei between N,Z=20
and N,Z=26 closed shell configurations results in a phase transition
which is responsible for the formation of the chasm. Because of the
large $Q$ values associated with $\alpha$ capture in the mass range
$42 \leqslant A \leqslant 53$ due to shell structure (Figure
\ref{fig:Q_value_charts}), these reactions are the first to break
the local equilibria and form a continuous passage of nuclear flow
from the Si-Ca-group to the Fe-group nuclei. The large equilibrium
cluster dissolves into two smaller ones, with \ux{44}{Ti} being
located within the upper mass limits of the Si-Ca cluster, while
\ux{56}{Ni} is centralized in the Fe-group. The flow transfer
between the two equilibrium clusters results in the depletion of
\ux{44}{Ti} and the rest of the isotopes in the Si-group by the end
of the thermodynamic evolution (second row of Figure
\ref{fig:mass_fractions_AD1_PL2_temp9}). On the contrary,
\ux{56}{Ni} is one of the Fe-group isotopes that benefit from this
transfer since the reaction equilibria in its neighborhood are
maintained until freeze-out. Equilibria estimates for this small
group of nuclei within the Fe-group are still a good approximation.
The formation of the chasm in Figure
\ref{fig:contour_ti44_AD1_ye0500_regimes} is a direct result of a
phase transition from the single cluster QSE configuration to a
double cluster QSE configuration and the subsequent flow leakage.

Region 3 corresponds to the conditions of $\alpha$-rich freeze-out
\citep{woosley_1973_aa}. As the plasma cools and rarefies, most
Si-group and Fe-group mass fractions acquire the topology of an
``arc'' in going from low values at high temperatures to a local
maximum and back to a local minimum at cooler temperatures, while in
QSE (third row of Figure \ref{fig:mass_fractions_AD1_PL2_temp9}).
The density at the $T_9$=5 threshold temperature within region 3
spans $10^{4}\lesssim\rho\lesssim 10^{7}$ g cm$^{-3}$, resulting in
less ef\-ficient particle captures compared to regions 1 and 2, and
a helium mass fraction \mr{X_{\alpha}\sim10^{-1}}. The excess of
free alpha particles allows the $3\alpha$ rate to dominate its
inverse photodisintegration, leading to a new phase transition.
Although the $3\alpha$ rate remains relatively slow
\citep{the_1998_aa}, it supplies external flow which breaks the
local equilibria in the neighborhood of \ux{12}{C}, \ux{16}{O} and
\ux{24}{Mg}. The subsequent energy release from alpha capture
reactions provides a significant nuclear flow towards heavier nuclei
by breaking successively other local equilibria. The QSE cluster
changes its shape and shifts gradually upwards in mass, instead of
dissolving into two clusters. \citet{meyer_1998_aa} identified this
cluster motion based on QSE calculations. The mass fractions of
nuclei which suddenly find themselves outside the QSE cluster begin
an ascending track. These are primarily the Si-group nuclei
(including \ux{44}{Ti}), and a few from the Fe-group. Close to
complete freeze-out, the yields for these nuclei are orders of
magnitude larger than their corresponding minimum value reached
prior to the phase transition. Because the $3\alpha$ reaction itself
is not very ef\-ficient, the process ends up with an excess of alpha
particles \mr{X_{\alpha}\sim10^{-2}}.

\begin{figure}[htp]
\includegraphics[width=0.8\textwidth]{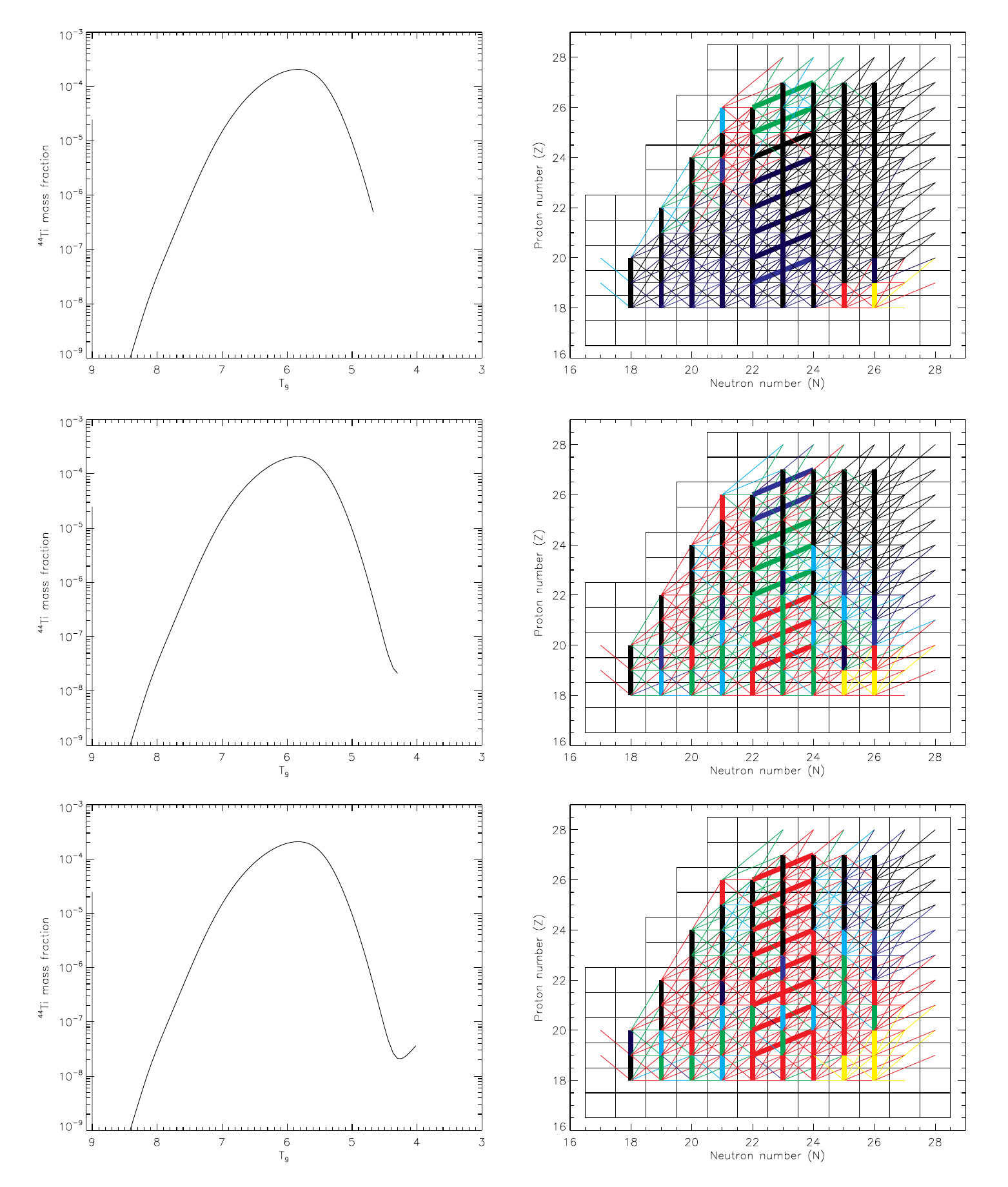}
\caption{Normalized nuclear flows at three dif\-ferent points in the
evolution of $^{44}$Ti from freeze-out; before the minimum (top
panel), at the minimum (middle panel), and after the minimum (bottom
panel). Normalized flows $\phi$ are colored black for $0 \le \phi <
0.01$, navy for $0.01 \le \phi < 0.05$, blue for   $0.05 \le \phi <
0.1$, cyan for $0.1 \le \phi < 0.4$, green for  $0.4 \le \phi <
0.8$, red for $0.8 \le \phi < 1.0$, yellow for $\phi$=1.0.
Normalized flows in the vertical direction, corresponding to
$(p,\gamma)$ reactions, and along diagonal directions corresponding
to $(\alpha,p)$ reactions have been drawn thicker for clarity. }
\label{fig:ti44_local_cluster_chart}
\end{figure}

Large scale QSE calculations cannot account for the increase of the
\ux{44}{Ti} mass fraction curve past the arc, since \ux{44}{Ti} and
other related isotopes have decoupled from the large scale
equilibrium cluster. However, the phase transition is not abrupt in
shifting from total equilibrium to total non-equilibrium. Nuclear
flow analysis shows that \ux{44}{Ti} participates in a smaller,
local equilibrium pattern which is responsible for its ascending
trend in the mass fraction curve at the end of freeze-out. This
transition is demonstrated in Figure
\ref{fig:ti44_local_cluster_chart} which displays the reaction links
between f$_{7/2}$-shell nuclei located between the Z,N=20 and Z,N=28
closed shells. The top panel in Figure
\ref{fig:ti44_local_cluster_chart} shows the network of reactions
prior to the $^{44}$Ti abundance minimum, which is characterized by
\mr{(p,\gamma)} equilibria along the N=22 and N=24 isotone chains
connected by \mr{(\alpha,p)} and \mr{(\alpha,\gamma)} channels in
equilibrium. These equilibria guarantee linkage of \ux{44}{Ti} to
the large scale QSE cluster, and hence, the downward portion of the
mass fraction curve is produced. The breakdown of the equilibrium
conditions for the \mr{\ux{44}{Ti}(\alpha,p)\ux{47}{V}} link signals
the phase transition for \ux{44}{Ti}. Soon afterward, the rest of
the \mr{(\alpha,p)} and \mr{(\alpha,\gamma)} equilibria connecting
the N=22 and N=24 isotone chains break, as reflected in the increase
of actual net flow shown in the middle panel in Figure
\ref{fig:ti44_local_cluster_chart}. \ux{44}{Ti} is left in
\mr{(p,\gamma)} equilibria along the N=22 isotone chain with
$^{45}$V, $^{46}$Cr, $^{47}$Mn and $^{48}$Fe and its mass fraction
starts to increase from the local minimum (bottom panel of Figure
\ref{fig:ti44_local_cluster_chart}). It is this \mr{(p,\gamma)}
equilibria chain which is responsible for the rising portion of the
mass fraction curve after the local minimum. This pattern persists
until complete freeze-out.

\begin{figure}[htp]
\includegraphics[width=0.9\textwidth]{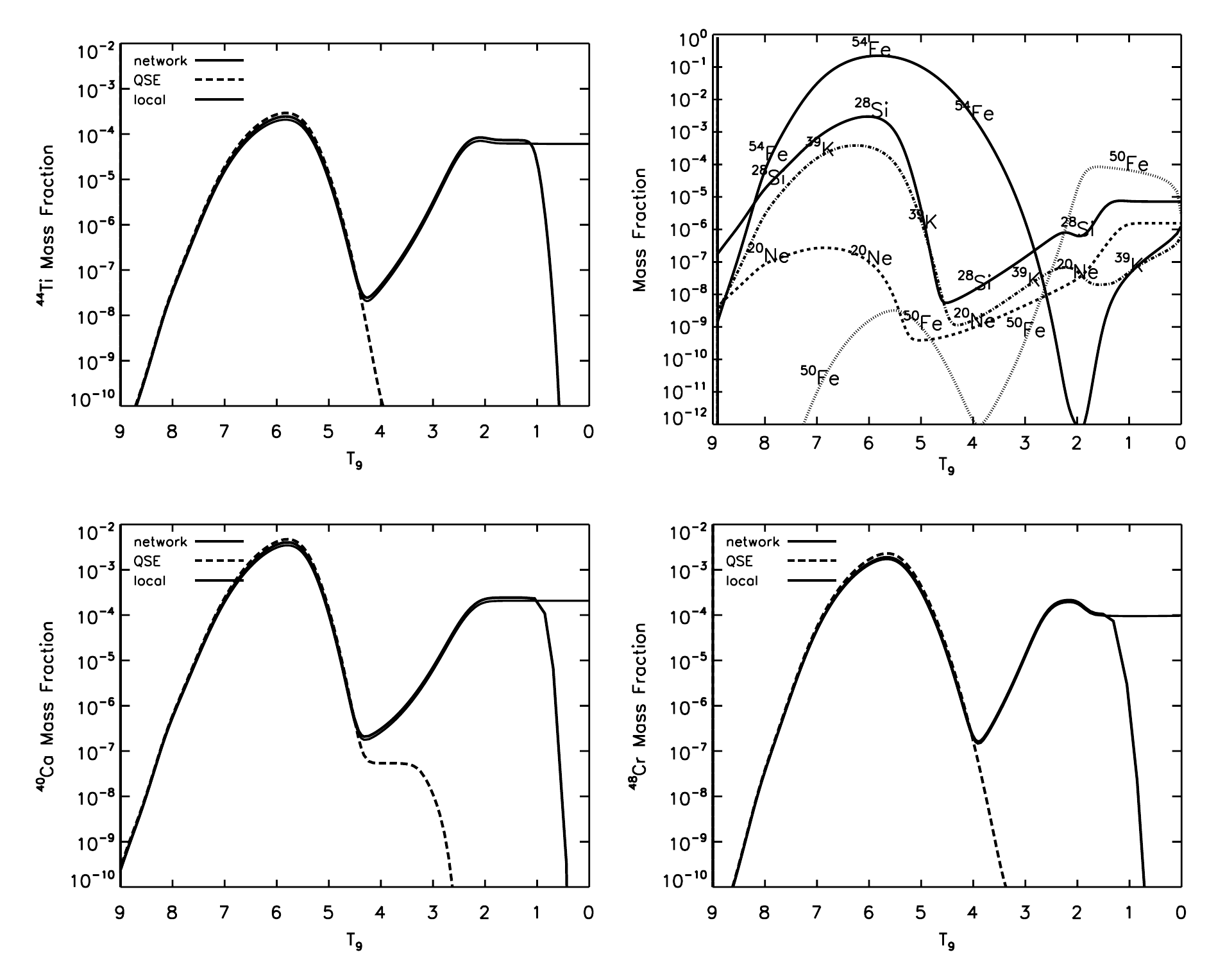}
\caption{ Reaction network (solid curves) and QSE solutions (dashed
and dot-dashed curves) for $^{44}$Ti (upper left), $^{40}$Ca (lower
left), and $^{48}$Cr (lower right). Each element shows the
characteristic arc driven by equilibrium $(\alpha,p)$ and
$(p,\gamma)$ reactions. Beginning at the local minima and onwards to
lower temperatures only $(p,\gamma)$ reactions remain in
equilibrium, which drive the mass fractions to larger values. This
general behavior applies to most of the elements within the silicon
and iron groups (small sample upper right). }
\label{fig:qse_vs_net}
\end{figure}

These reaction network flow study results can be verified by
localized QSE calculations. The advantage of QSE modeling is that
the abundances of all isotopes within a cluster may be expressed by
a semi-analytical formalism, in terms of the network abundances of
free protons, neutrons and an arbitrarily chosen reference isotope.
For this purpose, we adopt the \citet{hix_1996_aa,hix_1999_aa}
formalism. We model the cases of equilibrium (i) between the N=22
and N=24 isotone chains and (ii) only along the N=22 isotone chain
throughout the evolution, corresponding to the top and bottom panels
in Figure \ref{fig:ti44_local_cluster_chart} respectively. Both
cases reproduce the arc topology of the $^{44}$Ti mass fraction
curve. However, the first case does not reproduce the ascending part
of the curve beyond the local minimum in Figure
\ref{fig:qse_vs_net}. Instead, the $^{44}$Ti curve continues to
descend, expressing the trend of \ux{44}{Ti}, were it to remain in
global QSE. The second case on the other hand, which expresses only
$(p,\gamma)$ equilibria, fits the network results until the point
where complete freeze-out occurs. The discrepancy beyond this point
relies on the fact that nuclear reactions no longer take place.
Thus, mass fractions do not change any more and the curve from
network calculations acquires a plateau. This general behavior
applies to most of the elements within the silicon and iron groups,
as demonstrated for a small subset within these groups at the upper
right panel in Figure \ref{fig:qse_vs_net}. The mass fraction trends
of an isotope depend strongly on the local reaction equilibria
within its neighborhood. Further equilibria isotone chains are
readily identifiable in Figure \ref{fig:ti44_local_cluster_chart}.
For example, crucial equilibria reactions for \ux{40}{Ca} are
\mr{\ux{40}{Ca}(\alpha,\gamma)\ux{44}{Ti}} and
\mr{\ux{43}{Sc}(p,\gamma)\ux{44}{Ti}}. Its mass fraction profile in
the lower left of Figure \ref{fig:qse_vs_net} is in accordance with
the general mechanism. The increase of the \ux{40}{Ca} mass fraction
with cooling is maintained through the \mr{(p,\gamma)} equilibria
along the N=20 isotone chain. A similar development can be observed
for \ux{48}{Cr} as shown in the lower right of Figure
\ref{fig:qse_vs_net}, the crucial reaction now is
\mr{\ux{48}{Cr}(\alpha,p)\ux{51}{Mn}}.

Region 4 is a special case of an $\alpha$-rich freeze-out. Within
this region, the \mr{p(e^{-},\nu_{e})n} and
\mr{n(e^{+},\overline{\nu_{e}})p} reactions exert a greater
influence compared to the other regions.  These reactions drive the
composition slightly proton-rich near the beginning of the evolution
when temperature and density are still large.  Impacts to the
burning processes for proton-rich composition are described in more
detail in \S\ref{s.ye_sensitivity}, but some of the impacts include
a relatively high number of free protons and an enhanced
ef\-ficiency of proton captures
\citep{pruet_2005_aa,pruet_2006_ab,buras_2006_aa,frohlich_2006_aa}.
Consequently, this region is a proton-rich, $\alpha$-rich
freeze-out, henceforth an ``$\alpha$$p$-rich freeze-out''.  The
fourth row of Figure \ref{fig:mass_fractions_AD1_PL2_temp9} shows
the mass fraction profiles for \ux{44}{Ti} in region 4 have certain
similarities to the profiles in region 3.  A characteristic arc of
large scale QSE is formed, followed by the ascending track due to
the equilibrium chain connecting \ux{44}{Ti}, \ux{45}{V},
\ux{46}{Cr}, \ux{47}{Mn} and \ux{48}{Fe} via \mr{(p,\gamma)}
reactions along the N=22 isotone chain.  Among these linking
reactions \mr{\ux{45}{V}(p,\gamma)\ux{46}{Cr}} has the largest $Q$
value, and thus will break its equilibrium first as the plasma cools
and rarefies. When this reaction breaks equilibrium, the N=22
isotone chain dissolves into two smaller clusters, the first between
\ux{44}{Ti} and \ux{45}{V} and a second between \ux{46}{Cr},
\ux{47}{Mn} and \ux{48}{Fe}.  This is the second phase transition
that \ux{44}{Ti} sustains during its evolution. Similar transitions
occur along other isotone chains. Flows are now carried among such
isolated small scale clusters by out of equilibrium alpha and proton
captures. These flows favor mostly the proton-rich nuclei, resulting
in a decrease for \ux{44}{Ti} and other symmetric isotopes. Thus, a
second arc is clearly identifiable in the mass fraction curve for
most of the isotopes in the fourth row of Figure
\ref{fig:mass_fractions_AD1_PL2_temp9}. The ascending track beyond
the second arc for \ux{44}{Ti} and most of the symmetric isotopes is
a consequence of the flow transfer through weak interactions at the
expense of proton-rich nuclei, when the strong and electromagnetic
reactions become inef\-fective as freeze-out takes place.

In region 5 the temperatures are initially large enough to establish
equilibrium (NSE or QSE), but the initial densities are so low that
photodisintegrations soon dominate capture reactions. Long before
the complete freeze-out, all nuclei dissolve into neutrons, protons
and $\alpha$-particles. A slight recombination takes place during
the final stages of the freeze-out producing traces of \ux{12}{C},
\ux{16}{O} and \ux{28}{Si}. The recombination is driven mostly by
the 3$\alpha$, \mr{p(e^{-},\nu_{e})n} and
\mr{n(e^{+},\overline{\nu_{e}})p} reactions in $Y_{e}\geqslant 0.5$
matter. The products of this recombination set the seed for a
following chain of \mr{(p,\gamma)} and \mr{(n,p)} reactions that
produce heavier elements, including \ux{44}{Ti}, \ux{56}{Ni} and the
heaviest isotopes in the network used for the calculation. Similarly
to region 4, weak interactions at the close of the process carry
some flow from proton-rich nuclei to symmetric ones, enhancing this
way the mass fractions of \ux{44}{Ti} and \ux{56}{Ni}. However, the
contributions of the recombination and the chain of \mr{(p,\gamma)}
and \mr{(n,p)} reactions are not adequate to yield large production
factors for most of the isotopes. The final composition is dominated
by free alpha particles and protons, establishing this region to be
a photodisintegration driven regime.

Region 6 represents incomplete silicon burning, where \ux{28}{Si}
gradually dominates \ux{56}{Ni} from region 1 to the left of the
thin chasm line towards the inner part of this region. The peak
temperatures and densities are such that the timescale to reach a
single cluster QSE state is comparable or larger than the expansion
timescale. Multiple small scale QSE clusters are formed, but they do
not merge successfully into one large scale cluster. The mass
fractions freeze out from the established equilibrium state without
sustaining any phase transition. This resembles the mass fraction
trends within region 1, only that the freeze-out within region 6
originates from equilibria states which are sensitive to the number
and shape of clusters formed, and thus from the initial composition
for the burning process. The physical border between regions 3 and 1
is the thin chasm line oriented $\sim 70^{\circ}$ with respect to
the peak temperature axis.  Such a distinctive border does not exist
between regions 1 and 6, due to the lack of a phase transition in
both regions. However, an approximate border is the locus of points
given by $\tau_{{\rm QSE}}$ = 0.012 $\tau_{{\rm freeze}}$, where
$\tau_{{\rm QSE}}$ is the timescale to reach QSE
\citep{calder_2007_aa} and $\tau_{{\rm freeze}}$ is given by
equation \ref{eq:AD_timescale}. This locus is shown by the thin cyan
line in Figure \ref{fig:contour_ti44_AD1_ye0500_regimes}. The
relative dif\-ferences for \ux{44}{Ti} yields starting from pure
\ux{28}{Si} or \ux{12}{C} are less than 0.1 to the right of this
locus.

The case of \ux{56}{Ni} is simpler than \ux{44}{Ti}. The isotope
\ux{56}{Ni} tends to dominate the final composition for the majority
of the peak temperatures and peak densities for $Y_{e}=0.5$. The
topology of its final mass fractions in Figure
\ref{fig:contour_ti44_ni56_AD1_PL2_ye0500_sph} shows \ux{56}{Ni}
does not sustain any phase transitions like \ux{44}{Ti} because
\ux{56}{Ni} remains in equilibrium with its local neighborhood
\citep{woosley_1992_aa}.  While the macroscopic behavior of the
large QSE cluster changes in dif\-ferent regions, there are almost
no changes in \ux{56}{Ni}.

\begin{figure}[htp]
\includegraphics[width=0.8\textwidth]{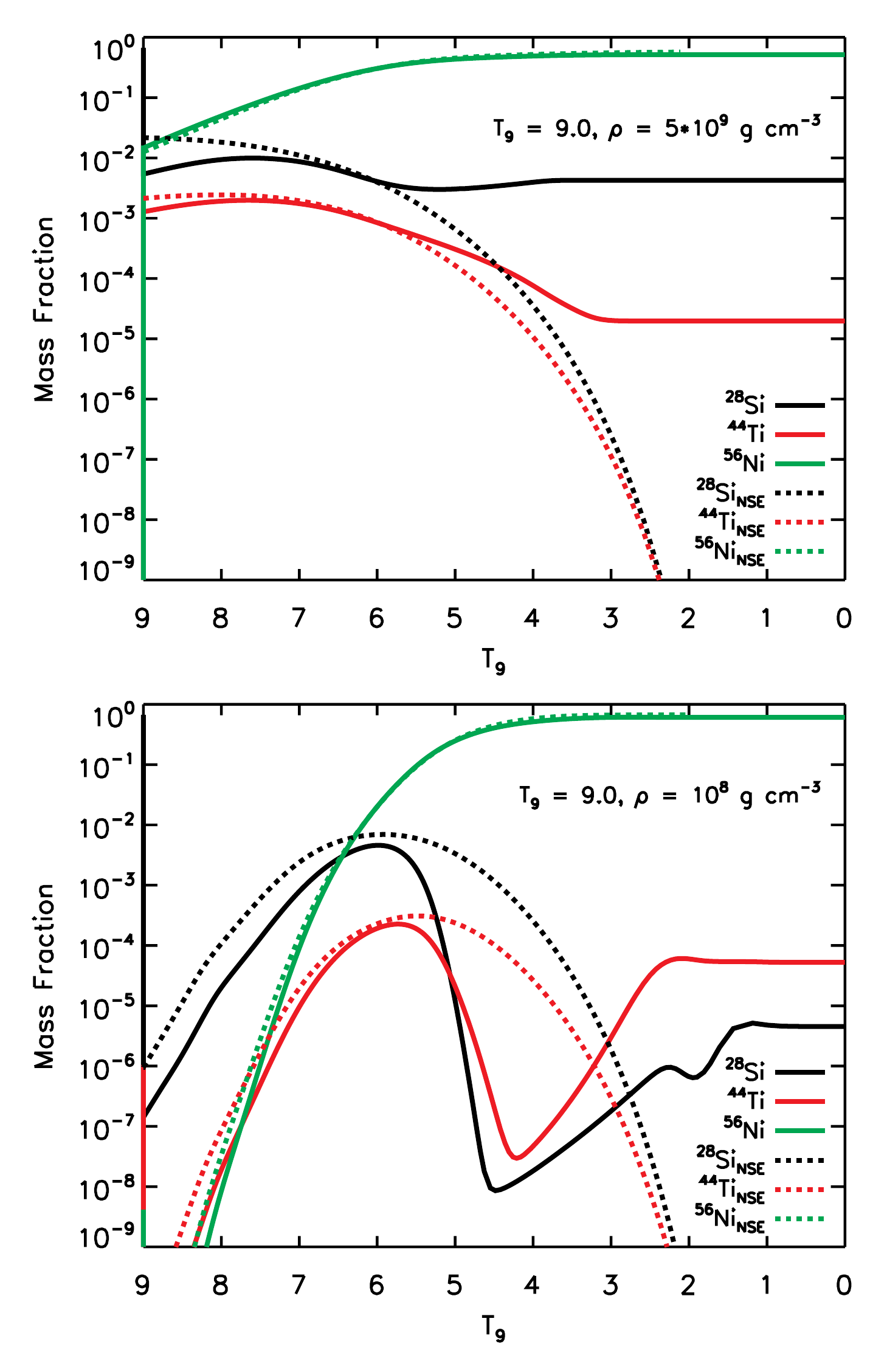}
\caption{ Mass fraction evolution of key isotopes from network and
NSE calculations for peak temperatures and peak densities
corresponding to a normal freeze-out (top panel) and an
$\alpha$-rich freeze-out (bottom panel). } \label{fig:nse_vs_net}
\end{figure}

In region 1, a single QSE cluster which includes $^{56}$Ni stays
intact until freeze-out. In region 2, the QSE cluster dissolves into
two smaller ones, with the cluster localized around the Fe-group
nuclei encompassing \ux{56}{Ni} at all times.  During the
$\alpha$-rich freeze-out of region 3 the QSE cluster shifts upwards
in mass and shrinks \citep{meyer_1998_aa}, but remains centralized
on Fe-group nuclei (including \ux{56}{Ni}). Near the end of the
evolution, the Fe-group nuclei are the most abundant in the network
with reaction equilibria maintained among them. Figure
\ref{fig:nse_vs_net} shows the mass fractions of \ux{28}{Si},
\ux{44}{Ti} and \ux{56}{Ni} for a normal and an $\alpha$-rich
freeze-out, accompanied by the corresponding NSE values for each
isotope were the NSE valid at all times.  For the normal freeze-out,
the network values are in good agreement with the corresponding NSE
values until the NSE threshold of $T_{9}\sim$5. For \ux{56}{Ni}, the
agreement between the network and NSE values persists until at least
$T_{9}\sim$2, at which point our NSE solver fails to converge.
During an $\alpha$-rich freeze-out the network values of \ux{28}{Si}
and \ux{44}{Ti} are quite different from their corresponding assumed
NSE values, while the NSE mass fraction of \ux{56}{Ni} still agrees
with reaction network values until $T_{9}\sim$2. Of course NSE at
$T_{9}\sim$2 does not exist, but the trends in Figure
\ref{fig:nse_vs_net} suggest that \ux{56}{Ni} may be considered to
be in a large scale equilibrium throughout the evolution for almost
every region on the temperature-density plane. That is, global
equilibrium estimates may interpret adequately the dominant trend of
\ux{56}{Ni} for an initially symmetric composition.

\subsection{Electron fraction sensitivity study}
\label{s.ye_sensitivity}

The electron fraction, or the total proton to nucleon ratio, $Y_{e}
= \sum_{i}Z_{i}Y_{i} = \sum_{i}\frac{Z_{i}}{A_{i}}X_{i}$ , is
equivalent to a weighted average of isotopic proton to nucleon
ratios, where each has a probability equal to $X_{i}$. Since the
distribution of isotopic ratios in a large network may be
approximated as continuous, the most abundant isotopes at any given
time in the thermodynamic evolution are generally the ones whose
individual proton to nucleon ratio is within a small range from the
current value of the electron fraction.  A small spread usually
exists due to nuclear structure ef\-fects for equilibrium states
(expressed primarily with $Q$ values), and reaction rate values for
non-equilibrium states \citep{arnett_1977_aa}.  The largest (major)
nuclear flows tend to be localized along the most abundant nuclei,
since the flows depend on multiplications of abundances.  During NSE
or QSE the major flows result in the most robust reaction
equilibria, while the same reactions typically become the most
ef\-ficient carriers of nuclear flow as soon as they depart from
equilibrium.  Almost all our $Y_e$ sensitivity results may be
explained by these guidelines for the major flows. An exception
exists for cases with initial $Y_{e}>0.5$ during large scale
equilibrium (NSE and QSE), where the equilibrium patterns are
configured according to a dif\-ferent principle
\citep{seitenzahl_2008_aa}.  Electron fraction variations alter the
nuclear composition and af\-fect the yields, the nucleosynthesis
mechanisms for each region in the peak temperature-density plane,
and change the regions topology.

\begin{figure}[htp]
\includegraphics[width=1.0\textwidth]{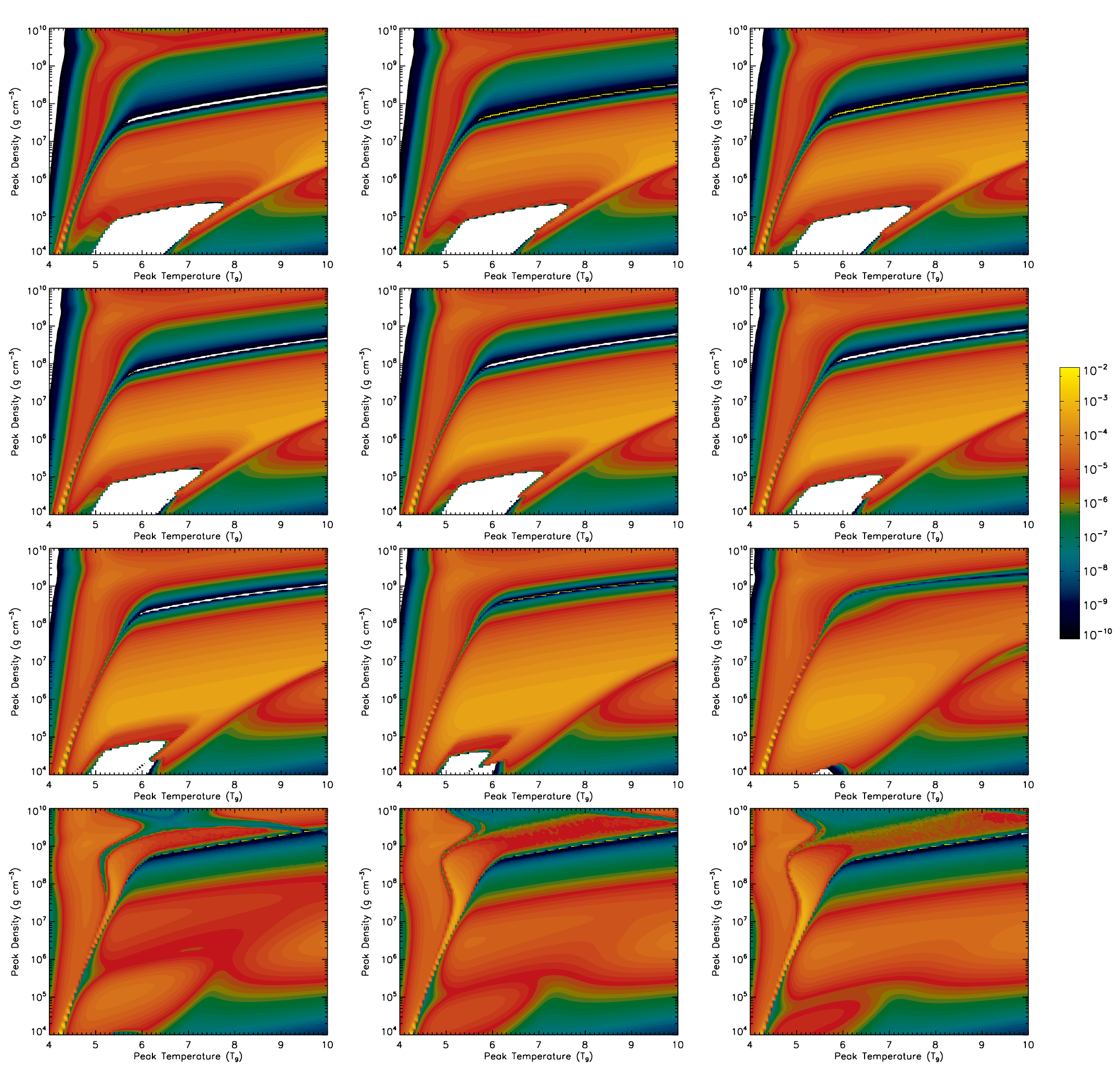}
\caption{ Final yield of $^{44}$Ti in the peak temperature-density
plane for dif\-ferent values of the initial electron fraction $Y_e$
under the exponential freeze-out profile. The top row, from left to
right, corresponds to $Y_e$ = 0.484, 0.486, and 0.488. The second
row, from left to right, corresponds to $Y_e$ = 0.490, 0.492, and
0.494. The third row, from left to right, corresponds to $Y_e$ =
0.496, 0.498, and 0.500. The bottom row, from left to right,
corresponds to $Y_e$ = 0.502, 0.504, and 0.506.
        }
\label{fig:contour_ti44_AD1_span}
\end{figure}

\begin{figure}[htp]
\includegraphics[width=1.0\textwidth]{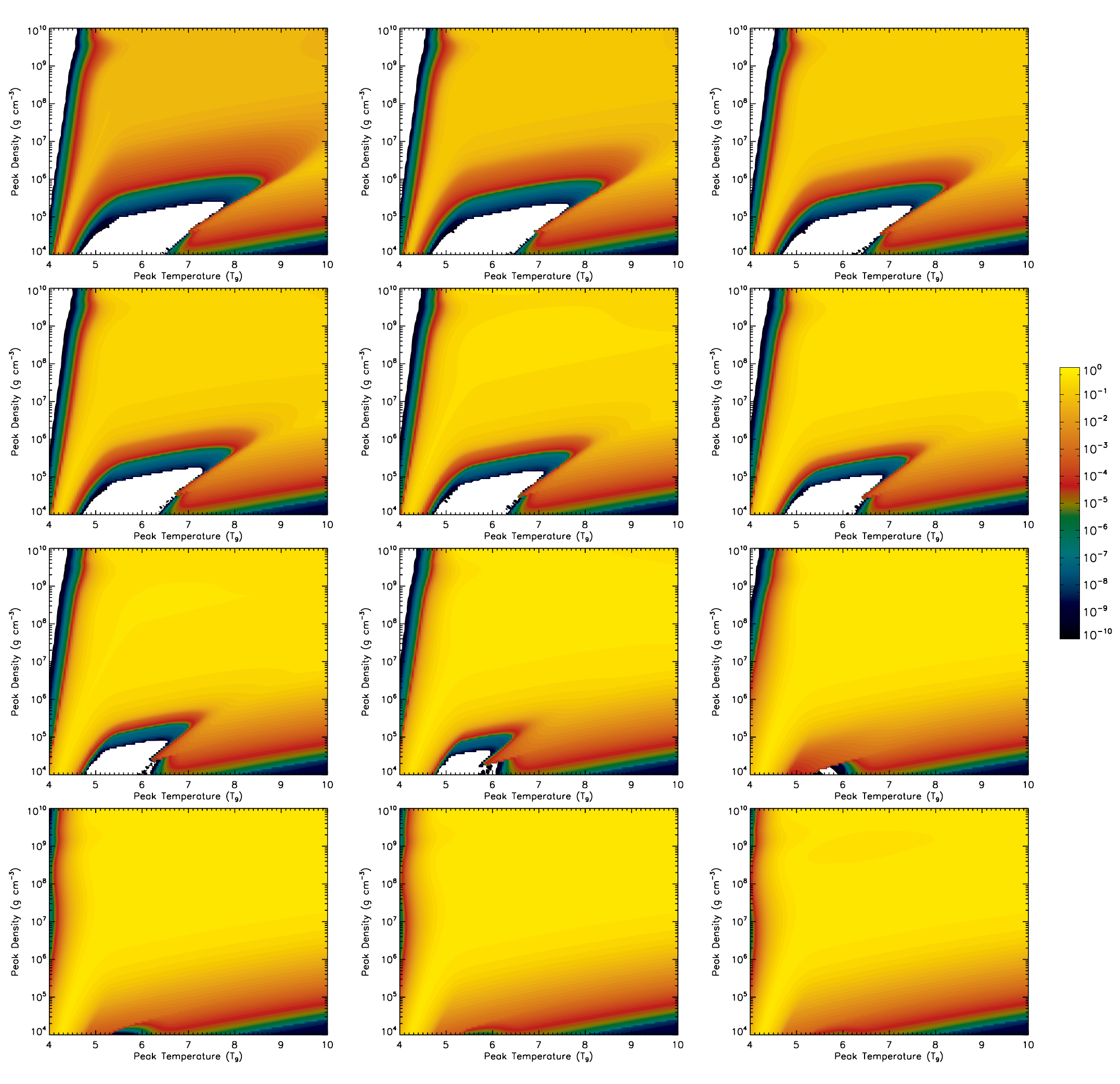}
\caption{ Same as Figure \ref{fig:contour_ti44_AD1_span}, but for
the final yield of $^{56}$Ni.
        }
\label{fig:contour_ni56_AD1_span}
\end{figure}

Figures \ref{fig:contour_ti44_AD1_span} and
\ref{fig:contour_ni56_AD1_span} show the final yields of $^{44}$Ti
and $^{56}$Ni, respectively, in the peak temperature-density plane
for $0.484\leqslant Y_{e}\leqslant 0.506$ under the exponential
freeze-out profile. Figure \ref{fig:contour_P44_AD1_span} shows the
$^{44}$Ti production factor P$_{44}$ for the same range of $Y_{e}$.
The production factor for a given species is defined as the final
mass fraction of the species in question divided by the mass
fraction to which it decays in the Sun. These production factors are
then normalized to the production factor of $^{56}$Fe
\citep{woosley_1991_aa,hoffman_2010_aa}. Within the electron
fraction values 0.498 and 0.5, yields for both isotopes are
maximized, resulting to the minimization of the chasm's width and
depth for \ux{44}{Ti}. For decreasing $Y_{e}$ values, both isotopes
tend to be under-produced compared to the symmetric case
\citep{woosley_1992_aa}. For increasing $Y_{e}$ values, \ux{56}{Ni}
is still favored by equilibria schemes and is produced at an amount
comparable to the symmetric case \citep{magkotsios_2008_aa}. The
temperature-density planes for \ux{56}{Ni} have similar featureless
structure to the corresponding plane for initially symmetric matter,
implying that this isotope is produced only by equilibria schemes
without sustaining any phase transition. The location of the border
between the regions of $\alpha$-rich and $\alpha$$p$-rich freeze-out
for \ux{44}{Ti} depends on the initial electron fraction value. The
lack of free protons for neutron-rich environments favors the
$\alpha$-rich freeze-out versus the $\alpha$$p$-rich one, until the
$\alpha$$p$-rich freeze-out is not manifested at all for $Y_e
\approx 0.46$. The situation is gradually reverted for increasing
$Y_{e}$, until the $\alpha$$p$-rich freeze-out dominates the
$\alpha$-rich freeze-out for $Y_e \approx 0.506$. These trends are
in accordance with the major flows guidelines discussed above, since
both isotopes are symmetric with an individual proton to nucleon
ratio equal to 0.5, and the amount of free protons increases
significantly for $Y_{e}>0.5$ \citep{seitenzahl_2008_aa}. Further
changes to the topological structure compared to the symmetric case
include the appearance of a depletion region for both \ux{44}{Ti}
and \ux{56}{Ni} for decreasing $Y_{e}$, and the appearance of a
physical border between regions 1 and 6 for increasing $Y_{e}$. This
physical border implies a new type of phase transition that
\ux{44}{Ti} sustains. These trends are the same for the power-law
freeze-out profile (not shown).

\begin{figure}[htp]
\includegraphics[width=1.0\textwidth]{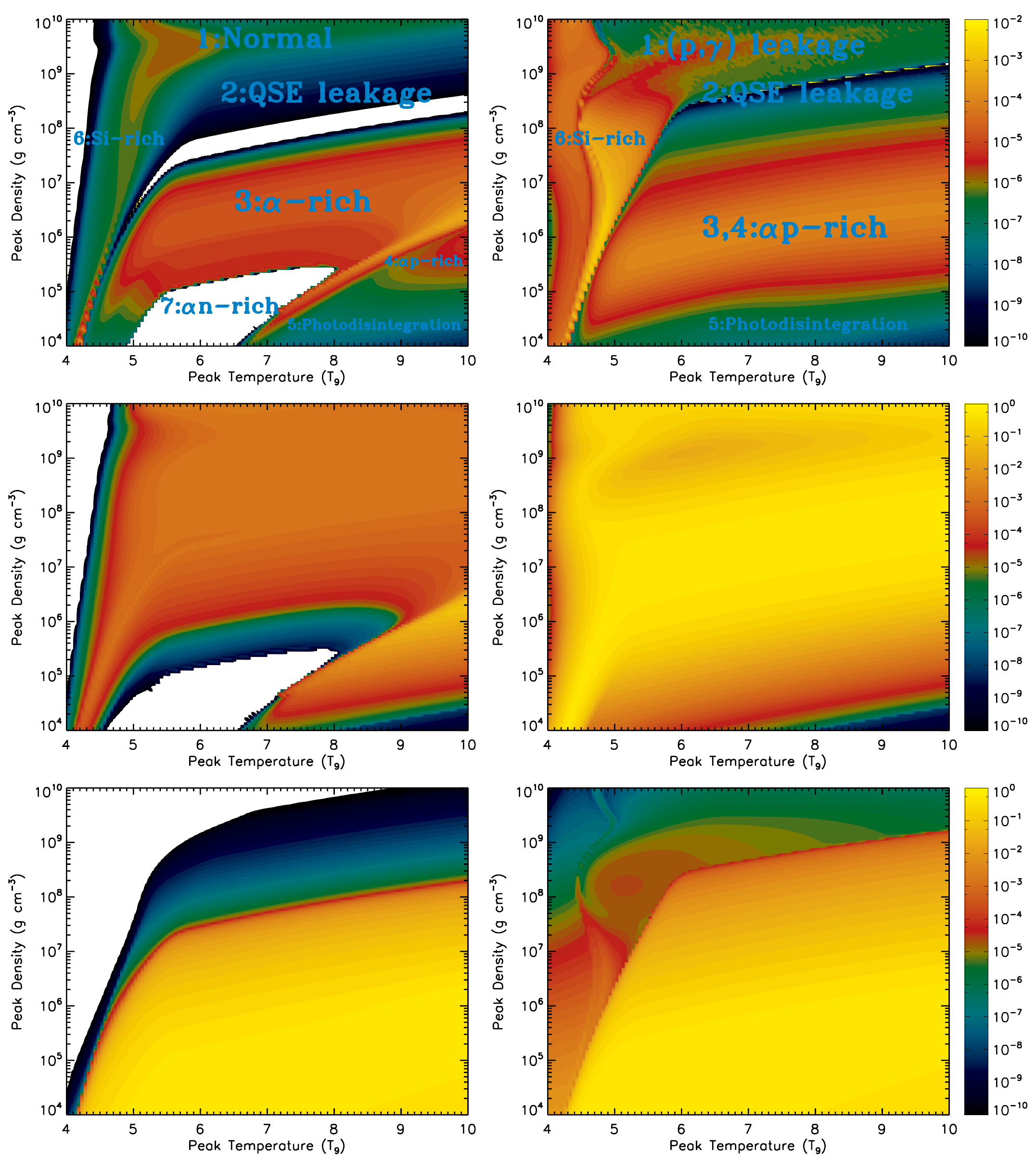}
\caption{ Final mass fraction of $^{44}$Ti (first row), $^{56}$Ni
(second row) and $^{4}$He (third row) in the peak
temperature-density plane for the exponential thermodynamic profile
at $Y_e$=0.48 (left column) and $Y_e$=0.52 (right column). The
distinct regions of $^{44}$Ti synthesis are labeled. Region 1:
normal freeze-out from NSE (left), (p,$\gamma$) leakage from
symmetric to proton-rich nuclei (right). Region 2: Chasm region,
passage from 1 QSE cluster to 2 QSE clusters. Region 3:
$\alpha$-rich freeze-out. Region 4: $\alpha$$p$-rich freeze-out.
Region 5: Photodisintegration regime, neutrons, protons, and
$\alpha$ dominate. Region 6: Incomplete silicon burning, \ux{28}{Si}
rich. Region 7: $\alpha$$n$-rich freeze-out. }
\label{fig:contour_ti44_ni56_AD1_ye0480_ye0520_regimes}
\end{figure}

We focus next on two relatively extreme $Y_{e}$ values, 0.48 for the
neutron-rich case and 0.52 for the proton-rich one. Both values are
adequately far from the standard value of symmetric matter, so that
the dif\-ferences in the trends for $^{44}$Ti and $^{56}$Ni are
emphasized and easily identified. The characteristic regions of the
\ux{44}{Ti} temperature-density planes with initial electron
fraction $Y_e$=0.48 and $Y_e$=0.52 are labeled in Figure
\ref{fig:contour_ti44_ni56_AD1_ye0480_ye0520_regimes} for the
exponential profile. Similarly to Figure
\ref{fig:contour_ti44_AD1_ye0500_regimes}, the only region sensitive
to the initial composition is region 6, the incomplete Si-burning
regime.

\begin{figure}[htp]
\includegraphics[width=0.9\textwidth]{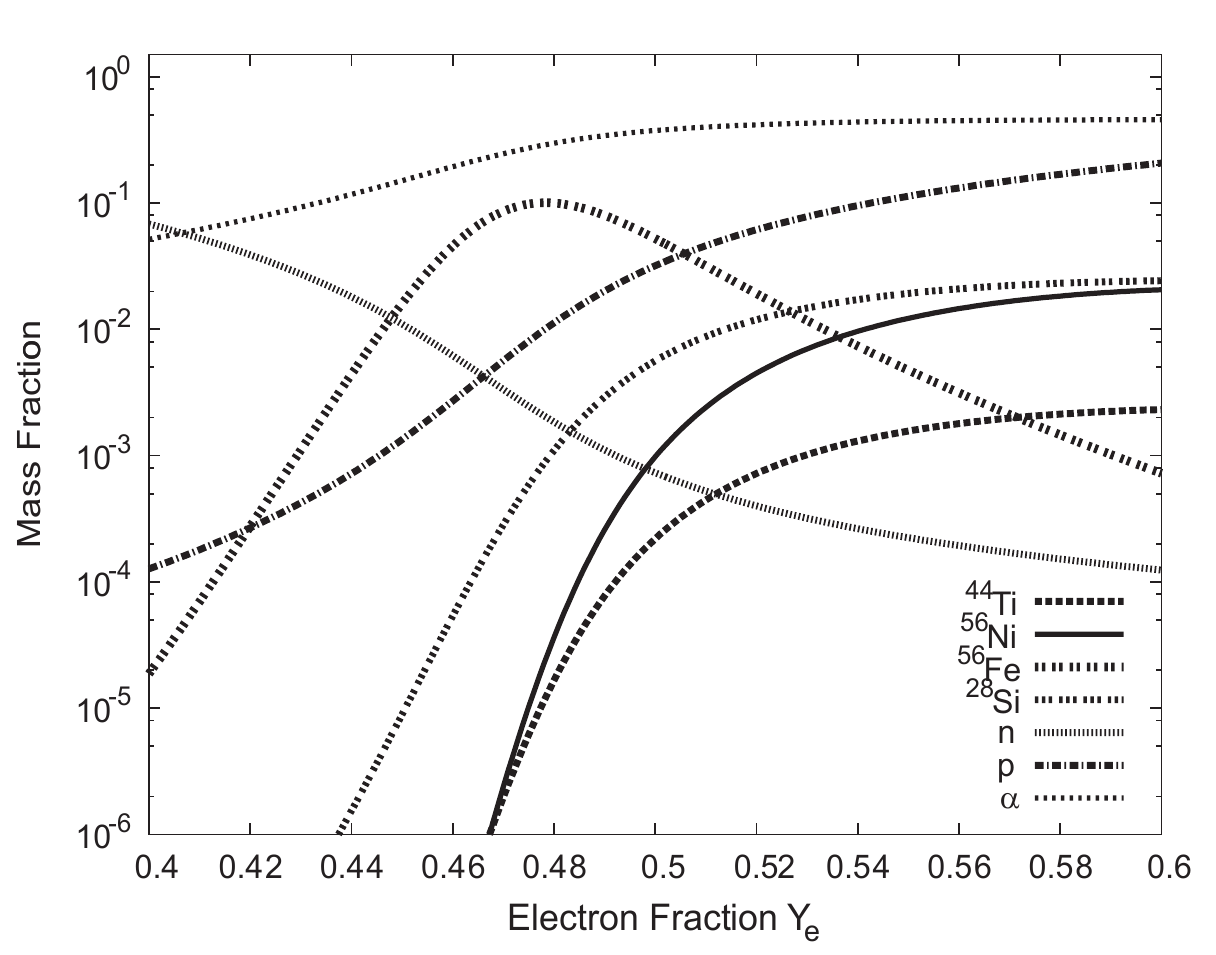}
\caption{ Mass fraction of select isotopes in NSE at $T_{9}=9$ and
$\rho=10^{9}$ g cm$^{-3}$ for dif\-ferent electron fractions. }
\label{fig:nse_rho1e10_temp1e10}
\end{figure}

For $Y_{e}$=0.48, NSE and QSE favor the formation of nuclei with a
proton to nucleon ratio around 0.48 during equilibrium, and the
major flows are localized in the neighborhood of the same nuclei
during the non-equilibrium parts of the evolution. Despite
\ux{56}{Ni} being marginally within the range of major flows, large
scale equilibrium patterns gradually favor \ux{56}{Fe} instead of
\ux{56}{Ni} for decreasing electron fraction (see Figure
\ref{fig:nse_rho1e10_temp1e10}). Regions 1-6 each have the same type
of physics compared to the corresponding ones for initially
symmetric matter. The chasm widening is an outcome of the overall
underproduction of \ux{44}{Ti}. The large scale equilibria patterns
do not favor \ux{44}{Ti} production (see Figure
\ref{fig:nse_rho1e10_temp1e10}), and the normal freeze-out region
merges with the chasm. The $\alpha$-rich freeze-out yields less
\ux{44}{Ti} compared to the symmetric case, ceding additional area
to the chasm region. The decreased ef\-ficiency of the $\alpha$-rich
freeze-out regarding \ux{44}{Ti} production is related to the manner
that the electron fraction value af\-fects the flow transfer by
\mr{(\alpha,\gamma)} reactions towards the $N=22$ isotone chain, and
the favor of the isotone chains towards neutron-rich isotopes rather
than \ux{44}{Ti}. The size reduction of the $\alpha$$p$-rich
freeze-out region is due to the absence of free protons in
neutron-rich equilibrium configurations.

Region 7 represents the case of neutron-rich, $\alpha$-rich
freeze-out, which barely appears for $Y_{e}$=0.5 (not labeled in
Figure \ref{fig:contour_ti44_AD1_ye0500_regimes}). It is also known
in the literature as the ``$\alpha$-process''
\citep{woosley_1992_aa}, but we term it henceforth as
``$\alpha$$n$-rich freeze-out''. This region combines the physics of
regions 3 and 5. A photodisintegration regime is established early
in the evolution and during the equilibrium stages, but contrary to
region 5, \pen\ and \nep\ tend to balance each other after the phase
transition imparted by the 3$\alpha$ forward rate dominance over its
inverse. Thus, the electron fraction value is maintained close to
its initial value, well below 0.5. Such values of the electron
fraction are prerequisite for the production of elements beyond the
Fe-group. Although the electron fraction has similar values within
region 3, there are traces of Si-group and Fe-group nuclei during
the equilibrium stages. The presence of \ux{56}{Ni} blocks the flows
towards heavier nuclei during the non-equilibrium stage and neutrons
are consumed among the Si-group and Fe-group. However, these traces
are absent for the equilibrium stages within region 7, and heavier
elements are produced during the non-equilibrium stage. Table
\ref{tab:dominant_yields} lists the dominant yields from freeze-outs
for $Y_e$ = 0.48, 0.50, and 0.52. The yields from the $\alpha$-rich
freeze-out are quite similar for $Y_e$ = 0.48 and 0.50
\citep{woosley_1992_aa}. Since the region for the $\alpha$$n$-rich
freeze-out increases in size for decreasing $Y_{e}$, it is expected
at some point to dominate all the area enclosed by the chasm.

For $Y_{e}$=0.52, NSE and large scale QSE are dominated by
\ux{56}{Ni} and free protons, with non-negligible abundances for
symmetric and proton-rich nuclei (also see Figure
\ref{fig:nse_rho1e10_temp1e10}), in accordance with a minimum of the
Helmholtz free energy. Despite the excess of free protons favored by
large scale equilibria patterns, neutrons are captured more
ef\-ficiently in a proton-rich environment. Thus, \nep \ reaction
dominates \pen, resulting in a slight increase to the value of the
electron fraction early in the evolution for the $\alpha$$p$-rich
freeze-out region. Since these reaction channels retain $Y_{e} > 0.5
$ during large scale equilibrium, major flows favor proton-rich
nuclei as soon as the QSE cluster dissolves. Thus, for regions 1-5
where NSE and large scale QSE are established, the \mr{(p,\gamma)}
reactions are all directed towards transferring the flow from
symmetric to proton-rich isotopes, which is equivalent to a phase
transition that all isotopes in the network sustain. Consequently,
there cannot be a normal freeze-out in region 1 (Figure
\ref{fig:contour_ti44_ni56_AD1_ye0480_ye0520_regimes}), because the
freeze-out does not take place from NSE (or large scale QSE). During
the non-equilibrium part of the freeze-out evolution, the weak
interactions decrease $Y_{e}$ by transferring flow towards more
stable isotopes, resulting in its non-monotonic evolution and the
reassemble of symmetric isotopes like \ux{44}{Ti}.

Within the $\alpha$$p$-rich freeze-out region (region 4 in Figure
\ref{fig:contour_ti44_ni56_AD1_ye0480_ye0520_regimes}) the
\ux{44}{Ti} mass fraction pattern resembles the corresponding one
with initial $Y_{e}=0.5$, with two arcs and an ascending track at
the end. A timescale dependent third arc is identifiable for the
power-law profile only. Its appearance relies on the equilibrium
state of the remaining \ux{44}{Ti}-\ux{45}{V} cluster and the net
flow towards this cluster by the interplay between neighboring
\mr{(p,\gamma)} and weak reactions.
\mr{\ux{45}{V}(p,\gamma)\ux{46}{Cr}} is the primary reaction to
control the flow leakage of\-f this cluster.

Within region 6, the initially formed small scale QSE clusters fail
to merge to a large-scale cluster. This results in randomly directed
flow supply among the small scale clusters and the absence of a
phase transition accompanied by complete consumption of fuel nuclei.
The physical border between regions 1 and 6 is an outcome of the
existence of a transition within region 1. For regions 1-5 the final
composition is always proton-rich.

\section{Reaction rate sensitivities}
\label{s.rate_sensitivity}

\begin{figure}[htp]
\includegraphics[width=1.0\textwidth]{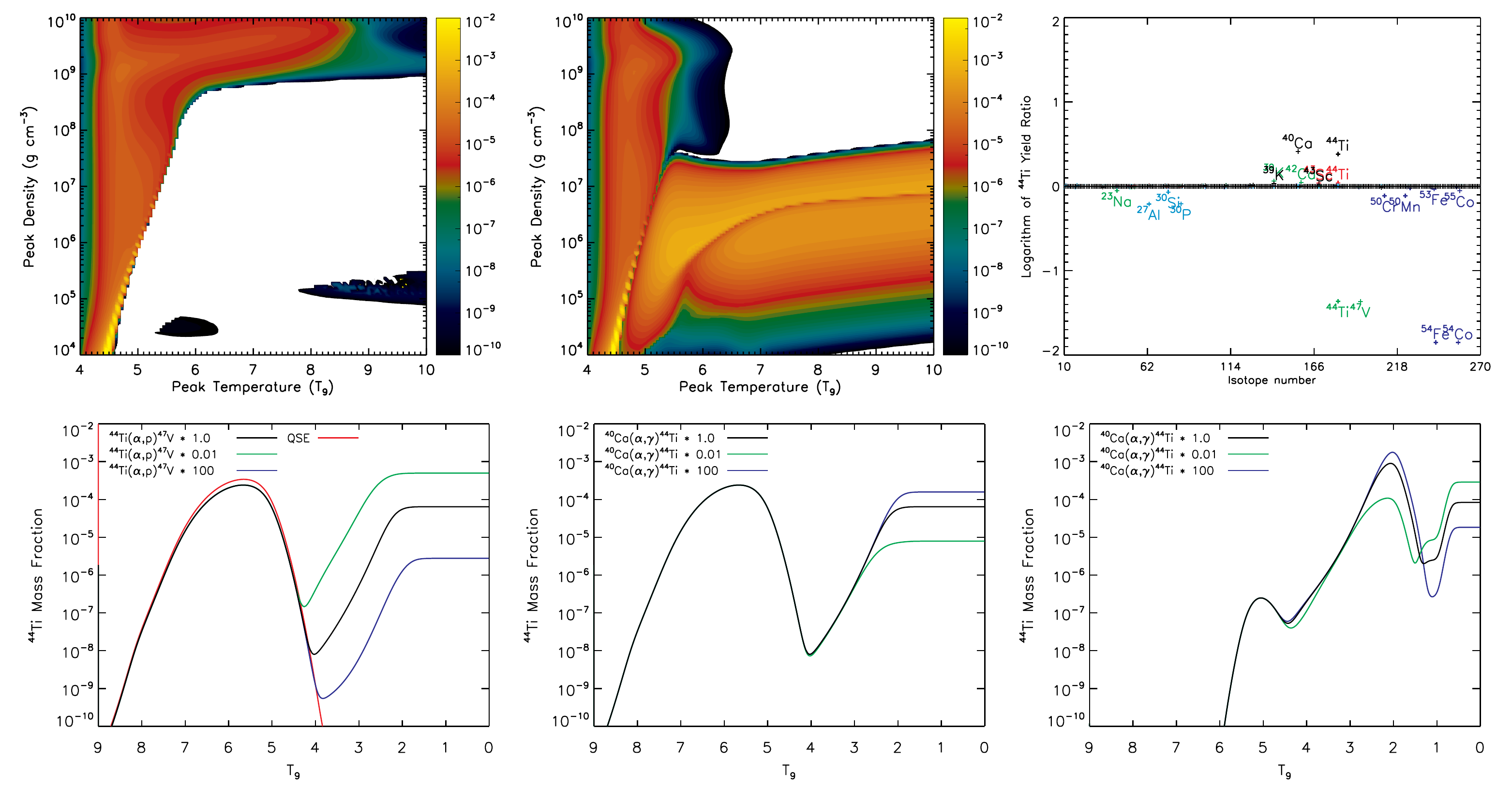}
\caption{ Examples of $^{44}$Ti sensitivity to reaction rates for
the power-law thermodynamic profile at $Y_e$=0.5. From left to right
in the first row, the contour plots show the ef\-fects of removing
the 3$\alpha$ and increasing the \pen \ + \nep\ by 1000, while the
third plot shows the ratio of the \ux{44}{Ti} yield with single
rates increased by 100 to the nominal \ux{44}{Ti} yield during
$\alpha$-rich freeze-out. Isotopes whose reaction rates produce
interesting variations are labeled. Black color is used for the
($\alpha$,$\gamma$) reactions, green for the ($\alpha$,p) reactions,
red for (p,$\gamma$) reactions, cyan for ($\alpha$,n), and blue for
the weak reactions. The second row shows from left to right the
\ux{44}{Ti} mass fraction sensitivity to
\mr{\ux{44}{Ti}(\alpha,p)\ux{47}{V}} and
\mr{\ux{40}{Ca}(\alpha,\gamma)\ux{44}{Ti}} for the $\alpha$-rich
freeze-out, and \mr{\ux{40}{Ca}(\alpha,\gamma)\ux{44}{Ti}} for the
$\alpha$$p$-rich freeze-out. Black curves are for the nominal rates,
red curves are for the QSE yields, green curves for rates multiplied
by 0.01 and blue curves for rates multiplied by 100.
        }
\label{fig:sensitivity_sample}
\end{figure}

The topology of the \ux{44}{Ti} and \ux{56}{Ni} contour plots is
af\-fected by certain key reactions, in combination with the
timescale of the expansion. We follow a three-stage method to
uncover the role and impact of these reactions. Figure
\ref{fig:sensitivity_sample} exemplifies the three stages of this
method. During the first stage, specific reaction channels are
either altered or removed from the network calculations for all
isotopes (e.g., all $(\alpha,\gamma)$ reactions) to assess the most
significant channels for every region. In addition, the $3\alpha$,
\pen\ and \nep\ reactions have their own brevet, since they may
af\-fect the reaction flows globally (first row in Figure
\ref{fig:sensitivity_sample}). Thus, the term ``weak reactions''
will imply all such reactions henceforth, excluding \pen \ and \nep.
We tabulate weak reactions by their dominant decay mode, although
all decay modes are considered in our calculations. The second stage
performs a sensitivity analysis on all groups of reactions by
increasing and decreasing reaction rates excessively one at a time
(third panel within first row in Figure
\ref{fig:sensitivity_sample}). Similarly to \citet{the_1998_aa},
reaction rates are either multiplied or divided by a factor of 100.
The exception are the weak reactions, where the factor is 1000. The
third stage conducts detailed nuclear flows and mass fraction
profile analysis to illustrate the impact of the final crucial
reactions that af\-fect \ux{44}{Ti} (second row in Figure
\ref{fig:sensitivity_sample}). Note the second and third stages are
applied independently for every distinctive thermodynamic region in
the temperature-density plane. The rates used for our calculations
are from the \citet{rath_2000_aa} compilation, updated with some
experimentally measured rates. Table \ref{tab:nuclear_reactions}
lists the most important reactions which our sensitivity study has
revealed to impact \ux{44}{Ti}. We rank reactions as ``primary'' or
``secondary'', depending on the dif\-ferences between the
\ux{44}{Ti} mass fraction curves for nominal and modified rates. A
reaction which involves dif\-ferences at any point of the evolution
by a factor of 10 or larger is ranked primary (Figure
\ref{fig:sensitivity_sample}). Reactions resulting in changes
smaller than a factor of 10 are ranked secondary. Reactions of
minimal impact are not tabulated.

It is important to clarify the advantages and disadvantages of our
methodology for the sensitivity studies. Within the regime of medium
mass nuclei where \ux{44}{Ti} and \ux{56}{Ni} belong, the nuclear
level densities are large enough, so that uncertainties to reaction
rates are expected to be constrained within a small range from their
nominal values \citep{iliadis_2007_aa}. However, such small changes
to the rates may not fully demonstrate the impact of individual
reactions to the burning process. Our goal is to understand the
microscopic mechanisms of explosive nucleosynthesis, which are
driven by the ef\-fect of individual reactions in combination with
localized equilibria patterns. Unrealistic changes to reaction rates
either by excessive factors or by removal from the network are
required to result in distinguishable changes to the dynamics of the
burning process. The changes to the burning process are related to
isolated microscopic components to the operation of explosive
nucleosynthesis. Our sensitivity study aims to identifying as many
as possible of the components related to \ux{44}{Ti} and \ux{56}{Ni}
synthesis. Thus, we add detail to our understanding of the process
for nominal values of the reaction rates. On the other hand, this
type of sensitivity study may not provide a numeric measure of the
importance of identified reactions. For this purpose, sensitivity
studies should be constrained within acceptable uncertainty limits
for the reaction rates. Such sensitivity studies have been performed
by \citet{hoffman_2010_aa} and \citet{tur_2010_aa}.

Overall, the (p,n), ($\alpha$,n) and (n,$\gamma$) reactions have
either secondary or minimal impact to the synthesis of \ux{44}{Ti}.
The reason is that \ux{44}{Ti} is produced mostly for
$Y_{e}\geqslant 0.5$, where neutrons tend to be depleted quite fast.
Moreover, reactions that emit neutrons usually have higher
thresholds than proton emitting ones, since neutron separation
energies are larger than proton separations energies for proton-rich
nuclei.

\subsection{The $3\alpha$ reaction}
\label{s.triple_alpha}

\begin{figure}[htp]
\includegraphics[width=0.9\textwidth]{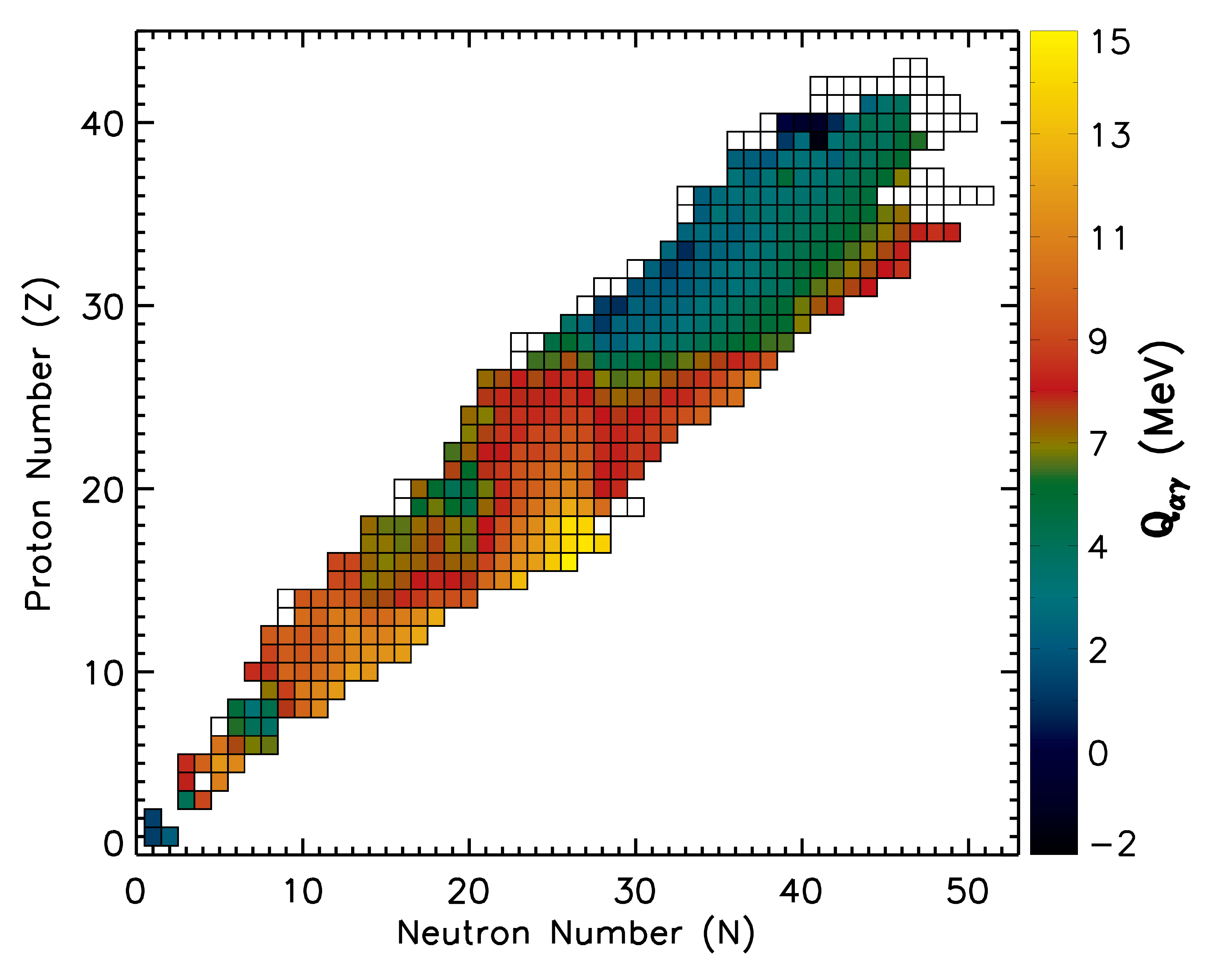}
\caption{Reaction $Q$ values for the \mr{(\alpha,\gamma)} channels
within our reference network of 489 isotopes. $Q_{\alpha\gamma}$ is
also equivalent to the alpha capture thresholds $S_{\alpha}$.
($N,Z$) boxes correspond to the $Q$ value of reaction
$^{A}_{Z}{X}(\alpha,\gamma)^{A+4}_{Z+2}{Y}$. White boxes imply the
absence of an \mr{(\alpha,\gamma)} reaction.}
\label{fig:Q_value_charts}
\end{figure}

One dif\-ference between the normal and $\alpha$-rich freeze-outs is
the behavior of the large scale QSE cluster.  For decreasing
temperatures and densities local equilibria successively break,
gradually dissolving the cluster. The most sensitive equilibria are
related to reactions with large $Q$ values. When no external flows
are applied to the QSE cluster, the first equilibria to break are
among isotopes with $42\leqslant A\leqslant53$, where the largest
reaction $Q$ values for alpha particle captures in the network are
localized (Figure \ref{fig:Q_value_charts}). This is the case for
the chasm, region 2 in Figure
\ref{fig:contour_ti44_AD1_ye0500_regimes}. When these local
equilibria break, the QSE cluster dissolves into two smaller QSE
clusters; the first localized within the Si-group nuclei, and the
second within the Fe-group nuclei. During this process, the
thermodynamic conditions dictate the net $3\alpha$ rate is always
dominated by its photodisintegration reverse rate.

In contrast, the forward flow dominates the net $3\alpha$ rate for
thermodynamic conditions conducive to an $\alpha$-rich or
$\alpha$$p$-rich freeze-out. Here the $3\alpha$ reaction supplies
the external flow to the large equilibrium cluster from the region
of light nuclei. Specifically, reactions in the neighborhood of
\ux{24}{Mg} have relatively large $Q$ values (Figure
\ref{fig:Q_value_charts}), although slightly lower compared to the
ones in the region $42\leqslant A\leqslant53$, and are the first
equilibria to break under contributions from the $3\alpha$ reaction.
The external flow supply results in a phase transition, leading to
the $\alpha$-rich freeze-out. Omission of the $3\alpha$ reaction
from the network calculations results in the severe underproduction
of the Si-group elements as shown in Figure
\ref{fig:sensitivity_sample} for \ux{44}{Ti}. This happens because
the phase transition is prohibited from taking place, and freeze-out
from QSE at these conditions favors only the Fe-group nuclei.
Omission of the $3\alpha$ reaction within the normal freeze-out
regime has little ef\-fect as the forward rate has no impact in this
regime.

\subsection{The ($\alpha$,$\gamma$) reactions}
\label{s.alpha_gamma}

Following the phase transition during an $\alpha$-rich freeze-out,
alpha captures break equilibrium and transfer nuclear flow between
\mr{(p,\gamma)} equilibria chains along isotone lines. Depending
primarily on (i) mass dif\-ferences between reactants and products
of a reaction and (i) the electron fraction value,
\mr{(\alpha,\gamma)} and \mr{(\alpha,p)} channels compete for the
dominance in flow transfer. The $Q$ value for
\mr{\ux{40}{Ca}(\alpha,p)\ux{43}{Sc}} ($Q\approx -3.522$ MeV) allows
the gradual dominance of \mr{\ux{40}{Ca}(\alpha,\gamma)\ux{44}{Ti}}
for decreasing temperature in both the $\alpha$-rich and
$\alpha$$p$-rich freeze-out regions. The impact of this reaction
appears as soon as \ux{44}{Ti} moves of\-f the large scale QSE
cluster, after the first dip in the mass fraction curves caused by
\mr{(\alpha,p)} and \mr{(\alpha,\gamma)} equilibrium breakages
(Figure \ref{fig:sensitivity_sample}).

In accordance with the major flow guidelines,
\mr{\ux{40}{Ca}(\alpha,\gamma)\ux{44}{Ti}} is the primary reaction
to supply flow along the N=22 isotone for symmetric matter. This
supply is responsible for maintaining the pattern of \mr{(p,\gamma)}
equilibria along that chain for decreasing conditions. Were this
flow supply absent, the \mr{(p,\gamma)} equilibria chain would break
and the ascending track of \ux{44}{Ti} mass fraction would cease.
Thus, \mr{\ux{40}{Ca}(\alpha,\gamma)\ux{44}{Ti}} regulates the
amplitude of the subsequent rise past the first dip. Breakage of
various \mr{(p,\gamma)} equilibria determines the formation of
additional such dips, and \mr{\ux{40}{Ca}(\alpha,\gamma)\ux{44}{Ti}}
regulates the amplitude of the formed arc in the mass fraction
curve. For the $\alpha$-rich freeze-out of Figure
\ref{fig:sensitivity_sample},
\mr{\ux{40}{Ca}(\alpha,\gamma)\ux{44}{Ti}} controls how high the
mass fraction curve rises once past the dip. Larger rates enhance
the flow into the N=22 isotone chain, resulting in an increase in
the \ux{44}{Ti} yield. However, a larger rate has the opposite
ef\-fect in the $\alpha$$p$-rich freeze-out of Figure
\ref{fig:sensitivity_sample}. The second phase transition is caused
by \mr{\ux{45}{V}(p,\gamma)\ux{46}{Cr}} breaking from equilibrium
(see \S \ref{s.p_gamma}). The depth of the second dip and the
magnitude of the subsequent ascent is controlled by
\mr{\ux{40}{Ca}(\alpha,\gamma)\ux{44}{Ti}}. A larger
\mr{\ux{40}{Ca}(\alpha,\gamma)\ux{44}{Ti}} rate enhances the depth
of second minimum, resulting in a smaller overall \ux{44}{Ti} yield.

Figure \ref{fig:sensitivity_sample} shows that
\mr{\ux{40}{Ca}(\alpha,\gamma)\ux{44}{Ti}} af\-fects the amplitude
of the second arc for the \ux{44}{Ti} mass fraction, but the slopes
of the ascending and descending tracks are relatively robust. These
slopes are determined by the \mr{(p,\gamma)} channels that
participate in the equilibrium chain. Their robustness for symmetric
matter is a direct consequence of the major flows guidelines.
However, proton captures are less ef\-ficient within a neutron-rich
environment, and the slopes are af\-fected by the net flow
transferred to the mildly connected equilibrium chain. The net
transfer is determined primarily by the flow supply from
\mr{\ux{40}{Ca}(\alpha,\gamma)\ux{44}{Ti}} and
\mr{\ux{40}{Ca}(\alpha,p)\ux{43}{Sc}} and the flow leakage from
\mr{\ux{44}{Ti}(\alpha,p)\ux{47}{V}} (see also \S \ref{s.alpha_p}).

For $Y_{e}<0.5$ the impact of \mr{\ux{44}{Ti}(\alpha,p)\ux{47}{V}}
on the chasm (see \S\ref{s.alpha_p}) is influenced by the secondary
\mr{\ux{42}{Ca}(\alpha,\gamma)\ux{46}{Ti}}. Further minimal
contributions from other ($\alpha$,$\gamma$) reactions are related
to the distribution of nuclear flow among the remaining equilibria
chains along various isotone lines past the QSE cluster dissolution.
In addition, \mr{\ux{12}{C}(\alpha,\gamma)\ux{16}{O}} is a secondary
reaction to af\-fect the flow supply to the QSE cluster after the
3$\alpha$ rate has dominated its inverse, with minimal contributions
from \mr{\ux{20}{Ne}(\alpha,\gamma)\ux{24}{Mg}}.

For $Y_{e}>0.5$, the amplitude regulation of the second arc by
\mr{\ux{40}{Ca}(\alpha,\gamma)\ux{44}{Ti}} has an impact to the
\ux{44}{Ti} yield only for the exponential profile, due to the
absence of a third arc in the \ux{44}{Ti} mass fraction for this
case. The rest of the \mr{(\alpha,\gamma)} reactions are secondary
to \ux{44}{Ti} synthesis for initially proton-rich composition.

\subsection{The ($\alpha$,p) reactions}
\label{s.alpha_p}

Of vital importance to \ux{44}{Ti} synthesis from this channel group
is \mr{\ux{44}{Ti}(\alpha,p)\ux{47}{V}}. This reaction is related
directly to the formation of the chasm, which is the border region
between the normal and the $\alpha$-rich freeze-outs (Figure
\ref{fig:contour_ti44_AD1_ye0500_regimes}), the depth of the chasm,
and the location of the chasm in the peak temperature-density plane
for dif\-ferent expansion profiles.  However, this reaction is not
responsible for the widening of the chasm; weak reactions discussed
in \S\ref{s.weak_interactions} largely control the chasm width. The
key feature of \mr{\ux{44}{Ti}(\alpha,p)\ux{47}{V}} is its small
negative $Q$ value ($Q\approx -410$ keV). This feature allows
\mr{\ux{44}{Ti}(\alpha,p)\ux{47}{V}} to dominate
\mr{\ux{44}{Ti}(\alpha,\gamma)\ux{48}{Cr}} even for low
temperatures. Based on the major flow guidelines,
\mr{\ux{44}{Ti}(\alpha,p)\ux{47}{V}} is the primary flow supplier
from the N=22 to the N=24 isotone for initially symmetric matter.
When this reaction is in equilibrium, \ux{44}{Ti} is considered to
belong in the large scale QSE cluster, a fact verified by QSE
calculations (Figures \ref{fig:ti44_local_cluster_chart} and
\ref{fig:qse_vs_net}). Its equilibrium breakage signals the phase
transition for \ux{44}{Ti}, leaving the isotope outside the QSE
cluster.

The \ux{44}{Ti} chasm is formed by dissolution of the large QSE
cluster into two smaller clusters, and the subsequent flow leakage
from one cluster to another.  For peak temperatures and densities
corresponding to the chasm region,
\mr{\ux{44}{Ti}(\alpha,p)\ux{47}{V}} is always in equilibrium until
the very end of freeze-out. The resulting mass fraction for
\ux{44}{Ti} ends up with a yield 2-3 orders of magnitude less than
its typical value in regions outside the chasm region, as shown in
Figure \ref{fig:mass_fractions_AD1_PL2_temp9}. For the $\alpha$-rich
and $\alpha$$p$-rich freeze-out regions,
\mr{\ux{44}{Ti}(\alpha,p)\ux{47}{V}} breaks equilibrium before the
end of the freeze-out. Equilibria transitions for \ux{44}{Ti} are
depicted in Figure \ref{fig:ti44_local_cluster_chart}. There is a
robust equilibrium between the two isotone chains initially, but
eventually \mr{\ux{44}{Ti}(\alpha,p)\ux{47}{V}} departs from
equilibrium. Although this is an endothermic reaction, the $\alpha$
capture dominates its inverse because free alpha particles are more
abundant than free protons ($X(\alpha)\gg X(p)$). The rest of the
equilibria connecting the N=22 and N=24 isotone chains break
sequentially. When no equilibria links connect the isotone chains,
the abundances of all related elements begin to increase. From this
perspective, the \mr{\ux{44}{Ti}(\alpha,p)\ux{47}{V}} reaction's
persistence until the end of freeze-out is important for all
isotopes along the N=22 isotone chain, not just \ux{44}{Ti}.

The chasm's depth is directly related to the minimum value of the
mass fraction curve for \ux{44}{Ti} prior to the equilibrium
breakage of \mr{\ux{44}{Ti}(\alpha,p)\ux{47}{V}}. Sensitivity
studies for this reaction reveal that the minimum value is
determined by the rate's strength. In Figure
\ref{fig:sensitivity_sample} the \ux{44}{Ti} mass fraction during an
$\alpha$-rich freeze-out is shown as a function of temperature for
various multiplicative factors to the
\mr{\ux{44}{Ti}(\alpha,p)\ux{47}{V}} rate. The minimum value is
smaller for larger reaction rates. Note that the slope of the mass
fraction curve after the minimum value is independent of the rate's
strength, showing this reaction has no impact on \ux{44}{Ti}
synthesis from the moment this reaction goes of\-f equilibrium.

One of the major dif\-ferences between the exponential and power-law
profiles is the location of the chasm in the peak
temperature-density plane. Figure
\ref{fig:contour_ti44_ni56_AD1_PL2_ye0500_sph} shows the chasm
occurs at smaller densities for the power-law profile. The reactions
that change the yield of \ux{44}{Ti} between these two profiles are
approximately the same. This excludes reactions alone as a reason
for the location of the chasm, implying timescale ef\-fects play a
key role. The power-law expansion always evolves slower than the
exponential one for the same initial peak temperature and peak
density. In an environment where nucleosynthesis is driven by
entropy changes, temperature sets to first order the threshold for a
particular phase transition to appear, but the density value at the
threshold temperature determines whether the transition takes place
or not. The contribution of timescale ef\-fects to the chasm shift
is related to the time spent by the plasma in between phase
transitions (Figure \ref{fig:regimes_cartoon}), resulting in
dif\-ferent density values at threshold temperatures. Thus, the
chasm shift is a density driven phenomenon. Specifically,
\mr{\ux{44}{Ti}(\alpha,p)\ux{47}{V}} departs from equilibrium
approximately at the same temperature \mr{T_{thr}\sim 4.3} GK for
both expansion profiles. When both the exponential and power-law
profiles reach \mr{T_{thr}}, the density associated with the
exponential profile is larger (and earlier in time) than the density
of the corresponding power-law profile (later in time).  Since both
profiles assume a constant radiation entropy, \mr{\rho\varpropto
T^{3}}, throughout the evolution, a larger (or smaller) density at
\mr{T_{thr}} translates directly into a larger (or smaller) initial
peak density. This causes the shift in the location of the chasm in
the peak temperature-density plane.

For neutron-rich environments, \mr{\ux{40}{Ca}(\alpha,p)\ux{43}{Sc}}
pipes flow from the major flows among neutron-rich isotopes to
\ux{44}{Ti}, increasing thus its mass fraction. The reaction's main
feature is a flow direction switch for conditions past the phase
transition. While the \ux{44}{Ti} mass fraction begins its ascending
track, the forward flow of \mr{\ux{40}{Ca}(\alpha,p)\ux{43}{Sc}}
dominates its inverse. When the flow direction for
\mr{\ux{40}{Ca}(\alpha,p)\ux{43}{Sc}} switches and the proton
capture dominates the alpha capture, part of the major flows is
supplied to \ux{40}{Ca}, and subsequently to the $N=22$ isotone
equilibrium chain through
\mr{\ux{40}{Ca}(\alpha,\gamma)\ux{44}{Ti}}. This pattern of
escalating flow exchange between the $N=20$ and $N=22$ isotone
chains has also an impact within a proton-rich environment. During
the formation of the second arc for the \ux{44}{Ti} mass fraction,
the dominant proton capture in \mr{\ux{40}{Ca}(\alpha,p)\ux{43}{Sc}}
results in an enhanced flow supply to \ux{44}{Ti}, which is lost
during the leakage to proton-rich nuclei by
\mr{\ux{45}{V}(p,\gamma)\ux{46}{Cr}}. Hence, the \ux{44}{Ti} mass
fraction decreases in this case.

\subsection{The (p,$\gamma$) reactions}
\label{s.p_gamma}

This group of channels is characteristic for the collective
contribution of reactions in the form of equilibria chains. The most
important proton captures are localized among symmetric and
proton-rich isotopes, because their inherently enhanced ef\-ficiency
may alter equilibria patterns and result in phase transitions. Their
ef\-fectiveness is enhanced significantly in proton-rich
environments, where they are favored by major flows and there is a
large availability of free protons. In practice, the weak reactions
set a proton-rich environment primarily with \pen \ and \nep.
Without this elegant combination of weak interactions and proton
captures, the $\alpha$$p$-rich freeze-out region in the contour
plots merges smoothly with the $\alpha$-rich freeze-out one. Proton
captures alter the local equilibria patterns, resulting in small
scale phase transitions. For initially symmetric matter, they sculpt
the $\alpha$$p$-rich freeze-out topology in the contour plots.

The \mr{(p,\gamma)} channels most relevant to \ux{44}{Ti}
nucleosynthesis operate along the $N=20$, $N=22$ and $N=24$ isotone
chains. The important isotone chain is the $N=22$ one, where
\ux{44}{Ti} resides. The specific chain is composed by \ux{44}{Ti},
\ux{45}{V}, \ux{46}{Cr}, \ux{47}{Mn} and \ux{48}{Fe}. The chain
terminates to \ux{44}{Ti} due to the early equilibrium break of
\mr{\ux{43}{Sc}(p,\gamma)\ux{44}{Ti}}, while the upper limit of
\ux{48}{Fe} appears due to the large negative $Q$ value of
\mr{\ux{48}{Fe}(p,\gamma)\ux{49}{Co}} close to the proton dripline.
For regions 3 and 4 in the temperature-density planes, the ascending
part of the second arc in the mass fraction profile for \ux{44}{Ti}
is formed when these isotopes are all in mutual equilibrium. For
$Y_{e}\geqslant0.5$, the major flows attribute a relative robustness
to the slope of the ascending track from single reaction
sensitivities. This robustness is gradually fading as material
becomes neutron-rich, and the equilibrium maintenance along the
chain depends on the net flow supply, which is configured primarily
by \mr{\ux{40}{Ca}(\alpha,\gamma)\ux{44}{Ti}},
\mr{\ux{40}{Ca}(\alpha,p)\ux{43}{Sc}} and
\mr{\ux{44}{Ti}(\alpha,p)\ux{47}{V}}.

Among the reactions connecting the isotopes within the $N=22$
isotone equilibrium chain, \mr{\ux{45}{V}(p,\gamma)\ux{46}{Cr}} has
the largest $Q$ value, rendering it the most sensitive equilibrium
link. Within the $\alpha$$p$-rich freeze-out region, it is the first
one to break, leaving \ux{44}{Ti} in equilibrium only with
\ux{45}{V} and prognosticating a phase transition where the mass
fraction of \ux{44}{Ti} decreases \citep{the_1998_aa}, due to the
flow transfer from the \ux{44}{Ti}-\ux{45}{V} cluster to the
\ux{46}{Cr}-\ux{47}{Mn}-\ux{48}{Fe} cluster. At the same time, the
equilibria patterns along the rest of the related isotone chains
change, contributing all together to the phase transition for
\ux{44}{Ti}. Furthermore, \mr{\ux{45}{V}(p,\gamma)\ux{46}{Cr}} is
secondary (see \S \ref{s.weak_interactions} below) to defining the
physical border between the regions of $\alpha$-rich and
$\alpha$$p$-rich freeze-outs. A stronger rate expands the
$\alpha$$p$-rich freeze-out region at the loss of the $\alpha$-rich
freeze-out region. The \mr{\ux{57}{Ni}(p,\gamma)\ux{58}{Cu}} is
another secondary reaction which contributes to the localization of
the physical border between regions 3 and 4. When the large scale
QSE cluster begins to dissolve, it is one of the primary reactions
to control the flow transfer within the remnant QSE cluster.

A couple of reactions with a sensible impact to the \ux{44}{Ti}
yield are \mr{\ux{41}{Sc}(p,\gamma)\ux{42}{Ti}} and
\mr{\ux{43}{Sc}(p,\gamma)\ux{44}{Ti}}. They are the primary
reactions to regulate the depth of the second dip in the \ux{44}{Ti}
mass fraction, by af\-fecting the flow supply to the equilibrium
chain by \mr{\ux{40}{Ca}(\alpha,\gamma)\ux{44}{Ti}}. The
\mr{\ux{44}{Ti}(p,\gamma)\ux{45}{V}} reaction is the immediate link
of \ux{44}{Ti} to the specific equilibrium chain. A stronger rate
maintains the existence of the \ux{44}{Ti}-\ux{45}{V} cluster,
resulting in further loss of flow via
\mr{\ux{45}{V}(p,\gamma)\ux{46}{Cr}}. Thus, the \ux{44}{Ti} yield is
decreased. A secondary reaction to af\-fect the ascending track
beyond the second arc is \mr{\ux{40}{Ca}(p,\gamma)\ux{41}{Sc}}.

For proton-rich environments, the \mr{(p,\gamma)} channels are
primary to the formation of the \mr{(p,\gamma)}-leakage region
(Figure \ref{fig:contour_ti44_ni56_AD1_ye0480_ye0520_regimes}).
Large scale equilibria patterns favor both symmetric and proton-rich
nuclei \citep{seitenzahl_2008_aa}. As soon as the large scale QSE
cluster begins to dissolve, \mr{(p,\gamma)} reactions transfer the
flow from symmetric nuclei to proton-rich ones, so that the major
flows are localized in the neighborhood of the latter, in accordance
with the major flows guidelines. Without the contribution of the
\mr{(p,\gamma)} reactions, the \mr{(p,\gamma)}-leakage region would
be equivalent to a normal freeze-out regime and would merge smoothly
with region 6, such as the $Y_{e}=0.5$ and $Y_{e}<0.5$ cases. The
flow transfer by \mr{(p,\gamma)} reactions is massive, where almost
all of them in the network participate. Thus, the initially
descending track of the \ux{44}{Ti} mass fraction due to this flow
transfer is relatively robust to single rate sensitivities. The
\mr{\ux{45}{V}(p,\gamma)\ux{46}{Cr}} reaction is the only one to
af\-fect the depth of the descending track, especially for the
power-law expansion profile.

In addition, \mr{\ux{45}{V}(p,\gamma)\ux{46}{Cr}} controls the flow
leakage of\-f the remaining \ux{44}{Ti}-\ux{45}{V} cluster during
the $\alpha$$p$-rich freeze-out, once this reaction breaks
equilibrium. In combination with the timescale of certain weak
reactions in the locality of \ux{44}{Ti} this results in the
formation of the third arc for the \ux{44}{Ti} mass fraction for the
power-law profile. Secondary reactions within this group (along with
the weak reactions) which regulate the amplitude of the third arc
are listed in Table \ref{tab:nuclear_reactions}.

\subsection{The \pen \ and \nep \ and weak interactions}
\label{s.weak_interactions}

The electron fraction $Y_{e}$ expresses the proton to baryon ratio
in the plasma. Assuming charge neutrality, the electron fraction is
also the electron per baryon ratio. Weak interactions are the only
group of channels to violate the lepton number conservation, while
preserving the baryon number. Thus, they are the only ones to change
$Y_{e}$ during the evolution, with \pen \ and \nep \ having a
special contribution to this configuration
\citep{mclaughlin_1995_aa,fuller_1995_aa,mclaughlin_1996_aa,surman_2005_aa,
liebendorfer_2008_aa,aprahamian_2005_aa}. Depending on the
competition between \pen \ and \nep, the electron fraction $Y_{e}$
may increase or decrease. For these two reactions to be ef\-fective,
relatively large temperature and density values are needed.  Thus,
their impact is usually constrained during the first stages of the
evolution. On the contrary, the lifetimes of the remaining weak
interactions ensure their impact appears during the last stages of
the evolution. These reactions tend to transfer material towards the
valley of stability. Our calculations use the FFN rates for the
\pen, \nep \ and other weak reactions
\citep{fuller_1980_aa,fuller_1982_aa,fuller_1982_ab,oda_1994_aa,langanke_2001_aa}.
Using the FFN weak rates for the other reactions has little ef\-fect
on the synthesis and yields of \ux{44}{Ti} and \ux{56}{Ni}. We thus
use temperature and density independent $\beta^-$-decay and
$\beta^+$-decay rates, where the parent nucleus is assumed to be on
its ground state.

\begin{figure}[htp]
\includegraphics[width=1.0\textwidth]{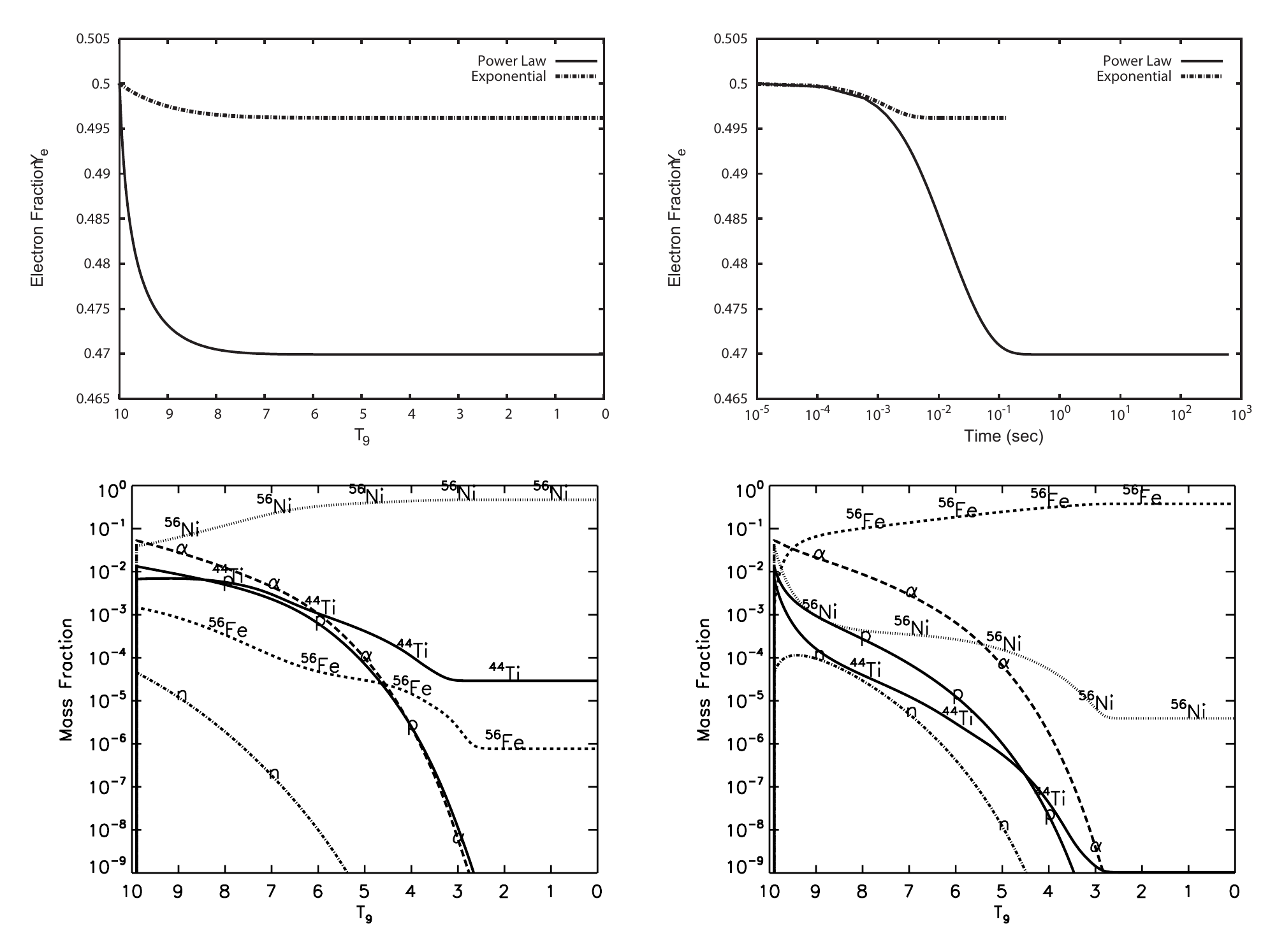}
\caption{ Evolution of the electron fraction $Y_e$ versus
temperature (upper left) and time (upper right) for the exponential
and power-law profiles starting from conditions, $T_{9}=10$ GK,
\hbox{$\rho=10^{10}$ g cm$^{-3}$} and $Y_{e}=0.5$, corresponding to
the normal freeze-out regime (region 1). The corresponding evolution
of the mass fractions are shown for the exponential profile (lower
left) and power-law profile (lower right).
         }
\label{fig:mass_fractions_Ye_AD1_PL2_temp1e10_den1e10}
\end{figure}

In combination with the nucleosynthesis trends for a varying
electron fraction (see \S \ref{s.ye_sensitivity} above) and
timescale ef\-fects, the \pen \ and \nep \ reactions are the key to
explaining the chasm's widening between the exponential and
power-law expansions. Depending on the expansion timescale, \pen \
and \nep \ reactions alter the electron fraction only for a limited
amount of time early in the evolution. The changes to $Y_{e}$ for
various peak temperatures and densities depend on the rate strengths
of these reactions and af\-fect the yields directly. Figure
\ref{fig:mass_fractions_Ye_AD1_PL2_temp1e10_den1e10} shows the
evolution of $Y_{e}$ and a few isotopes related to \ux{44}{Ti}
nucleosynthesis for a case of a normal freeze-out from initially
symmetric matter ($Y_{e}=0.5$) using nominal rates for both
expansion profiles. In this regime, \pen \ always dominates \nep \
and $Y_{e}$ decreases, while temperature and density still have
large values. Timescale ef\-fects are evident when using nominal
rate values, where the time spent in a high entropy environment is
larger for the power-law case and $Y_{e}$ decreases much more
compared to the exponential profile. The plasma adjusts to the $Y_e
\ne 0.5$ conditions while it is still in NSE and QSE subsequently.
Figure \ref{fig:nse_rho1e10_temp1e10} shows NSE mass fractions as a
function of the electron fraction where the production of
\ux{56}{Fe} is favored, while \ux{44}{Ti} and \ux{56}{Ni} are
under-produced
\citep{hartmann_1985_aa,woosley_1992_aa,seitenzahl_2008_aa}. On the
contrary, exponential expansion does not allow $Y_{e}$ to decrease
significantly, resulting in a final composition with significant
yields for \ux{44}{Ti} and \ux{56}{Ni}. Therefore, the chasm expands
only for the power-law profile.

The chasm width is regulated primarily by the strength of \pen \ and
\nep \ and secondarily by timescale ef\-fects. Both $^{44}$Ti and
$^{56}$Ni show a large chasm expansion for both thermodynamic
profiles when these two reactions are enhanced by a factor $10^{3}$.
The new chasm widths for the two profiles are similar, because the
reaction rate enhancement results in the same decrement to $Y_{e}$
for both thermodynamic profiles and diminishes the impact of
timescale ef\-fects. In Figure \ref{fig:sensitivity_sample}, the
normal freeze-out regime (region 1) for \ux{44}{Ti} merges with the
chasm (region 2) and the chasm expands into the area that belonged
to the $\alpha$-rich freeze-out regime (region 3).

The remaining weak interactions also assist in the decrement of the
electron fraction, and thus to the chasm expansion, but their
contributions are smaller than \pen \ and \nep \ due to their
lifetimes. The lifetime of any weak interactions that are primarily
responsible for the changes to the electron fraction must be smaller
than the expansion timescale. In Figure
\ref{fig:mass_fractions_Ye_AD1_PL2_temp1e10_den1e10} the changes to
$Y_e$ take place between $10^{-3}\lesssim t \lesssim 10^{-1}$ sec.
The exponential profile has a timescale of the order of 1 sec, and
the changes to $Y_{e}$ are moderate. On the contrary, the timescale
for the power-law is larger by two orders of magnitude, resulting in
dramatic changes to $Y_{e}$ due to the impact of \pen, \nep, and
weak interactions. Despite the initial identical configuration of
the two expansions, the equilibrium state that the normal freeze-out
begins is very dif\-ferent for the two expansion profiles. For the
exponential trajectory, the weak interactions do not have the time
to change the equilibrium state adequately, and the final yields
have a significant amount of \ux{44}{Ti}, with \ux{56}{Ni}
dominating the final composition. For the power-law profile
\ux{56}{Fe} is the dominant element and \ux{44}{Ti} is
under-produced, expanding the chasm region into the normal
freeze-out regime (see Figure
\ref{fig:mass_fractions_Ye_AD1_PL2_temp1e10_den1e10}). In addition
to \pen \ and \nep \ reactions, Table \ref{tab:nuclear_reactions}
lists the primary weak interactions related to the chasm widening,
all with a half life of the order of $10^{-1}$ sec.

Weak interactions assist in defining the \ux{44}{Ti} topology for
the $\alpha$$p$-rich freeze-out regime (region 4). In this region,
\nep \ dominates over \pen \ and $Y_{e}$ rises above 0.5, driving
the material proton-rich. The relative strength of the \pen \ and
\nep \ rates determines the area of the peak temperature-density
plane occupied by the $\alpha$$p$-rich freeze-out regime as shown by
Figure \ref{fig:sensitivity_sample}. Both \ux{44}{Ti} and
\ux{56}{Ni} have significant mass fraction values during equilibrium
states in such environments, despite the relatively large mass
fractions of proton-rich isotopes. These isotopes decay within the
expansion timescale and transfer additional nuclear flow to the
symmetric isotopes. Weak reactions partially regulate the second arc
for the \ux{44}{Ti} mass fraction, and the formation of the
ascending track at the end of the freeze-out process.

For $Y_{e}=0.48$ the weak interactions barely have an impact on the
\ux{44}{Ti} yield. Within the $\alpha$$p$-rich freeze-out regime
their action is similar to the symmetric case, but the area that
region 4 occupies on the temperature-density plane for neutron-rich
matter is limited. Within the $\alpha$-rich freeze-out regime the
major flows are localized mostly among stable nuclei, or nuclei with
decay timescales much longer than the expansion timescale.

For $Y_{e}=0.52$ the action of the weak interactions has been
outlined before. They transfer nuclear flow from proton-rich to
symmetric nuclei for most of the peak conditions within the
temperature-density plane. For the $\alpha$$p$-rich freeze-out
region with the exponential profile, the average half-life range for
the primary flow carriers is $90\lesssim t_{1/2}\lesssim 500$ ms.
For the power-law expansion, the corresponding range is $200\lesssim
t_{1/2}\lesssim 900$ ms. Weak reactions for the \mr{(p,\gamma)}
leakage regime (region 1 for $Y_{e}=0.52$) are classified according
to the way they impact. The first group includes reactions which
hinder the flow transfer by \mr{(p,\gamma)} reactions when their
rates are enhanced. The second group includes reactions which boost
the flow transfer by \mr{(p,\gamma)} reactions when their rates are
diminished. The third group includes reactions which combine the
action from the two previous groups. Reactions within the third
group make an impact only for the power-law expansion profile. In
addition, weak reactions with relatively long half-lives contribute
to the amplitude regulation of the third arc in the \ux{44}{Ti} mass
fraction for the $\alpha$$p$-rich freeze-out regime (see Table
\ref{tab:nuclear_reactions}).

\section{Network size ef\-fects}
\label{s.network_size}

Trends in the \ux{44}{Ti} yields are controlled by a limited number
of reactions. This raises a query about the minimum number of nuclei
that are necessary to include in a network calculation such that all
relevant physical phenomena are captured. Our reference reaction
network contains 489 isotopes, spanning the light nuclei, silicon
group, and iron group. To assess network size ef\-fects, we compared
the \ux{44}{Ti} yields in the peak temperature-density plane from
the 489 isotope network with the final yields generated by reaction
networks with 204, 1341, and 3304 isotopes. Table
\ref{tab:nuclear_networks} lists the isotopes used in each network.
In addition to the new elements introduced, the larger networks
expand into larger both neutron-rich and proton-rich regimes.  The
addition of new elements beyond the Fe-group has a minimal ef\-fect
on equilibrium clusters, since for the thermodynamic trajectories of
interest the largest partial flows are localized around the Si-group
and Fe-group.

For $Y_{e}=0.48$, the final mass fractions are essentially
independent of the network size since the major flows occur along
the valley of stability, which is modeled adequately by all
networks. For $Y_{e}\geqslant0.5$, the final mass fractions depend
on the network size as weak interactions have a larger role.  In
particular, the 204 isotope network does not include most of the
required proton-rich isotopes to accurately describe the
nucleosynthesis. The dif\-ferences compared to our reference network
are localized to the $\alpha$$p$-rich freeze-out region for
$Y_{e}=0.5$, but they span all the parameter space for $Y_{e}=0.52$.
Consequently, this 204 isotope network is inadequate to describe the
nucleosynthesis of proton-rich material.

\begin{figure}[htp]
\includegraphics[width=0.8\textwidth]{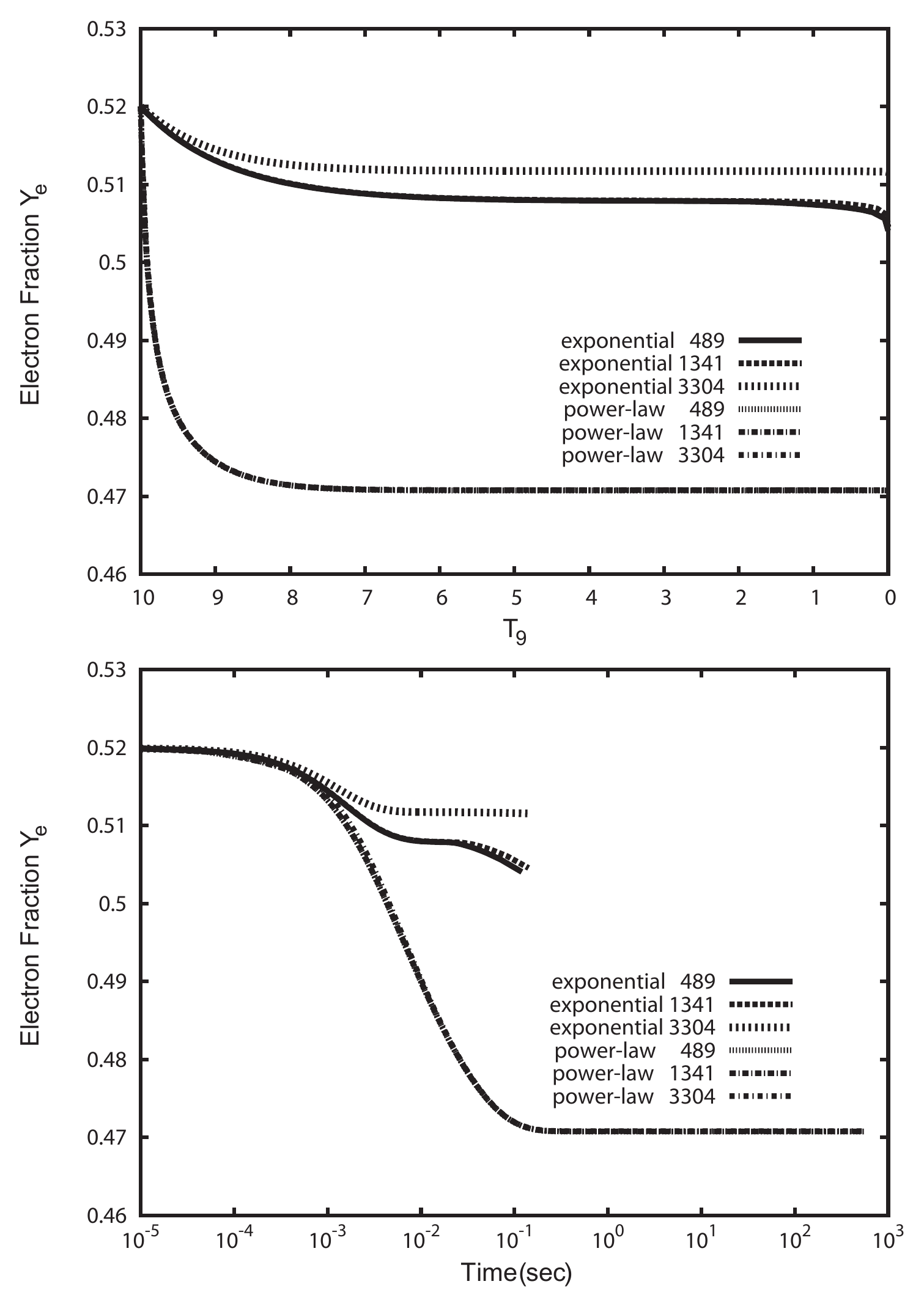}
\caption{ Evolution of the electron fraction $Y_e$ for the
exponential and power-law freeze-out profiles for a freeze-out
starting from $T_{9}=10$ GK, $\rho=10^{10}$ g cm$^{-3}$ and
$Y_{e}=0.52$, corresponding to the (p,$\gamma$) leakage regime
(region 1). The evolution is shown versus temperature (upper plot)
and time (lower plot).
         }
\label{fig:Ye_AD1_PL2_temp1e10_den1e10_489vs1341vs3304}
\end{figure}

There are mass fraction dif\-ferences in region 1 for
$Y_{e}\geqslant0.5$ between our reference network and the larger
networks, which is related to the dif\-ferences in the equilibrium
state configurations by the changes in temperature, density and
$Y_{e}$ during the freeze-out evolution. Specifically, the
dif\-ferences are due to the relationship between the expansion and
weak interaction timescales. Figure
\ref{fig:Ye_AD1_PL2_temp1e10_den1e10_489vs1341vs3304} shows the
temperature dependence of the electron fraction during a freeze-out
for the 489, 1341, and 3304 isotope networks for the exponential and
power-law trajectories. The 489 and 1341 isotope networks have
relatively similar numbers of isotopes per element, for those
elements that are common to both networks. Consequently, the
evolution of $Y_{e}$ is quite similar for both profiles. The 3304
isotope network, however, has a larger number of isotopes per
element for elements that are common among the three networks (see
Table \ref{tab:nuclear_networks}). The presence of more proton-rich
nuclei in the 3304 isotope network slows the electron fraction
decrease driven by \pen \ and \nep \ compared to our reference
network. This results in slightly dif\-ferent large scale
equilibrium states. For a short expansion timescale, such as the
exponential profile, $Y_{e}$ values remain above 0.5 for all
networks, resulting in the dif\-ferences in region 1. The long
timescale of the power-law expansion decreases the electron fraction
value below 0.5 quite early in the evolution. This results in all
three networks converging to the same $Y_e$ values since all three
networks include the necessary isotopes related to production of
\ux{44}{Ti} and \ux{56}{Ni}. Overall, our reference 489 isotope
network is adequate for describing the trends in the \ux{44}{Ti} and
\ux{56}{Ni} yield trends. This is relevant for ef\-ficient use of
computational resources.

\section{Post-Process Yields from Collapse Simulations}
\label{s.postprocessed_yields}

In this section we compare the $^{44}$Ti and $^{56}$Ni yields from
post-processing core-collapse supernovae models with the exponential
and power-law trajectories.  We use the same reference 489 isotope
reaction network for the post-processing and parameterized
trajectories. Our aim is to of\-fer a calibration of where
parameterized trajectories provide a reasonable approximation to the
final yields. Our analysis in the preceding sections allows an
explanation for the behavior of the $^{44}$Ti and $^{56}$Ni profiles
and any dif\-ferences between the post-processed and parameterized
yields. For this assessment we consider 3 of the supernova explosion
calculations whose tracks in the peak temperature-density plane are
shown in Figure \ref{fig:contour_ti44_ni56_AD1_PL2_ye0500_sph}. In
all three of these models the initial $Y_{e}$ profile as a function
of interior mass is very close to Ye = 0.5.

\subsection{A Cassioppeia A model}
\label{s.cassioppeia}

Our first supernova model uses a progenitor designed to match the
supernova remnant Cassioppeia A \citep{young_2006_aa,eriksen_2009_aa},
specifically model M16E1.1BinA from \citet{young_2008_aa}.  The
hydrogen envelope of this spherically symmetric 16 M$_\odot$
progenitor was removed by an assumed binary mass transfer event as the
progenitor evolved into a giant.  We use a multi-step collapse and
explosion process to model the explosion
\citep[for example, see][]{young_2007_aa}.  We model the entire star from
collapse through the formation and stall of the bounce shock.  At this
point, the proto-neutron star is removed from the simulation and
energy is injected just above the proto-neutron star to drive an
explosion.  The explosion is followed as the shock moves out of the
star and most of the fallback has accreted onto the newly-formed
neutron star.  We only calculate the yields of material that is
ejected after fallback.

\begin{figure}[htp]
\includegraphics[width=1.0\textwidth]{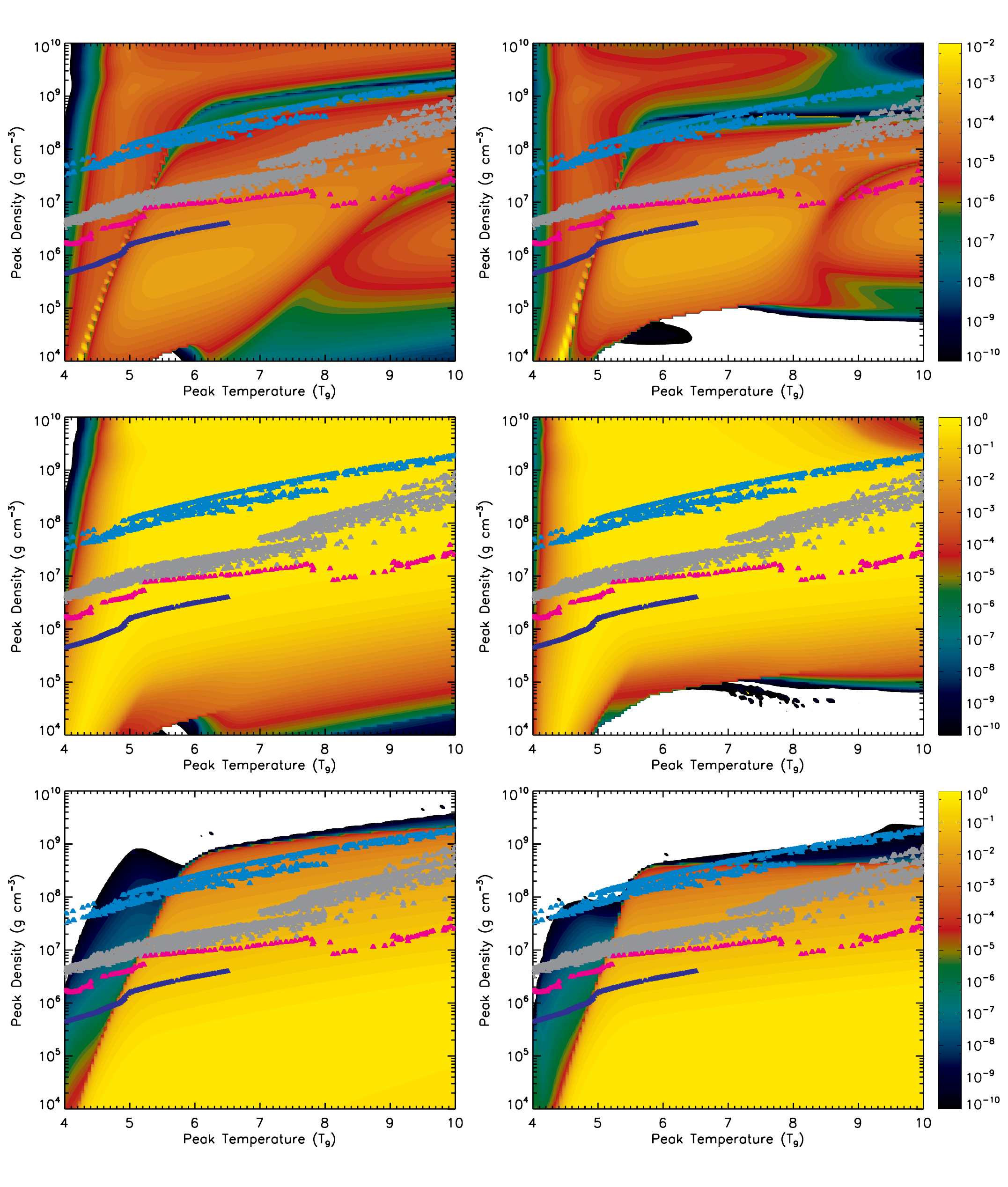}
\caption{Final mass fraction of $^{44}$Ti (first row), $^{56}$Ni
(second row) and $^{4}$He (third row) in the peak
temperature-density plane for the exponential thermodynamic profile
(first column) and power-law profile (second column) at $Y_e$=0.5.
Dif\-ferent colored triangles show the temperature-density positions
from dif\-ferent supernova and hypernova models - blue for a 1D Cas
A model \citep{young_2006_aa}, gray for the  2D rotating progenitor
E15B model \citep{fryer_2000_aa}, pink for a 1D hypernova model
\citep{fryer_2006_ab}, and cyan for a 2D magnetohydrodynamic
collapsar model.} \label{fig:contour_ti44_ni56_AD1_PL2_ye0500_sph}
\end{figure}

\begin{figure}[htp]
\includegraphics[width=0.8\textwidth]{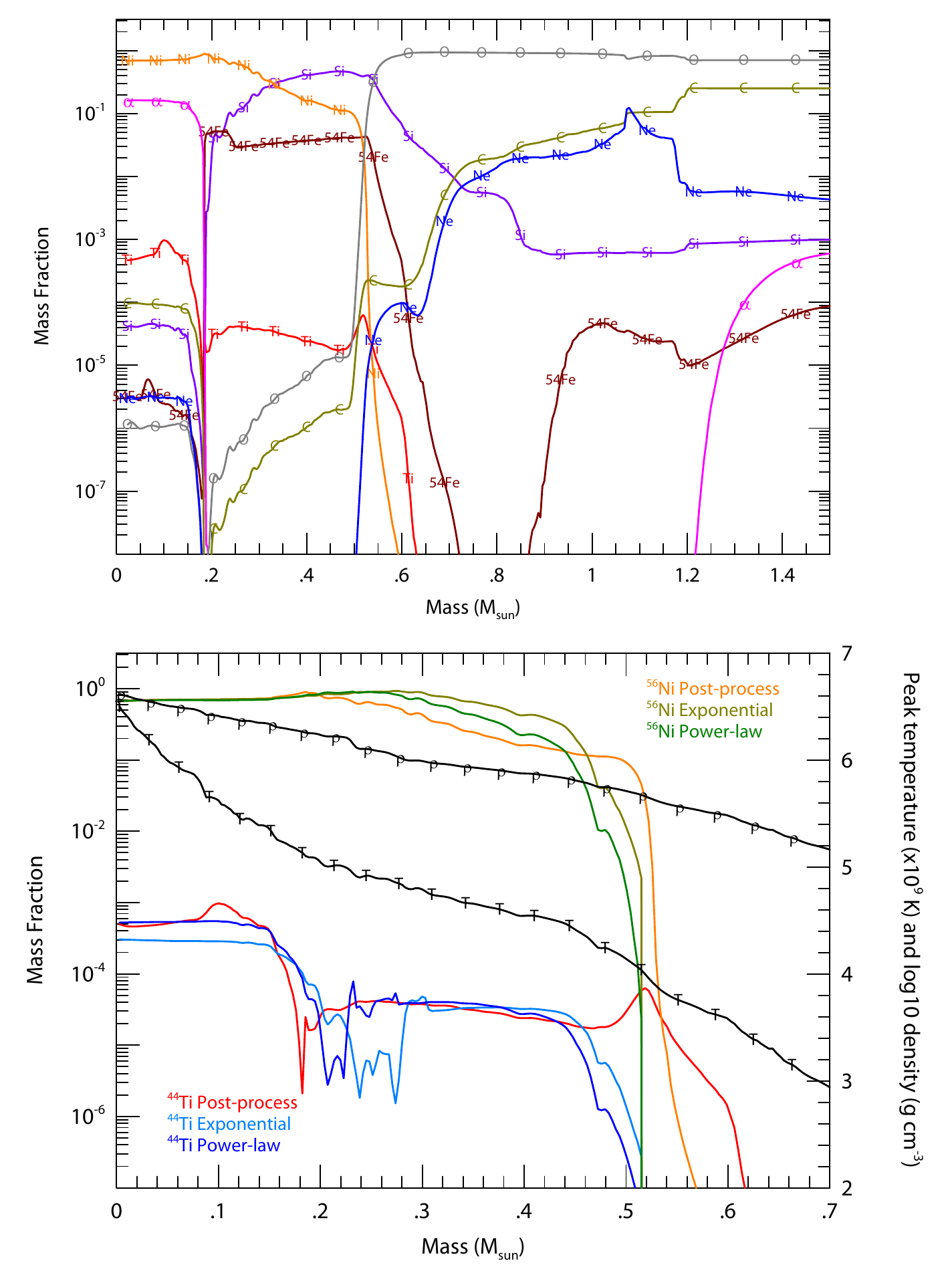}
\caption{ Mass fractions from post-processing the Lagrangian
thermodynamic trajectories (top panel) and post-processing vs
parameterized profiles (bottom panel) for the \hbox{16 M$_{\odot}$}
Cas A model \citep{young_2008_aa}. See text for a discussion of the
sharp dip in $^{44}$Ti $\approx$ 0.2 M$_{\odot}$, and the decline of
$^{44}$Ti and $^{56}$Ni $\approx$ 0.6 M$_{\odot}$.
        }
\label{fig:w16_final_profile_post_vs_param}
\end{figure}

Mass fraction profiles from post-processing the Lagrangian
thermodynamic trajectories with our reference 489 isotope network
are shown in the top panel of Figure
\ref{fig:w16_final_profile_post_vs_param}. The iron-group,
silicon-group, oxygen-rich shells are visible within the innermost
1.5 M$_{\odot}$. The bottom panel shows the mass fraction profiles
from the post-process, exponential, and power-law trajectories. Mass
fraction profiles correspond to the left-hand y-axis, and the peak
temperature and peak density curves correspond to the right-hand
y-axis.  Figure \ref{fig:contour_ti44_ni56_AD1_PL2_ye0500_sph} also
shows the peak temperatures and densities of this explosion model,
and explains the general trends in the $^{44}$Ti, $^{56}$Ni and
$^{4}$He mass fraction profiles of Figure
\ref{fig:w16_final_profile_post_vs_param}.

For interior masses less than $\approx$ 0.2 M$_{\odot}$, the
synthesis of $^{44}$Ti is due to the $\alpha$-rich freeze-out,
region 3 in Figure \ref{fig:contour_ti44_AD1_ye0500_regimes}.  In
this mass range, the $^{44}$Ti mass fractions from the power-law
profile are closer to the post-process values than the exponential
profile. The $^{56}$Ni mass fractions given by the power-law and
exponential profiles generally agree with the post-process mass
fractions.  Near 0.2 M$_{\odot}$, the thin chasm region separating
region 3 ($\alpha$-rich freeze-out) and region 6 (silicon-rich) is
traversed, which causes the downward spike in the $^{44}$Ti profile
(see Figures \ref{fig:contour_ti44_ni56_AD1_PL2_ye0500_sph} and
\ref{fig:contour_ti44_AD1_ye0500_regimes}).  Precisely where the
thin chasm is traversed depends sensitively on the exact values of
peak temperatures and densities, and the location of the thin chasm
line on the temperature-density plane due to the timescale of the
expansion profile. This explains why the downward spike occurs at
slightly dif\-ferent mass locations for the post-process,
exponential and power-law profiles. Between $\approx$ 0.2
M$_{\odot}$ and $\approx$ 0.5 M$_{\odot}$ the explosion is operating
in the silicon-rich, region 6 of Figure
\ref{fig:contour_ti44_ni56_AD1_PL2_ye0500_sph}.  In this region the
final yields of both $^{44}$Ti and $^{56}$Ni are sensitive to the
initial composition, which in this case is given by the model at the
time when energy is injected just above the proto-neutron star to
drive an explosion.  In this mass range, the $^{44}$Ti mass
fractions post-processing, exponential, and power-law profiles
generally agree. For $^{56}$Ni, the power-law profile generally
agrees with the post-process mass fractions better than the
exponential profile.  Both $^{44}$Ti and $^{56}$Ni abruptly decline
at $\approx$ 0.5 M$_{\odot}$ as the peak temperature drops below
4$\times$10$^9$ K.

Yields from the exponential and power-law trajectories are generally
within a factor $\sim$ 2 of the post-process yields, except in
region where the thin chasm is being crossed or the temperature
falls below 4$\times$10$^9$ K. Integrating the $^{44}$Ti mass
fraction profiles in Figure
\ref{fig:w16_final_profile_post_vs_param} over the interior mass
gives the total mass of $^{44}$Ti ejected by this model.  We find
1.04$\times$10$^{-4}$ M$_{\odot}$ for post-processing,
5.62$\times$10$^{-5}$ M$_{\odot}$ for the exponential profile, and
9.30$\times$10$^{-5}$ M$_{\odot}$ for the power-law profile.
Similarly for $^{56}$Ni, we find 2.46$\times$10$^{-1}$ M$_{\odot}$
for post-processing, 3.16$\times$10$^{-1}$ M$_{\odot}$ for the
exponential profile, and 2.78$\times$10$^{-1}$ M$_{\odot}$ for the
power-law profile. Overall, yields of $^{44}$Ti and $^{56}$Ni from
the power-law profile mimic the post-process values better than the
exponential profile for this Cas A model.

\subsection{A Weak-Strong Hypernova Model}
\label{s.weakstrong}

Our second massive star explosion model uses a similar multi-step
process.  In this case however, first a weak explosion is launched
and followed 1 s later by a strong $1.6\times10^{52} {\rm erg}$
hypernova explosion.  This model has a very dif\-ferent
thermodynamic evolution as the material is hit by two shocks. The
first, weaker shock ignites a substantial amount of burning within
the inner 0.6 M$_{\odot}$.  Most of the peak temperatures and peak
densities inside 0.6 M$_{\odot}$ are due to the weaker shock. Peak
conditions at larger masses are due to the second, stronger shock.
For additional details on this weak-strong double shock model, see
40WS1.0 from \citet{fryer_2006_ab}.

\begin{figure}[htp]
\includegraphics[width=0.8\textwidth]{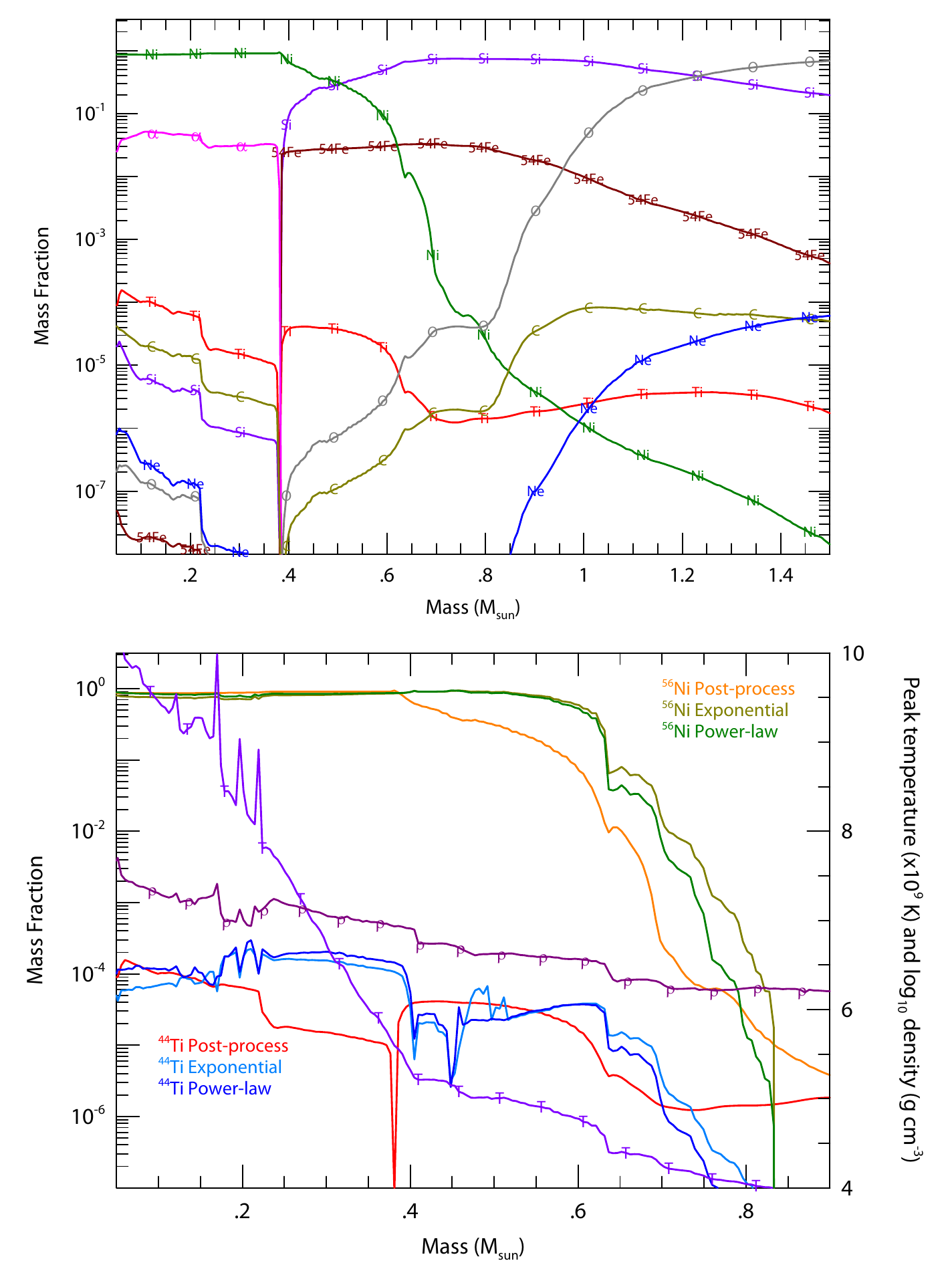}
\caption{ Mass fractions from post-processing the Lagrangian
thermodynamic trajectories (top panel) and  post-processing vs
parameterized profiles (bottom panel) for the spherically symmetric
\hbox{40 M$_{\odot}$} hypernova model \citep{fryer_2006_ab} that
features a weak shock followed by a strong shock, with our reference
489 isotope network. Note the sharp dip in $^{44}$Ti $\approx$ 0.4
M$_{\odot}$, and the extended $^{44}$Ti and $^{56}$Ni distributions
for the post-processed trajectories.
        }
\label{fig:ws623_final_profile_post_vs_param}
\end{figure}

Mass fraction profiles from post-processing the Lagrangian
thermodynamic trajectories with our reference 489 isotope network
are shown in the top panel of Figure
\ref{fig:ws623_final_profile_post_vs_param}. The iron-group and
silicon-group shells are visible within the innermost 1.5
M$_{\odot}$. The bottom panel of Figure
\ref{fig:ws623_final_profile_post_vs_param} shows the mass fraction
profiles from the post-process, exponential, and power-law
trajectories. Mass fraction profiles correspond to the left-hand
y-axis, and the peak temperature and peak density curves correspond
to the right-hand y-axis.  Figure
\ref{fig:contour_ti44_ni56_AD1_PL2_ye0500_sph} also shows the peak
temperatures and densities of this explosion model, and explains the
general trends in the $^{44}$Ti and $^{56}$Ni mass fraction profiles
of Figure \ref{fig:ws623_final_profile_post_vs_param}. For interior
masses less than $\approx$ 0.4 M$_{\odot}$, the synthesis of
$^{44}$Ti is due to the $\alpha$-rich and $\alpha$p-rich
freeze-outs, regions 3 and 4 respectively in Figure
\ref{fig:contour_ti44_AD1_ye0500_regimes}. In this mass range, the
$^{44}$Ti and $^{56}$Ni mass fractions given by the exponential and
power-law profiles have about the same level of agreement with the
post-process values.  Near 0.4 M$_{\odot}$, the thin chasm region
separating region 3 ($\alpha$-rich freeze-out) and region 6
(silicon-rich) is traversed, which causes the downward spike in the
$^{44}$Ti profile (see Figures
\ref{fig:contour_ti44_ni56_AD1_PL2_ye0500_sph} and
\ref{fig:contour_ti44_AD1_ye0500_regimes}).  The downward spike
occurs at dif\-ferent mass locations for the post-process,
exponential and power-law profiles because when the thin chasm is
traversed depends on the exact values of peak temperatures and
densities, and the location of the thin chasm line on the
temperature-density plane due to the timescale of the expansion
profile.

Between $\approx$ 0.4 M$_{\odot}$ and $\approx$ 0.7 M$_{\odot}$ the
explosion is operating in the silicon-rich, region 6 of Figure
\ref{fig:contour_ti44_ni56_AD1_PL2_ye0500_sph}. In this region the
final yields of $^{44}$Ti and $^{56}$Ni are sensitive to the initial
composition.  Beyond $\approx$ 0.7 M$_{\odot}$ there are two reasons
for the rapid decline of the $^{44}$Ti and $^{56}$Ni mass fraction
profiles from the parameterized trajectories while the post-process
mass fraction profiles remain essentially flat at $\approx$
10$^{-6}$ out to 1.5 M$_{\odot}$ First, the parameterized
trajectories assume an initial composition that is generally pure
$^{28}$Si (with perhaps some neutrons or protons to adjust $Y_e$;
see Section \ref{s.thermo_profiles}) which is dif\-ferent than the
initial composition of the hypernova model.  In region 6 and where
peak temperature drops below 4$\times$10$^9$ K  the yields from the
parameterized trajectories are initial composition dependent.
Second, the strong shock that follows the weak shock raises the
temperature to $\approx$ 3.5$\times$10$^{9}$ K at densities of
$\approx$ 10$^{6}$ g cm$^{-3}$ out to $\approx$ 1.5 M$_{\odot}$.
Over another 5 s of evolution, this is suf\-ficient to turn some of
the $^{28}$Si into $^{44}$Ti.

Yields from the exponential and power-law trajectories are generally
only within an order of magnitude of the post-process yields.
Integrating the $^{44}$Ti mass fraction profiles in Figure
\ref{fig:ws623_final_profile_post_vs_param} over the interior mass
gives the total mass of $^{44}$Ti ejected by this model.  We find
2.66$\times$10$^{-5}$ M$_{\odot}$ for post-processing,
5.23$\times$10$^{-5}$ M$_{\odot}$ for the exponential profile, and
6.34$\times$10$^{-5}$ M$_{\odot}$ for the power-law profile.
Similarly for $^{56}$Ni, we find 3.77$\times$10$^{-1}$ M$_{\odot}$
for post-processing, 4.83$\times$10$^{-1}$ M$_{\odot}$ for the
exponential profile, and 4.97$\times$10$^{-1}$ M$_{\odot}$ for the
power-law profile.  When the hydrodynamic evolution is complicated,
the post-process and parameterized profile yields will generally not
agree, but parameterized profiles still provide guidance on
interpreting the post-processed yields.

\subsection{A 2D Rotating Supernova Model}
\label{s.e15b}

Our third model is a 2D explosion of a rotating 15 M$_\odot$ star
\citep[model 1 from][]{fryer_2000_aa}.  This simulation follows the
collapse, bounce, explosion, and includes the entire proto-neutron
star throughout the evolution.  The dynamical trajectory of the
particles, and the time material spends at any location, can play a
major role in the final $^{44}$Ti and $^{56}$Ni yields.  In some
cases, the convective motion makes the matter undergo a series of
heating-cooling cycles.  However, for most matter, these cycles
occur well above NSE temperatures and, fortunately, the evolution in
the NSE regime does not af\-fect the final yields significantly
except through changes in the electron fraction.  This rotating
supernova model ends at 1.4 s after bounce which has two
implications.  First, some of the nominal ejecta may still fall back
onto the proto-neutron star. That is, some of the particles we
post-process may not ultimately be part of the nucleosynthetic
yield. Second, most of the particles have temperatures large enough
to interfere with a comparison of material that has undergone a
complete freeze-out via the exponential or power-law trajectory. To
facilitate this comparison we have appended exponential and
power-law thermodynamic trajectories to the final time point of the
dynamical model. In this manner we extend the thermodynamic
evolution to 4.2 s, by which time all the particles have
temperatures below 5$\times$10$^8$ K. The quantitative dif\-ferences
between the exponential tail and power-law tail appear small enough
that we will only show results for the power-law tail.

\begin{figure}[htp]
\includegraphics[width=0.9\textwidth]{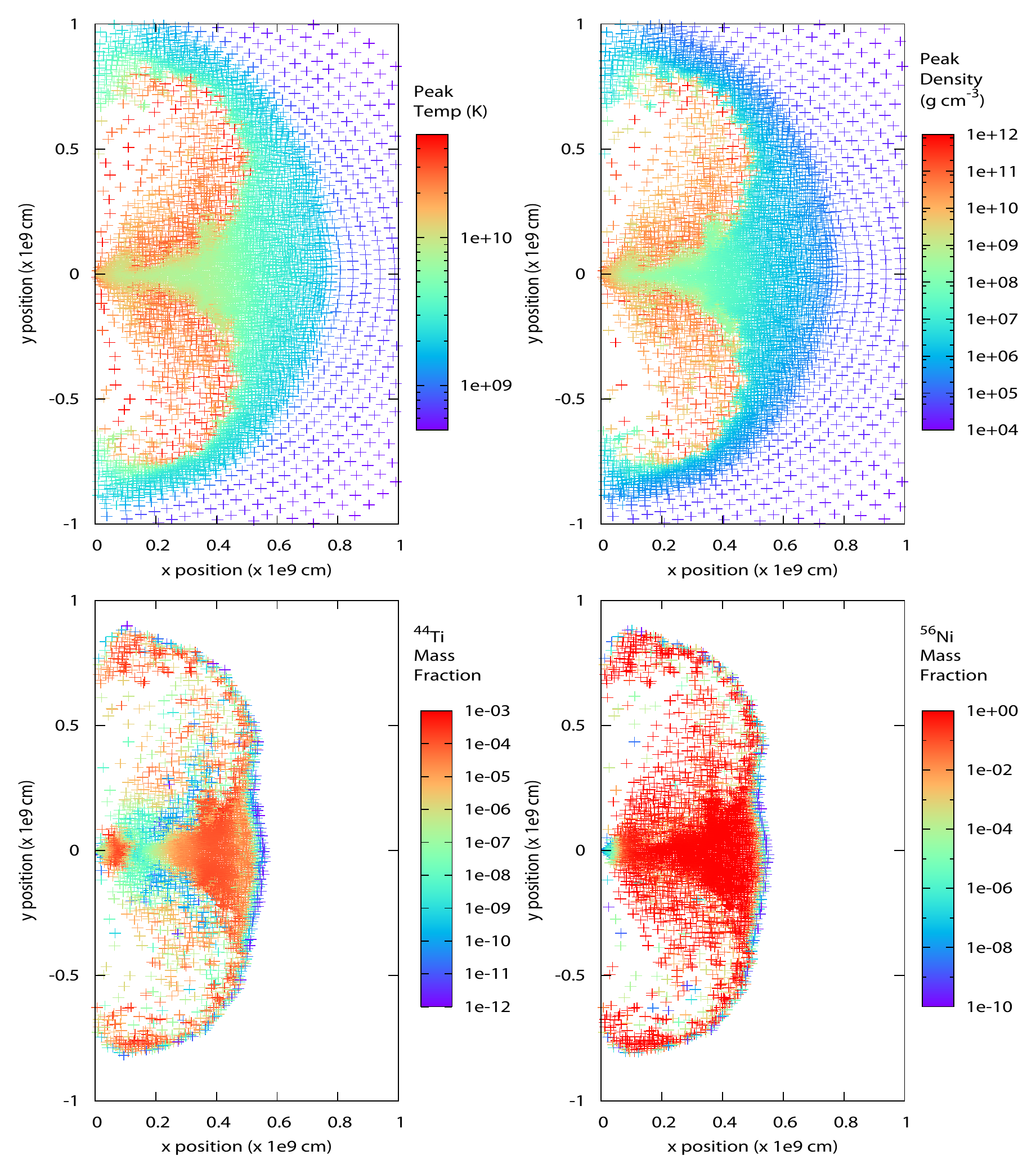}
\caption{ Peak temperatures and peak densities (top row) and mass
fractions of $^{44}$Ti and $^{56}$Ni (bottom row) generated by the
innermost regions in a 2D explosion of a rotating 15 M$_\odot$ star
\citep{fryer_2000_aa}. Particles coordinates are shown at 1.4 s, the
end of the dynamical model.  Note the y=0 equatorial plane, the
double lobbed structure, the general asymmetry of the model and
evidence for the $^{44}$Ti chasm along the equatorial regions.
        }
\label{fig:e15b_post_thermo_yields}
\end{figure}

The top row of Figure \ref{fig:e15b_post_thermo_yields} shows the
peak temperatures and peak densities within the innermost $\sim$ 1.2
M$_{\odot}$ at the coordinates reached by all particles at 4.2 s.
The equatorial plane is located at y=0, two lobes appear at roughly
$\pm$ 45$^{\circ}$, and the overall geometry is not symmetric due to
rotation and convective fluid motions. Most of the peak
thermodynamic conditions are within the bounds of our analysis. We
ignore those particles with peak temperatures above
1$\times$10$^{10}$ K or peak densities above 1$\times$10$^{10}$ g
cm$^{-3}$.  The post-process yields of $^{44}$Ti and $^{56}$Ni
within the innermost $\sim$ 1.2 M$_{\odot}$ are shown in the bottom
row of Figure \ref{fig:e15b_post_thermo_yields}. Most of the
$^{44}$Ti is created within roughly $\pm$ 15$^{\circ}$ of the
equatorial plane, but split into two distinct regions because of the
$^{44}$Ti chasm, the QSE-leakage region 2 of Figure
\ref{fig:contour_ti44_AD1_ye0500_regimes}.

\begin{figure}[htp]
\includegraphics[width=0.6\textwidth]{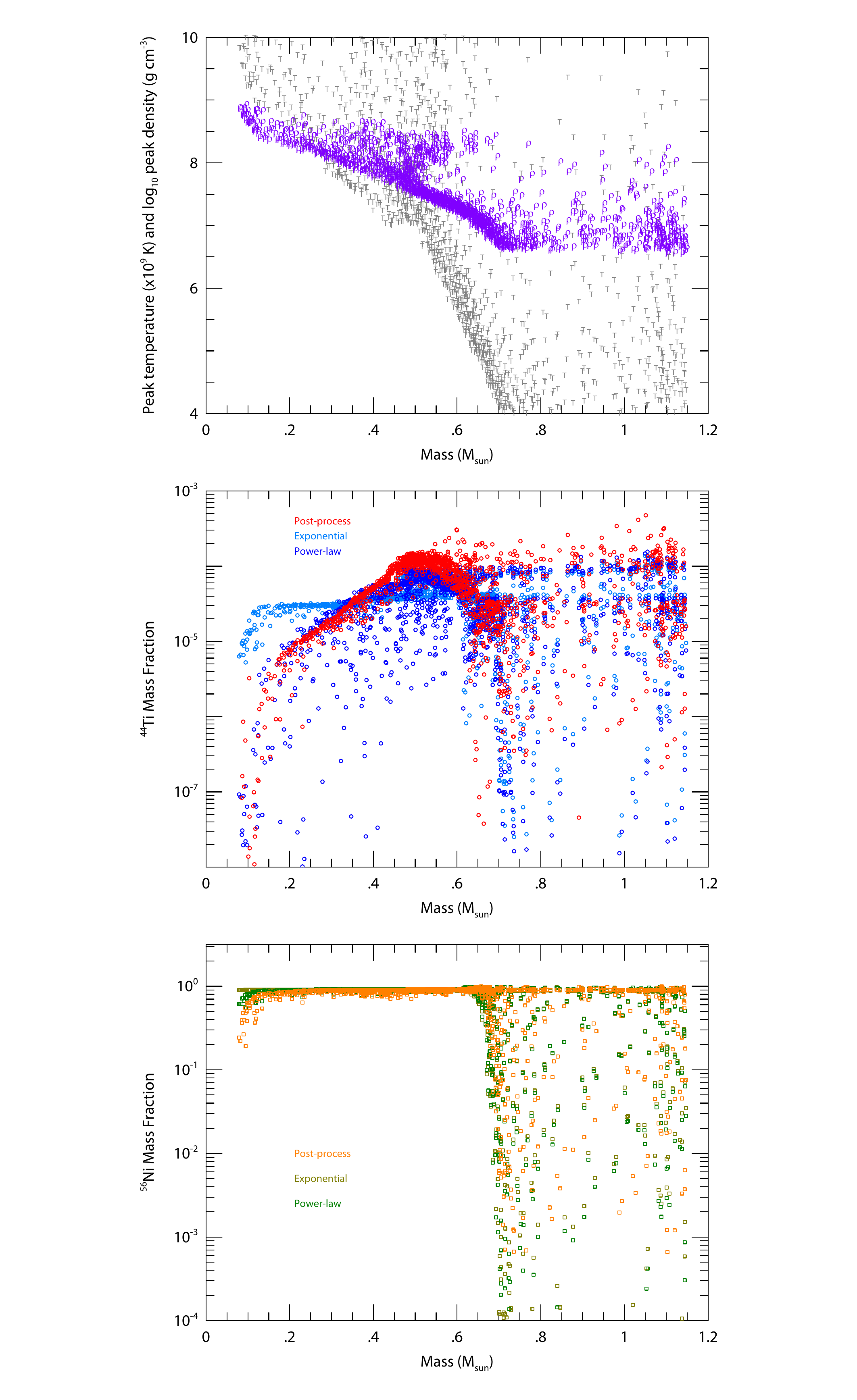}
\caption{ Peak temperatures in grey and peak densities in purple
(first row), and comparison of post-process vs parameterized
$^{44}$Ti (second row) and $^{56}$Ni (third row) profiles as a
function of interior mass for the 2D explosion of a rotating 15
M$_\odot$ star. The scatter is due to the asymmetries of the model.
} \label{fig:e15b_post_vs_param_temp_ti44_ni56}
\end{figure}

The first row of Figure \ref{fig:e15b_post_vs_param_temp_ti44_ni56}
shows the peak temperatures and peak densities as a function of
interior mass. There is considerable scatter at almost any mass
location due to the asymmetries inherent in the 2D model.  Mass
fraction profiles of $^{44}$Ti for the post-process, exponential,
and power-law trajectories are compared in the second row of the
figure. Between $\approx$ 0.1 M$_{\odot}$ and $\approx$ 0.3
M$_{\odot}$ the $^{44}$Ti mass fractions from all three
thermodynamic trajectories rapidly decrease because a subset of the
particles have peak temperatures and peak densities characteristic
of the chasm, the QSE-leakage region 2 of Figure
\ref{fig:contour_ti44_AD1_ye0500_regimes}.  The width of the chasm
associated with an exponential profile is narrower than the chasm of
a power-law profile, accounting for the power-law $^{44}$Ti mass
fraction profile decreasing more rapidly than the exponential
$^{44}$Ti mass fraction profile in this mass range. As expected from
our previous analysis, $^{56}$Ni undergoes no such phase transition,
with the result that all thermodynamic trajectories give values of
order unity in this mass range. Between $\approx$ 0.6 M$_{\odot}$
and $\approx$ 0.7 M$_{\odot}$ a large subset of the particles in the
equatorial regions have peak temperatures that drop below
4$\times$10$^9$ K (see Figure
\ref{fig:e15b_post_vs_param_temp_ti44_ni56}) and traverse the thin
chasm.  This results in the strong trend towards decreasing
$^{44}$Ti and $^{56}$Ni mass fraction profiles in this mass range.
Beyond $\approx$ 0.6 M$_{\odot}$ the two asymmetric lobes of Figure
\ref{fig:e15b_post_thermo_yields} are visible as the two horizonal
bands in the second row of Figure
\ref{fig:e15b_post_vs_param_temp_ti44_ni56}. The ``rain'' of points
in this mass range is due to the large scatter in the peak
thermodynamic conditions, where a number of particles have peak
temperatures that drop below 4$\times$10$^9$ K and traverse the thin
chasm.

Integrating the $^{44}$Ti mass fraction profiles over the interior
mass gives the total mass of $^{44}$Ti ejected by this model.  We find
6.98$\times$10$^{-5}$ M$_{\odot}$ for post-processing,
5.09$\times$10$^{-5}$ M$_{\odot}$ for the exponential profile, and
4.82$\times$10$^{-5}$ M$_{\odot}$ for the power-law profile.
Similarly for $^{56}$Ni, we find
3.89$\times$10$^{-1}$ M$_{\odot}$ for post-processing,
3.99$\times$10$^{-1}$ M$_{\odot}$ for the exponential profile, and
4.10$\times$10$^{-1}$ M$_{\odot}$ for the power-law profile.

Overall, the mass fraction profiles from exponential and power-law
trajectories are generally within a factor $\sim$ 4 of the
post-process values, except in regions where the chasm is traversed,
and the total yields of $^{44}$Ti and $^{56}$Ni from the
parameterized profiles mimic the post-process yields for this 2D,
rotating supernova model.

\section{Discussion}
\label{s.summary}

We have explored the trends in, and sensitivity to, the \ux{44}{Ti}
and \ux{56}{Ni} yields in the ejecta of three contrasting
core-collapse supernovae models. We used yields from two
parameterized expansion profiles and compared them to the yields
from post-processing trajectories from the core-collapse models.

Both parameterized profiles, the classic exponential and our new
power-law expressions, assume a constant $T^{3}/\rho$ adiabat
throughout the evolution.  For any given peak temperature and peak
density initial conditions, the power-law is slower compared to the
exponential and together they generally bound the trajectories from
core-collapse simulations. We find that \ux{44}{Ti} may be produced
by more than one type of freeze-out, depending on the peak
temperatures, densities and electron fraction values of the
thermodynamic trajectories. We have identified several distinct
regions in the peak temperature-density plane from the parameterized
profiles. Each region is characterized by dif\-ferent types of
transitions that the QSE cluster sustains during the evolution.
Reactions that break equilibrium are responsible for the flow
transfer to the remaining small scale clusters, maintaining their
structure until freeze-out. The result is unique mass fraction
profiles per region. The transitions are entropy driven, not
expansion timescale driven, where the temperature sets an
approximate threshold for a transition, while the density at the
threshold temperature determines whether the transition takes place
or not. The expansion timescale affects the locus of the borders
among different regions in the peak temperature-density plane.

For initially symmetric matter, region 1 is the normal freeze-out
regime, where no phase transition takes place and the yields from
the parameterized profiles are in good agreement with NSE or QSE
estimates. Region 2 is the $^{44}$Ti chasm, where \ux{44}{Ti} is
depleted as a result of the large scale QSE cluster dissolution to
two smaller ones, and the subsequent flow leakage from the Si-group
towards the Fe-group nuclei. Region 3 is the $\alpha$-rich
freeze-out regime, where the large scale QSE cluster shrinks and
shifts upward in mass due to the domination of the 3$\alpha$ forward
rate over its inverse.  Region 4 is the $\alpha$$p$-rich freeze-out
regime, where \pen\ and \nep\ drive the material slightly
proton-rich early in the evolution. Region 5 is the regime where
photodisintegrations dominate capture reactions. Region 6 is the
incomplete Si-burning regime, where the timescales for the plasma to
reach large scale QSE or NSE are comparable to the freeze-out
timescale, preventing in general their establishment.

Compared to the symmetric case, \ux{44}{Ti} and \ux{56}{Ni} are
gradually underproduced for initial $Y_{e} < 0.5$ and become less
sensitive to reaction rates. The basic structure of the
temperature-density plane is maintained, with the exception that the
$\alpha$$p$-rich freeze-out region decreases in size until complete
extinction for $Y_{e}\approx 0.46$ due to the absence of free
protons in neutron-rich environments, and the \ux{44}{Ti} chasm
region expands as a result of the \ux{44}{Ti} underproduction in the
neighboring regions in the temperature-density plane. In addition,
the region of $\alpha$$n$-rich freeze-out appears, where both
\ux{44}{Ti} and \ux{56}{Ni} are depleted since they are symmetric
isotopes.

For initial $Y_{e}>0.5$ \ux{44}{Ti} and \ux{56}{Ni} are favored by
large scale NSE and QSE equilibria due to the minimization of the
Helmholtz energy.  This results in \ux{56}{Ni} still being one of
the dominant yields, although symmetric isotopes are not favored by
the major flows once the large scale QSE cluster dissolves.  In this
electron fraction regime, weak interactions are crucial to
\ux{44}{Ti} production.  The dominance of \nep\ over \pen\ early in
the evolution and the rest of the weak interactions later on,
results in significant production of symmetric isotopes.  Regions 3
and 4 merge to become a regime of $\alpha$$p$-rich freeze-out due to
the large proton excess, and the $^{44}$Ti chasm expands. Region 1
is no longer a normal freeze-out regime, but it is characterized by
a phase transition due to the interplay between \mr{(p,\gamma)} and
weak reactions.

The three core-collapse models we post-processed were a 1D
Cassioppeia A model, a 1D double-shock hypernova model, and a 2D
rotating 15 M$_{\odot}$ model.  Mass fractions of \ux{44}{Ti} and
\ux{56}{Ni} from the exponential and power-law trajectories were
shown to generally lie within a factor $\sim$ 8 or less of the
post-process yields, except in regions where the thin chasm is being
crossed or the temperature fell below 4$\times$10$^9$ K. The total
ejected masses of \ux{44}{Ti} and \ux{56}{Ni} were shown to be
within a factor of $\sim$ 2 or less for all three models.  When the
thermodynamic trajectories of a core-collapse model have an
expansion profile similar to a parameterized expansion of any form,
it is generally safe to trust the yields from the parameterized
profiles.  For more complicated thermodynamic trajectories, the
yields from the parameterized profiles should not be trusted,
although the parameterized profiles may provide useful information
about the underlying physics.

\begin{figure}[htp]
\includegraphics[width=1.0\textwidth]{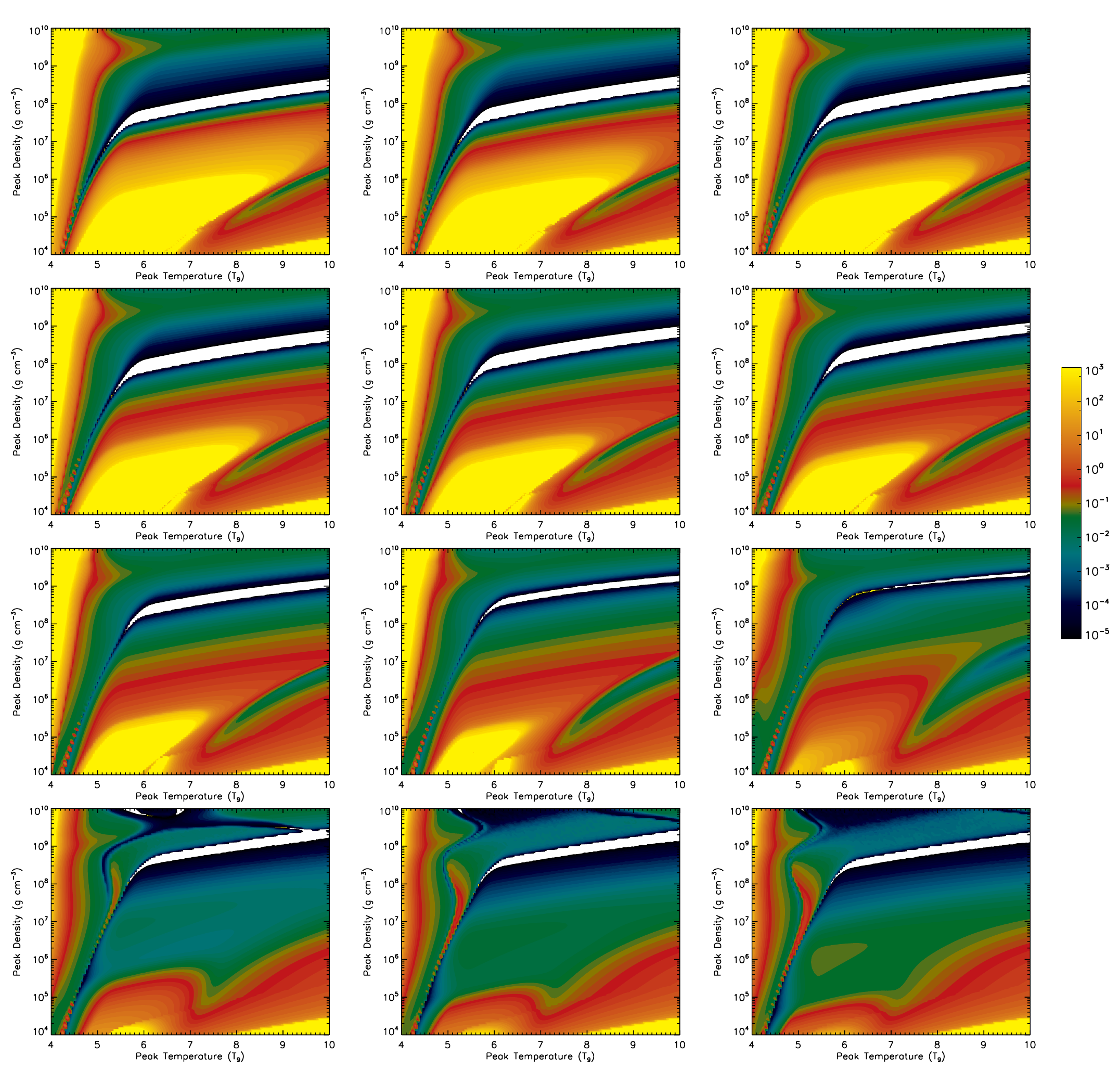}
\caption{ Same as Figure \ref{fig:contour_ti44_AD1_span}, but for
the \ux{44}{Ti} normalized production factor P$_{44}$.
        }
\label{fig:contour_P44_AD1_span}
\end{figure}

The location of the \ux{44}{Ti} chasm region is profile dependent,
and its width is minimized for initially symmetric matter but expand
dramatically for $Y_e \ne$0.5. These trends could account in part
for the observed paucity of supernova detected in the light of
radioactive \ux{44}{Ti}. A mass cut in supernovae models where the
electron fraction begins to fall below 0.5 may not be the most
suitable choice, since the layers above the mass cut are biased to
initially symmetric compositions and can produce ample \ux{44}{Ti}.
We find that variations in the \ux{44}{Ti}/\ux{56}{Ni} ratio
originate from variations in \ux{44}{Ti}, since \ux{56}{Ni} is
produced in large quantities over most of the peak
temperature-density plane (Figure \ref{fig:contour_P44_AD1_span}).
Although the massive production of $^{56}$Ni and its decay to
$^{56}$Co and $^{56}$Fe outshine every other decay during their
lifetime, the decay of $^{44}$Ti to $^{44}$Sc and $^{44}$Ca has a
longer lifetime. This implies that measurements of the yield from
$^{44}$Ti may be used to estimate the yield of $^{56}$Ni (e.g.
measure $^{44}$Ti in Cas A and deduce Cas A's $^{56}$Ni yield),
assuming the supernova models' thermodynamic trajectories
approximate one of the parameterized profiles and $^{44}$Ti
originates from regions where $^{56}$Ni was dominant.

\citet{woosley_1973_aa} first defined the notions of the normal and
$\alpha$-rich freeze-outs. Using an exponential profile they
describe how the various types of freeze-out are driven by the
ef\-fect of the 3$\alpha$ reaction. They identify 3 regions in the
peak temperature-density plane -- normal and $\alpha$-rich
freeze-outs and the incomplete silicon burning regime -- based on
the availability of $\alpha$-particles. These 3 regions are bordered
with thin lines based on semi-analytical relationships for the
timescale required to reach QSE. Their description is constrained to
the interplay between the 3$\alpha$ reaction and the large-scale QSE
cluster.

\citet{woosley_1992_aa} explored cases of freeze-outs starting at
large neutron excesses, resulting in the production of isotopes
heavier than the Fe-group and illustrating the smooth merging of the
freeze-out process into the r-process. They used the exponential
profile for a peak temperature $T_{9}=10$, three dif\-ferent values
for the peak density and a grid of initial neutron excesses within
the range $0\leqslant\eta\leqslant0.21$ (corresponding to a range
for the electron fraction $0.395\leqslant Y_{e}\leqslant0.5$). For
the specific case of initial $Y_{e}=0.48$ they mentioned that yields
for the Fe-group nuclei were not very dif\-ferent from the symmetric
case, but the production of heavier nuclei is also possible. This is
in agreement with our results (Table \ref{tab:dominant_yields}),
where the yields for \ux{44}{Ti} and \ux{56}{Ni} are decreased but
are not negligible compared to the symmetric case, and a region of
$\alpha$$n$-rich freeze-out appears in the peak temperature-density
planes. Their argument for flows being inhibited beyond \ux{56}{Ni}
for nearly symmetric compositions is functionally the same as ours.
Their argument is based on $Q$ values and other nuclear structure
ef\-fects around the doubly magic nucleus \ux{56}{Ni}, which is
equivalent to a persistent small equilibrium cluster localized
around \ux{56}{Ni} until complete freeze-out. Finally, they
suggested the possibility of freeze-out expansions with negative
neutron excesses ($Y_{e}>0.5$) due to the interplay between \pen\
and \nep, which we have discussed in detail within this study.

\citet{the_1998_aa} conducted the first detailed sensitivity study
of reaction rates on the \ux{44}{Ti} mass fraction by using the
exponential profile and one pair of peak conditions ($T_{9}=5.5$,
$\rho=10^7$ g cm$^{-3}$). They identified a significant amount of
reactions af\-fecting \ux{44}{Ti}, and suggested the impact of
\mr{\ux{45}{V}(p,\gamma)\ux{46}{Cr}} and
\mr{\ux{12}{C}(\alpha,\gamma)\ux{16}{O}} for $Y_{e}<0.5$. However,
the choice of only one set pair of peak conditions and one expansion
profile doesn't allow all crucial reactions identified in all
regions.

Recent sensitivity studies have varied rates within their
experimental uncertainty limits. \citet{hoffman_2010_aa} used
one-zone calculations with results for \ux{44}{Ti}, \ux{57}{Ni},
\ux{58}{Ni} and \ux{56}{Ni} presented in proportion to solar
\ux{56}{Fe}, considering in addition yields for these isotopes from
published supernovae models. They varied all published rate
compilations of \mr{\ux{44}{Ti}(\alpha,p)\ux{47}{V}} and
\mr{\ux{40}{Ca}(\alpha,\gamma)\ux{44}{Ti}} for peak conditions taken
from various points within the temperature-density plane given in
\citet{magkotsios_2008_aa}. They also varied the expansion timescale
of the exponential profile, concluding that the yield of \ux{44}{Ti}
is affected by the time it remains within successive burning stages.
\citet{tur_2010_aa} have utilized stellar evolution models to
address the impact of the $3\alpha$ and
\mr{\ux{12}{C}(\alpha,\gamma)\ux{16}{O}} reactions on \ux{26}{Al},
\ux{44}{Ti} and \ux{60}{Fe}. They conclude that \ux{44}{Ti} is
relatively insensitive to these rates, although they comment that
this result depends on the explosion physics and supernovae rate
assumed. Variations to the $3\alpha$ rate within our study also
result in relatively insensitive yields, because only a limited flow
is required by this reaction to impact the QSE cluster. In order to
illustrate the reaction's role, it is necessary to remove it
completely from network calculations.
\mr{\ux{12}{C}(\alpha,\gamma)\ux{16}{O}} has an impact on
\ux{44}{Ti} mostly for neutron-rich compositions (Table
\ref{tab:nuclear_reactions}).

Our study adds to these works by considering all the freeze-out
regions in the peak temperature-density plane over a broad range of
$Y_e$. The various types of freeze-out may be understood as
dif\-fering equilibrium patterns during the evolution, where a
change to the pattern is usually signaled by individual reactions
dropping out of equilibrium. The use of two expansion profiles and
detailed sensitivity studies for all regions within the peak
temperature-density plane reveals the importance of additional
reactions crucial to \ux{44}{Ti} synthesis beyond the set first
identified by \citet{the_1998_aa}.

\acknowledgments The authors thank Raphael Hix, Lih-Sin The, Rob
Hof\-fman, and Hendrik Schatz for useful discussions and test
calculations, and the anonymous referee for suggestions that
improved the manuscript. This work is supported by the NSF under
Grant PHY 02-16783 for the Frontier Center ``Joint Institute for
Nuclear Astrophysics'' (JINA), under US Government Contract
DE-AC52-06NA25396 for Los Alamos National Laboratory, which is
operated by the Los Alamos National Security, LLC (LANS) for the
U.S. Department of Energy.

\clearpage

\bibliographystyle{apj}
\bibliography{ti44_ni56}

\begin{thebibliography}{64}
\expandafter\ifx\csname natexlab\endcsname\relax\def\natexlab#1{#1}\fi

\bibitem[{{Aprahamian} {et~al.}(2005){Aprahamian}, {Langanke}, \&
  {Wiescher}}]{aprahamian_2005_aa}
{Aprahamian}, A., {Langanke}, K., \& {Wiescher}, M. 2005, {Progress in Particle
  and Nuclear Physics}, 54, 535

\bibitem[{{Arnett} {et~al.}(2008){Arnett}, {Meakin}, \&
  {Young}}]{arnett_2008_aa}
{Arnett}, D., {Meakin}, C., \& {Young}, P.~A. 2008, \apj

\bibitem[{{Arnett}(1977)}]{arnett_1977_aa}
{Arnett}, W.~D. 1977, \apjs, 35, 145

\bibitem[{{Arnett} {et~al.}(1989){Arnett}, {Bahcall}, {Kirshner}, \&
  {Woosley}}]{arnett_1989_aa}
{Arnett}, W.~D., {Bahcall}, J.~N., {Kirshner}, R.~P., \& {Woosley}, S.~E. 1989,
  \araa, 27, 629

\bibitem[{{Bodansky} {et~al.}(1968){Bodansky}, {Clayton}, \&
  {Fowler}}]{bodansky_1968_aa}
{Bodansky}, D., {Clayton}, D.~D., \& {Fowler}, W.~A. 1968, \apjs, 16, 299

\bibitem[{{Bruenn} {et~al.}(2006){Bruenn}, {Dirk}, {Mezzacappa}, {Hayes},
  {Blondin}, {Hix}, \& {Messer}}]{bruenn_2006_aa}
{Bruenn}, S.~W., {Dirk}, C.~J., {Mezzacappa}, A., {Hayes}, J.~C., {Blondin},
  J.~M., {Hix}, W.~R., \& {Messer}, O.~E.~B. 2006, Journal of Physics
  Conference Series, 46, 393

\bibitem[{{Buras} {et~al.}(2006){Buras}, {Janka}, {Rampp}, \&
  {Kifonidis}}]{buras_2006_aa}
{Buras}, R., {Janka}, H.-T., {Rampp}, M., \& {Kifonidis}, K. 2006, \aap, 457,
  281

\bibitem[{{Calder} {et~al.}(2007){Calder}, {Townsley}, {Seitenzahl}, {Peng},
  {Messer}, {Vladimirova}, {Brown}, {Truran}, \& {Lamb}}]{calder_2007_aa}
{Calder}, A.~C., {Townsley}, D.~M., {Seitenzahl}, I.~R., {Peng}, F., {Messer},
  O.~E.~B., {Vladimirova}, N., {Brown}, E.~F., {Truran}, J.~W., \& {Lamb},
  D.~Q. 2007, \apj, 656, 313

\bibitem[{{Eriksen} {et~al.}(2009){Eriksen}, {Arnett}, {McCarthy}, \&
  {Young}}]{eriksen_2009_aa}
{Eriksen}, K.~A., {Arnett}, D., {McCarthy}, D.~W., \& {Young}, P. 2009, \apj,
  697, 29

\bibitem[{{Fowler} \& {Hoyle}(1964)}]{fowler_1964_aa}
{Fowler}, W.~A., \& {Hoyle}, F. 1964, \apjs, 9, 201

\bibitem[{{Fr{\"o}hlich} {et~al.}(2006){Fr{\"o}hlich}, {Hauser},
  {Liebend{\"o}rfer}, {Mart{\'{\i}}nez-Pinedo}, {Thielemann}, {Bravo},
  {Zinner}, {Hix}, {Langanke}, {Mezzacappa}, \& {Nomoto}}]{frohlich_2006_aa}
{Fr{\"o}hlich}, C., {Hauser}, P., {Liebend{\"o}rfer}, M.,
  {Mart{\'{\i}}nez-Pinedo}, G., {Thielemann}, F.-K., {Bravo}, E., {Zinner},
  N.~T., {Hix}, W.~R., {Langanke}, K., {Mezzacappa}, A., \& {Nomoto}, K. 2006,
  \apj, 637, 415

\bibitem[{{Fryer} \& {Heger}(2000)}]{fryer_2000_aa}
{Fryer}, C.~L., \& {Heger}, A. 2000, \apj, 541, 1033

\bibitem[{{Fryer} \& {Young}(2007)}]{fryer_2007_aa}
{Fryer}, C.~L., \& {Young}, P.~A. 2007, \apj, 659, 1438

\bibitem[{{Fryer} {et~al.}(2006){Fryer}, {Young}, \&
  {Hungerford}}]{fryer_2006_ab}
{Fryer}, C.~L., {Young}, P.~A., \& {Hungerford}, A.~L. 2006, \apj, 650, 1028

\bibitem[{{Fryxell} {et~al.}(2000){Fryxell}, {Olson}, {Ricker}, {Timmes},
  {Zingale}, {Lamb}, {MacNeice}, {Rosner}, {Truran}, \&
  {Tufo}}]{fryxell_2000_aa}
{Fryxell}, B., {Olson}, K., {Ricker}, P., {Timmes}, F.~X., {Zingale}, M.,
  {Lamb}, D.~Q., {MacNeice}, P., {Rosner}, R., {Truran}, J.~W., \& {Tufo}, H.
  2000, \apjs, 131, 273

\bibitem[{{Fuller} {et~al.}(1980){Fuller}, {Fowler}, \&
  {Newman}}]{fuller_1980_aa}
{Fuller}, G.~M., {Fowler}, W.~A., \& {Newman}, M.~J. 1980, \apjs, 42, 447

\bibitem[{{Fuller} {et~al.}(1982{\natexlab{a}}){Fuller}, {Fowler}, \&
  {Newman}}]{fuller_1982_ab}
---. 1982{\natexlab{a}}, \apj, 252, 715

\bibitem[{{Fuller} {et~al.}(1982{\natexlab{b}}){Fuller}, {Fowler}, \&
  {Newman}}]{fuller_1982_aa}
---. 1982{\natexlab{b}}, \apjs, 48, 279

\bibitem[{{Fuller} \& {Meyer}(1995)}]{fuller_1995_aa}
{Fuller}, G.~M., \& {Meyer}, B.~S. 1995, \apj, 453, 792

\bibitem[{{Hartmann} {et~al.}(1985){Hartmann}, {Woosley}, \& {El
  Eid}}]{hartmann_1985_aa}
{Hartmann}, D., {Woosley}, S.~E., \& {El Eid}, M.~F. 1985, \apj, 297, 837

\bibitem[{{Hix} \& {Thielemann}(1996)}]{hix_1996_aa}
{Hix}, W.~R., \& {Thielemann}, F.-K. 1996, \apj, 460, 869

\bibitem[{{Hix} \& {Thielemann}(1999)}]{hix_1999_aa}
---. 1999, \apj, 511, 862

\bibitem[{{Hof\-fman} {et~al.}(2010){Hof\-fman}, {Sheets}, {Burke}, {Scielzo},
  {Rauscher}, {Norman}, {Tumey}, {Brown}, {Grant}, {Hurst}, {Phair}, {Stoyer},
  {Wooddy}, {Fisker}, \& {Bleuel}}]{hoffman_2010_aa}
{Hof\-fman}, R.~D., {Sheets}, S.~A., {Burke}, J.~T., {Scielzo}, N.~D.,
  {Rauscher}, T., {Norman}, E.~B., {Tumey}, S., {Brown}, T.~A., {Grant}, P.~G.,
  {Hurst}, A.~M., {Phair}, L., {Stoyer}, M.~A., {Wooddy}, T., {Fisker}, J.~L.,
  \& {Bleuel}, D. 2010, \apj, 715, 1383

\bibitem[{{Hoyle} {et~al.}(1964){Hoyle}, {Fowler}, {Burbidge}, \&
  {Burbidge}}]{hoyle_1964_aa}
{Hoyle}, F., {Fowler}, W.~A., {Burbidge}, G.~R., \& {Burbidge}, E.~M. 1964,
  \apj, 139, 909

\bibitem[{{Hungerford} {et~al.}(2005){Hungerford}, {Fryer}, \&
  {Rockefeller}}]{hungerford_2005_ab}
{Hungerford}, A.~L., {Fryer}, C.~L., \& {Rockefeller}, G. 2005, \apj, 635, 487

\bibitem[{{Iliadis}(2007)}]{iliadis_2007_aa}
{Iliadis}, C. 2007, {Nuclear Physics of Stars} (Wiley-VCH Verlag)

\bibitem[{{Kifonidis} {et~al.}(2006){Kifonidis}, {Plewa}, {Scheck}, {Janka}, \&
  {M{\"u}ller}}]{kifonidis_2006_aa}
{Kifonidis}, K., {Plewa}, T., {Scheck}, L., {Janka}, H.-T., \& {M{\"u}ller}, E.
  2006, \aap, 453, 661

\bibitem[{{Knie} {et~al.}(2004){Knie}, {Korschinek}, {Faestermann}, {Dorfi},
  {Rugel}, \& {Wallner}}]{knie_2004_aa}
{Knie}, K., {Korschinek}, G., {Faestermann}, T., {Dorfi}, E.~A., {Rugel}, G.,
  \& {Wallner}, A. 2004, Physical Review Letters, 93, 171103

\bibitem[{{Langanke} \& {Mart{\'{\i}}nez-Pinedo}(2001)}]{langanke_2001_aa}
{Langanke}, K., \& {Mart{\'{\i}}nez-Pinedo}, G. 2001, Atomic Data and Nuclear
  Data Tables, 79, 1

\bibitem[{{Liebend{\"o}rfer} {et~al.}(2008){Liebend{\"o}rfer}, {Fischer},
  {Fr{\"o}hlich}, {Thielemann}, \& {Whitehouse}}]{liebendorfer_2008_aa}
{Liebend{\"o}rfer}, M., {Fischer}, T., {Fr{\"o}hlich}, C., {Thielemann}, F.-K.,
  \& {Whitehouse}, S. 2008, Journal of Physics G Nuclear Physics, 35, 014056

\bibitem[{{Lodders}(2003)}]{lodders_2003_aa}
{Lodders}, K. 2003, \apj, 591, 1220

\bibitem[{{Lunardini} {et~al.}(2008){Lunardini}, {M{\"u}ller}, \&
  {Janka}}]{lunardini_2008_aa}
{Lunardini}, C., {M{\"u}ller}, B., \& {Janka}, H.-T. 2008, \prd, 78, 023016

\bibitem[{{Magkotsios} {et~al.}(2008){Magkotsios}, {Timmes}, {Wiescher},
  {Fryer}, {Hungerford}, {Young}, {Bennett}, {Diehl}, {Herwig}, {Hirschi},
  {Pignatari}, \& {Rockefeller}}]{magkotsios_2008_aa}
{Magkotsios}, G., {Timmes}, F.~X., {Wiescher}, M., {Fryer}, C.~L.,
  {Hungerford}, A., {Young}, P., {Bennett}, M., {Diehl}, S., {Herwig}, F.,
  {Hirschi}, R., {Pignatari}, M., \& {Rockefeller}, G. 2008, ArXiv e-prints

\bibitem[{{McLaughlin} \& {Fuller}(1995)}]{mclaughlin_1995_aa}
{McLaughlin}, G.~C., \& {Fuller}, G.~M. 1995, \apj, 455, 202

\bibitem[{{McLaughlin} {et~al.}(1996){McLaughlin}, {Fuller}, \&
  {Wilson}}]{mclaughlin_1996_aa}
{McLaughlin}, G.~C., {Fuller}, G.~M., \& {Wilson}, J.~R. 1996, \apj, 472, 440

\bibitem[{{Messer} {et~al.}(2008){Messer}, {Bruenn}, {Blondin}, {Hix}, \&
  {Mezzacappa}}]{messer_2008_aa}
{Messer}, O.~E.~B., {Bruenn}, S.~W., {Blondin}, J.~M., {Hix}, W.~R., \&
  {Mezzacappa}, A. 2008, Journal of Physics Conference Series, 125, 012010

\bibitem[{{Meyer}(1994)}]{meyer_1994_aa}
{Meyer}, B.~S. 1994, \araa, 32, 153

\bibitem[{{Meyer} {et~al.}(1998){Meyer}, {Krishnan}, \&
  {Clayton}}]{meyer_1998_aa}
{Meyer}, B.~S., {Krishnan}, T.~D., \& {Clayton}, D.~D. 1998, \apj, 498, 808

\bibitem[{{Nagataki} {et~al.}(1997){Nagataki}, {Hashimoto}, {Sato}, \&
  {Yamada}}]{nagataki_1997_aa}
{Nagataki}, S., {Hashimoto}, M.-A., {Sato}, K., \& {Yamada}, S. 1997, \apj,
  486, 1026

\bibitem[{{Oda} {et~al.}(1994){Oda}, {Hino}, {Muto}, {Takahara}, \&
  {Sato}}]{oda_1994_aa}
{Oda}, T., {Hino}, M., {Muto}, K., {Takahara}, M., \& {Sato}, K. 1994, Atomic
  Data and Nuclear Data Tables, 56, 231

\bibitem[{{Ott} {et~al.}(2008){Ott}, {Burrows}, {Dessart}, \&
  {Livne}}]{ott_2008_aa}
{Ott}, C.~D., {Burrows}, A., {Dessart}, L., \& {Livne}, E. 2008, \apj, 685,
  1069

\bibitem[{{Pruet} {et~al.}(2006){Pruet}, {Hoffman}, {Woosley}, {Janka}, \&
  {Buras}}]{pruet_2006_ab}
{Pruet}, J., {Hoffman}, R.~D., {Woosley}, S.~E., {Janka}, H.-T., \& {Buras}, R.
  2006, \apj, 644, 1028

\bibitem[{{Pruet} {et~al.}(2005){Pruet}, {Woosley}, {Buras}, {Janka}, \&
  {Hoffman}}]{pruet_2005_aa}
{Pruet}, J., {Woosley}, S.~E., {Buras}, R., {Janka}, H.-T., \& {Hoffman}, R.~D.
  2005, \apj, 623, 325

\bibitem[{{Rauscher} \& {Thielemann}(2000)}]{rath_2000_aa}
{Rauscher}, T., \& {Thielemann}, F.~K. 2000, Atomic Data and Nuclear Data
  Tables, 75, 1

\bibitem[{{Renaud} {et~al.}(2006){Renaud}, {Vink}, {Decourchelle}, {Lebrun},
  {Hartog}, {Terrier}, {Couvreur}, {Kn{\"o}dlseder}, {Martin}, {Prantzos},
  {Bykov}, \& {Bloemen}}]{renaud_2006_aa}
{Renaud}, M., {Vink}, J., {Decourchelle}, A., {Lebrun}, F., {Hartog}, P.~R.~d.,
  {Terrier}, R., {Couvreur}, C., {Kn{\"o}dlseder}, J., {Martin}, P.,
  {Prantzos}, N., {Bykov}, A.~M., \& {Bloemen}, H. 2006, \apjl, 647, L41

\bibitem[{{Seitenzahl} {et~al.}(2008){Seitenzahl}, {Timmes},
  {Marin-Lafl{\`e}che}, {Brown}, {Magkotsios}, \&
  {Truran}}]{seitenzahl_2008_aa}
{Seitenzahl}, I.~R., {Timmes}, F.~X., {Marin-Lafl{\`e}che}, A., {Brown}, E.,
  {Magkotsios}, G., \& {Truran}, J. 2008, \apjl, 685, L129

\bibitem[{{Surman} \& {McLaughlin}(2005)}]{surman_2005_aa}
{Surman}, R., \& {McLaughlin}, G.~C. 2005, \apj, 618, 397

\bibitem[{{The} {et~al.}(2006){The}, {Clayton}, {Diehl}, {Hartmann}, {Iyudin},
  {Leising}, {Meyer}, {Motizuki}, \& {Sch{\"o}nfelder}}]{the_2006_aa}
{The}, L.-S., {Clayton}, D.~D., {Diehl}, R., {Hartmann}, D.~H., {Iyudin},
  A.~F., {Leising}, M.~D., {Meyer}, B.~S., {Motizuki}, Y., \&
  {Sch{\"o}nfelder}, V. 2006, \aap, 450, 1037

\bibitem[{{The} {et~al.}(1998){The}, {Clayton}, {Jin}, \&
  {Meyer}}]{the_1998_aa}
{The}, L.-S., {Clayton}, D.~D., {Jin}, L., \& {Meyer}, B.~S. 1998, \apj, 504,
  500

\bibitem[{{Timmes}(1999)}]{timmes_1999_ab}
{Timmes}, F.~X. 1999, \apjs, 124, 241

\bibitem[{{Timmes} {et~al.}(1996){Timmes}, {Woosley}, {Hartmann}, \&
  {Hoffman}}]{timmes_1996_ab}
{Timmes}, F.~X., {Woosley}, S.~E., {Hartmann}, D.~H., \& {Hoffman}, R.~D. 1996,
  \apj, 464, 332

\bibitem[{{Tominaga} {et~al.}(2007){Tominaga}, {Umeda}, \&
  {Nomoto}}]{tominaga_2007_aa}
{Tominaga}, N., {Umeda}, H., \& {Nomoto}, K. 2007, \apj, 660, 516

\bibitem[{{Tur} {et~al.}(2010){Tur}, {Heger}, \& {Austin}}]{tur_2010_aa}
{Tur}, C., {Heger}, A., \& {Austin}, S.~M. 2010, \apj, 718, 357

\bibitem[{{Umeda} \& {Nomoto}(2008)}]{umeda_2008_aa}
{Umeda}, H., \& {Nomoto}, K. 2008, \apj, 673, 1014

\bibitem[{{Vink} {et~al.}(2001){Vink}, {Laming}, {Kaastra}, {Bleeker},
  {Bloemen}, \& {Oberlack}}]{vink_2001_aa}
{Vink}, J., {Laming}, J.~M., {Kaastra}, J.~S., {Bleeker}, J.~A.~M., {Bloemen},
  H., \& {Oberlack}, U. 2001, \apjl, 560, L79

\bibitem[{{Wadhwa} {et~al.}(2007){Wadhwa}, {Amelin}, {Davis}, {Lugmair},
  {Meyer}, {Gounelle}, \& {Desch}}]{wadhwa_2007_aa}
{Wadhwa}, M., {Amelin}, Y., {Davis}, A.~M., {Lugmair}, G.~W., {Meyer}, B.,
  {Gounelle}, M., \& {Desch}, S.~J. 2007, in Protostars and Planets V, ed.
  B.~{Reipurth}, D.~{Jewitt}, \& K.~{Keil}, 835--848

\bibitem[{{Wallerstein} {et~al.}(1997){Wallerstein}, {Iben}, {Parker},
  {Boesgaard}, {Hale}, {Champagne}, {Barnes}, {K{\"a}ppeler}, {Smith},
  {Hoffman}, {Timmes}, {Sneden}, {Boyd}, {Meyer}, \&
  {Lambert}}]{wallerstein_1997_aa}
{Wallerstein}, G., {Iben}, I.~J., {Parker}, P., {Boesgaard}, A.~M., {Hale},
  G.~M., {Champagne}, A.~E., {Barnes}, C.~A., {K{\"a}ppeler}, F., {Smith},
  V.~V., {Hoffman}, R.~D., {Timmes}, F.~X., {Sneden}, C., {Boyd}, R.~N.,
  {Meyer}, B.~S., \& {Lambert}, D.~L. 1997, Reviews of Modern Physics, 69, 995

\bibitem[{{Woosley} {et~al.}(1973){Woosley}, {Arnett}, \&
  {Clayton}}]{woosley_1973_aa}
{Woosley}, S.~E., {Arnett}, W.~D., \& {Clayton}, D.~D. 1973, \apjs, 26, 231

\bibitem[{{Woosley} \& {Hoffman}(1991)}]{woosley_1991_aa}
{Woosley}, S.~E., \& {Hoffman}, R.~D. 1991, \apjl, 368, L31

\bibitem[{{Woosley} \& {Hoffman}(1992)}]{woosley_1992_aa}
---. 1992, \apj, 395, 202

\bibitem[{{Young} {et~al.}(2008){Young}, {Ellinger}, {Timmes}, {Arnett},
  {Fryer}, {Rockefeller}, {Hungerford}, {Diehl}, {Bennett}, {Hirschi},
  {Pignatari}, {Herwig}, \& {Magkotsios}}]{young_2008_aa}
{Young}, P., {Ellinger}, C.~I., {Timmes}, F.~X., {Arnett}, D., {Fryer}, C.~L.,
  {Rockefeller}, G., {Hungerford}, A., {Diehl}, S., {Bennett}, M., {Hirschi},
  R., {Pignatari}, M., {Herwig}, F., \& {Magkotsios}, G. 2008, ArXiv e-prints

\bibitem[{{Young} \& {Fryer}(2007)}]{young_2007_aa}
{Young}, P.~A., \& {Fryer}, C.~L. 2007, \apj, 664, 1033

\bibitem[{{Young} {et~al.}(2006){Young}, {Fryer}, {Hungerford}, {Arnett},
  {Rockefeller}, {Timmes}, {Voit}, {Meakin}, \& {Eriksen}}]{young_2006_aa}
{Young}, P.~A., {Fryer}, C.~L., {Hungerford}, A., {Arnett}, D., {Rockefeller},
  G., {Timmes}, F.~X., {Voit}, B., {Meakin}, C., \& {Eriksen}, K.~A. 2006,
  \apj, 640, 891

\bibitem[{{Zinner}(1998)}]{zinner_1998_aa}
{Zinner}, E. 1998, Annual Review of Earth and Planetary Sciences, 26, 147

\end{thebibliography}

\clearpage

\begin{deluxetable}{lcccccccc}

\tablecaption{NUCLEAR NETWORKS} \tablewidth{293pt}
\tablehead{ &
\multicolumn{2}{c}{204} & \multicolumn{2}{c}{489} & \multicolumn{2}{c}{1341} &
\multicolumn{2}{c}{3304}\\ \hline Z & $A_{min}$ & $A_{max}$ &
$A_{min}$ & $A_{max}$ & $A_{min}$ & $A_{max}$ & $A_{min}$ & $A_{max}$}
\startdata
H  &  2       &  3 &  2    &  3       &  2  &  3       & 2   & 3   \\
He &  3       &  3 &  3    &  3       &  3  &  3       & 3   & 3   \\
Li &  6       &  7 &  6    &  7       &  6  &  9       & 6   & 9   \\
Be &  7       &  9 &  7    &  9       &  7  &  12      & 7   & 12  \\
B  &  8       &  11 &  8    &  11      &  8  &  14      & 8   & 14  \\
C  &  11      &  14 &  11      &  14      &  11  &  17      & 9   & 18  \\
N  &  13      &  15 &  12      &  15      &  12  &  20      & 11  & 21  \\
O  &  14      &  19 &  14      &  19      &  14  &  21      & 13  & 22  \\
F  &  17      &  19 &  17      &  21      &  18  &  22      & 16  & 26  \\
Ne &  18      &  23 &  17      &  24      &  18  &  29      & 16  & 31  \\
Na &  21      &  24 &  19      &  27      &  19  &  32      & 17  & 34  \\
Mg &  22      &  27 &  20      &  29      &  20  &  37      & 18  & 37  \\
Al &  25      &  28 &  22      &  31      &  22  &  40      & 20  & 40  \\
Si &  27      &  32 &  23      &  34      &  23  &  41      & 22  & 43  \\
P  &  29      &  34 &  27      &  38      &  27  &  44      & 24  & 46  \\
S  &  31      &  37 &  28      &  42      &  28  &  47      & 26  & 49  \\
Cl &  33      &  38 &  31      &  45      &  31  &  50      & 28  & 51  \\
Ar &  36      &  41 &  32      &  46      &  32  &  53      & 30  & 54  \\
K  &  37      &  42 &  35      &  49      &  34  &  58      & 32  & 56  \\
Ca &  40      &  49 &  36      &  49      &  36  &  59      & 34  & 59  \\
Sc &  41      &  50 &  40      &  51      &  40  &  64      & 36  & 64  \\
Ti &  44      &  51 &  41      &  53      &  41  &  55      & 38  & 67  \\
V  &  45      &  52 &  43      &  55      &  43  &  68      & 40  & 72  \\
Cr &  48      &  55 &  44      &  58      &  44  &  69      & 42  & 75  \\
Mn &  51      &  57 &  46      &  61      &  46  &  74      & 44  & 76  \\
Fe &  52      &  61 &  47      &  63      &  48  &  74      & 46  & 78  \\
Co &  55      &  62 &  50      &  65      &  50  &  78      & 48  & 80  \\
Ni &  56      &  65 &  51      &  67      &  51  &  80      & 50  & 83  \\
Cu &  57      &  66 &  55      &  69      &  57  &  85      & 52  & 86  \\
Zn &  60      &  69 &  57      &  72      &  59  &  86      & 54  & 89  \\
Ga &  61      &  70 &  59      &  75      &  59  &  94      & 56  & 92  \\
Ge &  64      &  71 &  62      &  78      &  62  &  97      & 58  & 95  \\
As & $\cdots$ & $\cdots$ &  65      &  79      &  68  &  103     & 60  & 100 \\
Se & $\cdots$ & $\cdots$ &  67      &  83      &  63  &  103     & 63  & 103 \\
Br & $\cdots$ & $\cdots$ &  68      &  83      &  69  &  106     & 65  & 105 \\
Kr & $\cdots$ & $\cdots$ &  69      &  87      &  72  &  109     & 68  & 108 \\
Rb & $\cdots$ & $\cdots$ &  73      &  85      &  74  &  113     & 74  & 111 \\
Sr & $\cdots$ & $\cdots$ &  74      &  84      &  76  &  118     & 73  & 114 \\
Y  & $\cdots$ & $\cdots$ &  75      &  87      &  78  &  121     & 75  & 119 \\
Zr & $\cdots$ & $\cdots$ &  78      &  90      &  80  &  122     & 77  & 122 \\
Nb & $\cdots$ & $\cdots$ &  82      &  90      &  81  &  123     & 80  & 124 \\
Mo & $\cdots$ & $\cdots$ &  83      &  90      &  82  &  125     & 82  & 125 \\
Tc & $\cdots$ & $\cdots$ &  89      &  91      &  87  &  127     & 85  & 126 \\
Ru & $\cdots$ & $\cdots$ & $\cdots$ & $\cdots$ &  90  &  130     & 86  & 128 \\
Rh & $\cdots$ & $\cdots$ & $\cdots$ & $\cdots$ &  93  &  131     & 89  & 130 \\
Pd & $\cdots$ & $\cdots$ & $\cdots$ & $\cdots$ &  94  &  132     & 91  & 132 \\
Ag & $\cdots$ & $\cdots$ & $\cdots$ & $\cdots$ &  97  &  133     & 93  & 134 \\
Cd & $\cdots$ & $\cdots$ & $\cdots$ & $\cdots$ &  98  &  136     & 95  & 134 \\
In & $\cdots$ & $\cdots$ & $\cdots$ & $\cdots$ &  99  &  149     & 97  & 137 \\
Sn & $\cdots$ & $\cdots$ & $\cdots$ & $\cdots$ &  102  &  152     & 99  & 140 \\
Sb & $\cdots$ & $\cdots$ & $\cdots$ & $\cdots$ & $\cdots$ & $\cdots$ & 105 & 151 \\
Te & $\cdots$ & $\cdots$ & $\cdots$ & $\cdots$ & $\cdots$ & $\cdots$ & 109 & 154 \\
I  & $\cdots$ & $\cdots$ & $\cdots$ & $\cdots$ & $\cdots$ & $\cdots$ & 112 & 158 \\
Xe & $\cdots$ & $\cdots$ & $\cdots$ & $\cdots$ & $\cdots$ & $\cdots$ & 113 & 161 \\
Cs & $\cdots$ & $\cdots$ & $\cdots$ & $\cdots$ & $\cdots$ & $\cdots$ & 118 & 165 \\
Ba & $\cdots$ & $\cdots$ & $\cdots$ & $\cdots$ & $\cdots$ & $\cdots$ & 119 & 168 \\
La & $\cdots$ & $\cdots$ & $\cdots$ & $\cdots$ & $\cdots$ & $\cdots$ & 122 & 172 \\
Ce & $\cdots$ & $\cdots$ & $\cdots$ & $\cdots$ & $\cdots$ & $\cdots$ & 122 & 175 \\
Pr & $\cdots$ & $\cdots$ & $\cdots$ & $\cdots$ & $\cdots$ & $\cdots$ & 126 & 178 \\
Nd & $\cdots$ & $\cdots$ & $\cdots$ & $\cdots$ & $\cdots$ & $\cdots$ & 127 & 178 \\
Pm & $\cdots$ & $\cdots$ & $\cdots$ & $\cdots$ & $\cdots$ & $\cdots$ & 130 & 185 \\
Sm & $\cdots$ & $\cdots$ & $\cdots$ & $\cdots$ & $\cdots$ & $\cdots$ & 133 & 188 \\
Eu & $\cdots$ & $\cdots$ & $\cdots$ & $\cdots$ & $\cdots$ & $\cdots$ & 136 & 190 \\
Gd & $\cdots$ & $\cdots$ & $\cdots$ & $\cdots$ & $\cdots$ & $\cdots$ & 139 & 191 \\
Tb & $\cdots$ & $\cdots$ & $\cdots$ & $\cdots$ & $\cdots$ & $\cdots$ & 142 & 192 \\
Dy & $\cdots$ & $\cdots$ & $\cdots$ & $\cdots$ & $\cdots$ & $\cdots$ & 143 & 193 \\
Ho & $\cdots$ & $\cdots$ & $\cdots$ & $\cdots$ & $\cdots$ & $\cdots$ & 146 & 196 \\
Er & $\cdots$ & $\cdots$ & $\cdots$ & $\cdots$ & $\cdots$ & $\cdots$ & 148 & 198 \\
Tm & $\cdots$ & $\cdots$ & $\cdots$ & $\cdots$ & $\cdots$ & $\cdots$ & 150 & 198 \\
Yb & $\cdots$ & $\cdots$ & $\cdots$ & $\cdots$ & $\cdots$ & $\cdots$ & 152 & 200 \\
Lu & $\cdots$ & $\cdots$ & $\cdots$ & $\cdots$ & $\cdots$ & $\cdots$ & 156 & 209 \\
Hf & $\cdots$ & $\cdots$ & $\cdots$ & $\cdots$ & $\cdots$ & $\cdots$ & 159 & 212 \\
Ta & $\cdots$ & $\cdots$ & $\cdots$ & $\cdots$ & $\cdots$ & $\cdots$ & 161 & 217 \\
W  & $\cdots$ & $\cdots$ & $\cdots$ & $\cdots$ & $\cdots$ & $\cdots$ & 163 & 220 \\
Re & $\cdots$ & $\cdots$ & $\cdots$ & $\cdots$ & $\cdots$ & $\cdots$ & 167 & 225 \\
Os & $\cdots$ & $\cdots$ & $\cdots$ & $\cdots$ & $\cdots$ & $\cdots$ & 169 & 226 \\
Ir & $\cdots$ & $\cdots$ & $\cdots$ & $\cdots$ & $\cdots$ & $\cdots$ & 172 & 230 \\
Pt & $\cdots$ & $\cdots$ & $\cdots$ & $\cdots$ & $\cdots$ & $\cdots$ & 175 & 232 \\
Au & $\cdots$ & $\cdots$ & $\cdots$ & $\cdots$ & $\cdots$ & $\cdots$ & 178 & 236 \\
Hg & $\cdots$ & $\cdots$ & $\cdots$ & $\cdots$ & $\cdots$ & $\cdots$ & 178 & 239 \\
Tl & $\cdots$ & $\cdots$ & $\cdots$ & $\cdots$ & $\cdots$ & $\cdots$ & 182 & 245 \\
Pb & $\cdots$ & $\cdots$ & $\cdots$ & $\cdots$ & $\cdots$ & $\cdots$ & 185 & 246 \\
Bi & $\cdots$ & $\cdots$ & $\cdots$ & $\cdots$ & $\cdots$ & $\cdots$ & 188 & 251 \\
Po & $\cdots$ & $\cdots$ & $\cdots$ & $\cdots$ & $\cdots$ & $\cdots$ & 193 & 237 \\
At & $\cdots$ & $\cdots$ & $\cdots$ & $\cdots$ & $\cdots$ & $\cdots$ & 210 & 211 \\
\enddata
\label{tab:nuclear_networks}
\end{deluxetable}

\clearpage

\begin{deluxetable}{cccc}

\tablecaption{FREEZE-OUT DOMINANT YIELDS} \tablewidth{450pt}
\tablehead{\colhead{Freeze-out} & \colhead{$\rho_{peak}$ (g
cm$^{-3}$)} & \colhead{Exponential} & \colhead{Power-Law}}
\startdata
\multicolumn{4}{c}{$Y_{e}=0.48$, $T_{9}=9$}\\
\hline Normal & $5\times10^{9}$ & \ux{54}{Fe}, \ux{58}{Ni},
\ux{56}{Fe}, \ux{55}{Fe}, \ux{60}{Ni} & \ux{54}{Fe}, \ux{56}{Fe},
\ux{58}{Ni}, \ux{60}{Ni},
\ux{55}{Fe}\\
$\alpha$-rich & $10^{7}$ & \ux{58}{Ni}, \ux{4}{He}, \ux{60}{Ni},
\ux{64}{Zn}, \ux{62}{Zn} & \ux{58}{Ni}, \ux{4}{He}, \ux{60}{Ni},
\ux{64}{Zn}, \ux{62}{Zn}\\
$\alpha$$p$-rich & $\cdots$ & $\cdots$ & $\cdots$\\
$\alpha$$n$-rich & $5\times10^{4}$ ($T_{9}=6$) & \ux{4}{He},
\ux{87}{Kr}, n, \ux{49}{Ca}, \ux{83}{Se} & \ux{4}{He}, n,
\ux{49}{Ca}, \ux{87}{Kr},
\ux{72}{Zn} \\
\hline
\multicolumn{4}{c}{$Y_{e}=0.50$, $T_{9}=9$}\\
\hline Normal & $5\times10^{9}$ & \ux{56}{Ni}, \ux{54}{Fe},
\ux{52}{Fe}, \ux{55}{Co}, \ux{57}{Ni} & \ux{54}{Fe},
\ux{58}{Ni}, \ux{56}{Ni}, \ux{55}{Fe}, \ux{56}{Fe}\\
$\alpha$-rich & $10^{8}$ & \ux{56}{Ni}, \ux{4}{He}, \ux{60}{Zn},
\ux{57}{Ni}, \ux{58}{Cu} & \ux{56}{Ni}, \ux{60}{Cu}, \ux{4}{He},
\ux{57}{Ni}, \ux{58}{Ni}\\
$\alpha$$p$-rich & $10^{6}$ & \ux{4}{He}, \ux{56}{Ni}, p,
\ux{60}{Zn}, \ux{52}{Fe} &
\ux{4}{He}, \ux{56}{Ni}, p, \ux{60}{Cu}, \ux{52}{Fe}\\
$\alpha$$n$-rich & $\cdots$ & $\cdots$ & $\cdots$\\
\hline
\multicolumn{4}{c}{$Y_{e}=0.52$, $T_{9}=9$}\\
\hline \mr{(p,\gamma)} leakage & $10^{8}$ ($T_{9}=5$) & \ux{56}{Ni},
\ux{58}{Cu}, \ux{59}{Cu}, \ux{58}{Ni}, \ux{57}{Ni} &
\ux{56}{Ni}, \ux{60}{Cu}, \ux{59}{Ni}, \ux{60}{Zn}, \ux{60}{Ni}\\
$\alpha$-rich & $\cdots$ & $\cdots$ & $\cdots$ \\
$\alpha$$p$-rich & $10^{7}$ & \ux{56}{Ni}, \ux{4}{He}, p,
\ux{60}{Zn}, \ux{57}{Ni} & \ux{56}{Ni}, \ux{4}{He}, p, \ux{57}{Ni}, \ux{60}{Cu}\\
$\alpha$$n$-rich & $\cdots$ & $\cdots$ & $\cdots$\\
\enddata
\label{tab:dominant_yields}
\end{deluxetable}

\clearpage

\begin{deluxetable}{cccccc}

\tablecaption{NUCLEAR REACTIONS RELEVANT TO \ux{44}{Ti} SYNTHESIS}
\tablewidth{500pt}

\tablehead{\colhead{Reaction} &\colhead{Contribution} &
\colhead{Rank} & \colhead{$Y_{e}$} & \colhead{Region} &
\colhead{Profile}} \startdata
\multicolumn{6}{c}{Global scope reactions}\\
\hline
3$\alpha$   &   flow transfer to QSE cluster    &   primary &   0.48-0.52   &   2-5 &   both    \\
\pen\ / \nep    &   $Y_e$ adjustment    &   primary &   0.48-0.52   &   1-5 &   both    \\
\pen\ / \nep    &   chasm widening  &   primary &   0.50-0.52   &   2   &   both    \\
\hline
\multicolumn{6}{c}{\mr{(\alpha,\gamma)} reactions}\\
\hline
\mr{\ux{40}{Ca}(\alpha,\gamma)\ux{44}{Ti}}  &   2nd arc amplitude/slope &   primary &   0.48-0.52   &   3-4 &   both    \\
\mr{\ux{12}{C}(\alpha,\gamma)\ux{16}{O}}    &   flow transfer to QSE cluster    &   secondary   &   0.48    &   2-5 &   both    \\
\mr{\ux{7}{Be}(\alpha,\gamma)\ux{11}{C}}    &   chasm depth &   secondary   &   0.5 &   2   &   both    \\
\mr{\ux{24}{Mg}(\alpha,\gamma)\ux{28}{Si}}  &   chasm depth &   secondary   &   0.5 &   2   &   both    \\
\mr{\ux{42}{Ca}(\alpha,\gamma)\ux{46}{Ti}}  &   chasm depth, 1st arc dip    &   secondary   &   0.48    &   2-3 &   both    \\
\hline
\multicolumn{6}{c}{\mr{(\alpha,p)} reactions}\\
\hline
\mr{\ux{44}{Ti}(\alpha,p)\ux{47}{V}}    &   chasm formation, depth, shift   &   primary &   0.48-0.52   &   2   &   both    \\
\mr{\ux{44}{Ti}(\alpha,p)\ux{47}{V}}    &   1st arc dip / 2nd arc slope &   primary &   0.48-0.50   &   3-4 &   both    \\
\mr{\ux{44}{Ti}(\alpha,p)\ux{47}{V}}    &   1st arc dip &   primary &   0.52    &   1   &   both    \\
\mr{\ux{40}{Ca}(\alpha,p)\ux{43}{Sc}}   &   2nd arc dip/slope   &   primary &   0.48-0.50   &   3-4 &   both    \\
\mr{\ux{17}{F}(\alpha,p)\ux{20}{Ne}}    &   1st arc dip &   primary &   0.52    &   1   &   both    \\
\mr{\ux{21}{Na}(\alpha,p)\ux{24}{Mg}}   &   1st arc dip &   primary &   0.52    &   1   &   both    \\
\mr{\ux{40}{Ca}(\alpha,p)\ux{43}{Sc}}   &   3rd arc amplitude   &   secondary   &   0.52    &   4   &   power-law   \\
\mr{\ux{27}{Al}(\alpha,p)\ux{30}{Si}}   &   chasm depth &   secondary   &   0.5 &   2   &   exponential \\
\mr{\ux{55}{Co}(\alpha,p)\ux{58}{Ni}}   &   chasm depth &   secondary   &   0.5 &   2   &   exponential \\
\mr{\ux{48}{Cr}(\alpha,p)\ux{51}{Mn}}   &   chasm depth &   secondary   &   0.5 &   2   &   exponential \\
\mr{\ux{17}{F}(\alpha,p)\ux{20}{Ne}}    &   chasm depth &   secondary   &   0.5 &   2   &   exponential \\
\mr{\ux{52}{Fe}(\alpha,p)\ux{55}{Co}}   &   chasm depth &   secondary   &   0.5 &   2   &   exponential \\
\mr{\ux{54}{Fe}(\alpha,p)\ux{57}{Co}}   &   chasm depth &   secondary   &   0.5 &   2   &   exponential \\
\mr{\ux{21}{Na}(\alpha,p)\ux{24}{Mg}}   &   chasm depth &   secondary   &   0.5 &   2   &   exponential \\
\mr{\ux{56}{Ni}(\alpha,p)\ux{59}{Cu}}   &   chasm depth &   secondary   &   0.5 &   2   &   exponential \\
\mr{\ux{6}{Li}(\alpha,p)\ux{9}{Be}} &   chasm depth &   secondary   &   0.5 &   2   &   power-law   \\
\mr{\ux{13}{N}(\alpha,p)\ux{16}{O}} &   chasm depth &   secondary   &   0.5 &   2   &   power-law   \\
\mr{\ux{42}{Ca}(\alpha,p)\ux{45}{Sc}}   &   1st arc dip / 2nd arc slope &   secondary   &   0.48    &   3   &   both    \\
\mr{\ux{43}{Sc}(\alpha,p)\ux{46}{Ti}}   &   1st arc dip &   secondary   &   0.48    &   3   &   both    \\
\mr{\ux{58}{Ni}(\alpha,p)\ux{61}{Cu}}   &   2nd arc amplitude   &   secondary   &   0.48    &   3   &   both    \\
\mr{\ux{38}{Ca}(\alpha,p)\ux{41}{Sc}}   &   1st arc dip &   secondary   &   0.52    &   1   &   both    \\
\mr{\ux{34}{Ar}(\alpha,p)\ux{37}{K}}    &   3rd arc amplitude   &   secondary   &   0.52    &   4   &   power-law   \\
\mr{\ux{38}{Ca}(\alpha,p)\ux{41}{Sc}}   &   3rd arc amplitude   &   secondary   &   0.52    &   4   &   power-law   \\
\hline
\multicolumn{6}{c}{\mr{(p,\gamma)} reactions}\\
\hline
\mr{\ux{45}{V}(p,\gamma)\ux{46}{Cr}}    &   2nd arc formation/dip   &   primary &   0.50-0.52   &   1,3,4   &   both    \\
\mr{\ux{45}{V}(p,\gamma)\ux{46}{Cr}}    &   3rd arc formation   &   primary &   0.52    &   4   &   power-law   \\
\mr{\ux{41}{Sc}(p,\gamma)\ux{42}{Ti}}   &   2nd arc dip &   primary &   0.5 &   4   &   both    \\
\mr{\ux{43}{Sc}(p,\gamma)\ux{44}{Ti}}   &   2nd arc dip &   primary &   0.5 &   4   &   both    \\
\mr{\ux{44}{Ti}(p,\gamma)\ux{45}{V}}    &   2nd arc dip / 3rd arc formation &   primary &   0.5 &   4   &   both    \\
\mr{\ux{44}{Ti}(p,\gamma)\ux{45}{V}}    &   \mr{\ux{44}{Ti}-\ux{45}{V}} cluster &   primary &   0.5 &   4   &   both    \\
\mr{\ux{41}{Sc}(p,\gamma)\ux{42}{Ti}}   &   2nd arc dip &   primary &   0.5, 0.52   &   4   &   both    \\
\mr{\ux{57}{Ni}(p,\gamma)\ux{58}{Cu}}   &   flow transfer within QSE cluster    &   primary &   0.5 &   3   &   both    \\
\mr{\ux{45}{V}(p,\gamma)\ux{46}{Cr}}    &   regions 3-4 borderline  &   secondary   &   0.5 &   3-4 &   both    \\
\mr{\ux{40}{Ca}(p,\gamma)\ux{41}{Sc}}   &   post-2nd dip track  &   secondary   &   0.5 &   4   &   both    \\
\mr{\ux{44}{V}(p,\gamma)\ux{45}{Cr}}    &   post-2nd dip track  &   secondary   &   0.52    &   1,4 &   both    \\
\mr{\ux{43}{Ti}(p,\gamma)\ux{44}{V}}    &   post-2nd dip track  &   secondary   &   0.52    &   1,4 &   both    \\
\mr{\ux{42}{Sc}(p,\gamma)\ux{43}{Ti}}   &   post-2nd dip track  &   secondary   &   0.52    &   1,4 &   both    \\
\mr{\ux{57}{Cu}(p,\gamma)\ux{58}{Zn}}   &   post-2nd dip track  &   secondary   &   0.52    &   1,4 &   both    \\
\mr{\ux{20}{Ne}(p,\gamma)\ux{21}{Na}}   &   post-2nd dip track  &   secondary   &   0.52    &   1,4 &   both    \\
\mr{\ux{47}{V}(p,\gamma)\ux{48}{Cr}}    &   post-2nd dip track  &   secondary   &   0.52    &   1,4 &   both    \\
\mr{\ux{43}{Sc}(p,\gamma)\ux{44}{Ti}}   &   post-2nd dip track  &   secondary   &   0.48, 0.52  &   1,3,4   &   both    \\
\mr{\ux{43}{Ti}(p,\gamma)\ux{44}{V}}    &   3rd arc amplitude   &   secondary   &   0.52    &   4   &   power-law   \\
\mr{\ux{41}{Sc}(p,\gamma)\ux{42}{Ti}}   &   3rd arc amplitude   &   secondary   &   0.52    &   4   &   power-law   \\
\mr{\ux{43}{Sc}(p,\gamma)\ux{44}{Ti}}   &   3rd arc amplitude   &   secondary   &   0.52    &   4   &   power-law   \\
\mr{\ux{40}{Ca}(p,\gamma)\ux{41}{Sc}}   &   post-2nd dip track  &   secondary   &   0.52    &   4   &   exponential \\
\mr{\ux{40}{Ca}(p,\gamma)\ux{41}{Sc}}   &   3rd arc amplitude   &   secondary   &   0.52    &   4   &   power-law   \\
\mr{\ux{42}{Ca}(p,\gamma)\ux{43}{Sc}}   &   2nd arc slope   &   secondary   &   0.48    &   3   &   both    \\
\mr{\ux{39}{K}(p,\gamma)\ux{40}{Ca}}    &   2nd arc slope   &   secondary   &   0.48    &   3   &   exponential \\
\mr{\ux{57}{Co}(p,\gamma)\ux{58}{Ni}}   &   chasm depth &   secondary   &   0.5 &   2   &   exponential \\
\mr{\ux{54}{Fe}(p,\gamma)\ux{55}{Co}}   &   chasm depth &   secondary   &   0.5 &   2   &   exponential \\
\mr{\ux{52}{Mn}(p,\gamma)\ux{53}{Fe}}   &   chasm depth &   secondary   &   0.5 &   2   &   exponential \\
\mr{\ux{57}{Ni}(p,\gamma)\ux{58}{Cu}}   &   regions 3-4 borderline  &   secondary   &   0.5 &   3-4 &   both    \\
\hline
\multicolumn{6}{c}{weak reactions}\\
\hline
\mr{\ux{54}{Co}(e^{-},\nu_{e})\ux{54}{Fe}}  &   chasm widening  &   primary &   0.50-0.52   &   2   &   both    \\
\mr{\ux{50}{Mn}(e^{-},\nu_{e})\ux{50}{Cr}}  &   chasm widening  &   primary &   0.50-0.52   &   2   &   both    \\
\mr{\ux{55}{Ni}(e^{-},\nu_{e})\ux{55}{Co}}  &   chasm widening  &   primary &   0.50-0.52   &   2   &   both    \\
\mr{\ux{57}{Cu}(e^{-},\nu_{e})\ux{57}{Ni}}  &   chasm widening  &   primary &   0.50-0.52   &   2   &   both    \\
\mr{\ux{53}{Co}(e^{-},\nu_{e})\ux{53}{Fe}}  &   chasm widening  &   primary &   0.50-0.52   &   2   &   both    \\
\mr{\ux{51}{Fe}(e^{-},\nu_{e})\ux{51}{Mn}}  &   chasm widening  &   primary &   0.50-0.52   &   2   &   both    \\
\mr{\ux{58}{Cu}(e^{-},\nu_{e})\ux{58}{Ni}}  &   chasm widening  &   primary &   0.50-0.52   &   2   &   power-law   \\
\mr{\ux{59}{Cu}(e^{-},\nu_{e})\ux{59}{Ni}}  &   chasm widening  &   primary &   0.50-0.52   &   2   &   power-law   \\
\mr{\ux{42}{Ti}(e^{-},\nu_{e})\ux{42}{Sc}}  &   2nd arc dip / post-2nd dip track    &   primary &   0.50-0.52   &   4   &   both    \\
\mr{\ux{41}{Sc}(e^{-},\nu_{e})\ux{41}{Ca}}  &   2nd arc dip / post-2nd dip track    &   primary &   0.50-0.52   &   4   &   both    \\
\mr{\ux{43}{Ti}(e^{-},\nu_{e})\ux{43}{Sc}}  &   2nd arc dip / post-2nd dip track    &   primary &   0.50-0.52   &   4   &   both    \\
\mr{\ux{44}{V}(e^{-},\nu_{e})\ux{44}{Ti}}   &   2nd arc dip / post-2nd dip track    &   primary &   0.50-0.52   &   4   &   both    \\
\mr{\ux{44}{V}(e^{-},\nu_{e})\ux{44}{Ti}}   &   2nd arc dip / post-2nd dip track    &   primary &   0.52    &   4   &   exponential \\
\mr{\ux{38}{Ca}(e^{-},\nu_{e})\ux{38}{K}}   &   2nd arc dip / post-2nd dip track    &   primary &   0.52    &   4   &   power-law   \\
\mr{\ux{39}{Ca}(e^{-},\nu_{e})\ux{39}{K}}   &   2nd arc dip / post-2nd dip track    &   primary &   0.52    &   4   &   power-law   \\
\mr{\ux{30}{S}(e^{-},\nu_{e})\ux{30}{P}}    &   flow transfer to symmetric nuclei   &   primary &   0.52    &   1   &   both    \\
\mr{\ux{57}{Cu}(e^{-},\nu_{e})\ux{57}{Ni}}  &   flow transfer to symmetric nuclei   &   primary &   0.52    &   1   &   both    \\
\mr{\ux{34}{Ar}(e^{-},\nu_{e})\ux{34}{Cl}}  &   flow transfer to symmetric nuclei   &   primary &   0.52    &   1   &   both    \\
\mr{\ux{42}{Ti}(e^{-},\nu_{e})\ux{42}{Sc}}  &   flow transfer to symmetric nuclei   &   primary &   0.52    &   1   &   both    \\
\mr{\ux{44}{V}(e^{-},\nu_{e})\ux{44}{Ti}}   &   flow transfer to symmetric nuclei   &   primary &   0.52    &   1   &   exponential \\
\mr{\ux{45}{Cr}(e^{-},\nu_{e})\ux{45}{V}}   &   flow transfer to symmetric nuclei   &   primary &   0.52    &   1   &   power-law   \\
\mr{\ux{55}{Ni}(e^{-},\nu_{e})\ux{55}{Co}}  &   flow transfer to symmetric nuclei   &   secondary   &   0.52    &   1   &   power-law   \\
\mr{\ux{58}{Zn}(e^{-},\nu_{e})\ux{58}{Cu}}  &   flow transfer to symmetric nuclei   &   secondary   &   0.52    &   1   &   both    \\
\mr{\ux{43}{Ti}(e^{-},\nu_{e})\ux{43}{Sc}}  &   flow transfer to symmetric nuclei   &   secondary   &   0.52    &   1   &   power-law   \\
\mr{\ux{41}{Ti}(e^{-},\nu_{e})\ux{41}{Sc}}  &   flow transfer to symmetric nuclei   &   secondary   &   0.52    &   1   &   power-law   \\
\mr{\ux{45}{V}(e^{-},\nu_{e})\ux{45}{Ti}}   &   flow transfer to symmetric nuclei   &   secondary   &   0.52    &   1   &   power-law   \\
\mr{\ux{37}{Ca}(e^{-},\nu_{e})\ux{37}{K}}   &   flow transfer to symmetric nuclei   &   secondary   &   0.52    &   1   &   power-law   \\
\mr{\ux{37}{Ca}(e^{-},\nu_{e})\ux{37}{K}}   &   2nd arc dip / post-2nd dip track    &   secondary   &   0.52    &   4   &   power-law   \\
\mr{\ux{34}{Ar}(e^{-},\nu_{e})\ux{34}{Cl}}  &   2nd arc dip / post-2nd dip track    &   secondary   &   0.52    &   4   &   power-law   \\
\mr{\ux{57}{Cu}(e^{-},\nu_{e})\ux{57}{Ni}}  &   2nd arc dip / post-2nd dip track    &   secondary   &   0.52    &   4   &   power-law   \\
\mr{\ux{34}{Ar}(e^{-},\nu_{e})\ux{34}{Cl}}  &   3rd arc amplitude   &   secondary   &   0.52    &   4   &   power-law   \\
\mr{\ux{38}{Ca}(e^{-},\nu_{e})\ux{38}{K}}   &   3rd arc amplitude   &   secondary   &   0.52    &   4   &   power-law   \\
\mr{\ux{39}{Ca}(e^{-},\nu_{e})\ux{39}{K}}   &   3rd arc amplitude   &   secondary   &   0.52    &   4   &   power-law   \\
\mr{\ux{42}{Ti}(e^{-},\nu_{e})\ux{42}{Sc}}  &   3rd arc amplitude   &   secondary   &   0.52    &   4   &   power-law   \\
\mr{\ux{57}{Cu}(e^{-},\nu_{e})\ux{57}{Ni}}  &   3rd arc amplitude   &   secondary   &   0.52    &   4   &   power-law   \\
\hline
\multicolumn{6}{c}{\mr{(p,n)} and \mr{(\alpha,n)} reactions}\\
\hline
\mr{\ux{57}{Co}(p,n)\ux{57}{Ni}}    &   chasm widening  &   secondary   &   0.5 &   2   &   both    \\
\mr{\ux{56}{Co}(p,n)\ux{56}{Ni}}    &   chasm widening  &   secondary   &   0.5 &   2   &   both    \\
\mr{\ux{27}{Al}(p,n)\ux{27}{Si}}    &   chasm widening  &   secondary   &   0.5 &   2   &   both    \\
\mr{\ux{11}{B}(p,n)\ux{11}{C}}  &   chasm widening  &   secondary   &   0.5 &   2   &   both    \\
\mr{\ux{10}{B}(\alpha,n)\ux{13}{N}} &   chasm widening  &   secondary   &   0.5 &   2   &   both    \\
\mr{\ux{11}{B}(\alpha,n)\ux{14}{N}} &   chasm widening  &   secondary   &   0.5 &   2   &   both    \\
\mr{\ux{20}{Ne}(\alpha,n)\ux{23}{Mg}}   &   chasm widening  &   secondary   &   0.5 &   2   &   both    \\
\mr{\ux{9}{Be}(\alpha,n)\ux{12}{C}} &   chasm widening  &   secondary   &   0.5 &   2   &   both    \\
\mr{\ux{42}{Ca}(\alpha,n)\ux{45}{Ti}}   &   1st arc dip / 2nd arc slope &   secondary   &   0.48    &   3   &   both    \\
\mr{\ux{34}{S}(\alpha,n)\ux{37}{Ar}}    &   2nd arc amplitude   &   secondary   &   0.48    &   3   &   both    \\
\enddata
\label{tab:nuclear_reactions}
\end{deluxetable}

\clearpage

\clearpage

\clearpage

\clearpage

\clearpage

\clearpage

\clearpage

\clearpage

\clearpage

\clearpage

\clearpage

\clearpage

\clearpage

\clearpage

\clearpage

\clearpage

\clearpage

\clearpage

\clearpage

\clearpage

\end{document}